\newif\ifanon

\newif\ifsubmission

\newif\ifconference

\PassOptionsToPackage{hyphens,spaces,obeyspaces}{url}

\documentclass[10pt,a4paper]{article}
\usepackage[utf8]{inputenc}

\usepackage{geometry}
\usepackage{amsthm}
\newtheorem{theorem}{Theorem}
\newtheorem{lemma}{Lemma}
\newtheorem{corollary}{Corollary}
\newtheorem{definition}{Definition}

\usepackage{amsmath}
\usepackage{amssymb}
\usepackage{xcolor}
\usepackage{multirow}
\usepackage[normalem]{ulem}

\usepackage[colorlinks=true,citecolor=black,linkcolor=black,urlcolor=black,bookmarksopen=true]{hyperref}

\theoremstyle{remark}

\usepackage{xifthen}
\usepackage{paralist}
\usepackage[capitalise,nameinlink,noabbrev,compress]{cleveref}

\newcommand{\TZ}[1]{\textcolor{magenta}{\textbf{Thomas: }{#1}}}

\newcommand{\revision}[1]{{#1}}
\newcommand{\ignore}[1]{}
\newcommand{\mc}{\mathcal}
\newcommand{\msf}{\mathsf}

\newcommand{\eg}{e.g., }
\newcommand{\ie}{i.e., }
\newcommand{\wrt}{w.r.t. }

\newcommand{\st}{s.t. }

\newcommand{\secparam}{\kappa}
\newcommand{\negl}{\msf{negl}}

\newcommand{\env}{\mc{Z}}
\newcommand{\totalParties}{n}
\newcommand{\party}{\mc{P}}
\newcommand{\partySet}{\mathbb{P}}
\newcommand{\observer}{\Omega}
\newcommand{\rounds}{N}
\newcommand{\leader}{\Lambda}
\newcommand{\committee}{\mathbb{C}}

\newcommand{\router}{\mc{R}}

\newcommand{\adversary}{\mc{A}}

\newcommand{\infractionPartySet}{\mathbb{A}}

\newcommand{\power}{\mu}

\newcommand{\chain}{C}
\newcommand{\block}{B}
\newcommand{\tx}{\msf{tx}}
\newcommand{\ledger}{L}
\newcommand{\hashPointer}{s}
\newcommand{\genesis}{\block_G}

\newcommand{\livenessParam}{u}
\newcommand{\threshold}{\alpha}

\newcommand{\prefix}{\preceq}

\newcommand{\msgNumber}{M}

\newcommand{\execution}{\mc{E}}

\newcommand{\trace}{\Im}
\newcommand{\proto}{\Pi}
\newcommand{\strategy}{S}
\newcommand{\strategySet}{\mathbb{S}}
\newcommand{\profile}{\sigma}

\newcommand{\equilibriaSet}{\mathbb{EQ}}

\newcommand{\messageValidityPredicate}{V}
\newcommand{\infractionPredicate}{\msf{IP}}
\newcommand{\compliant}{\infractionPredicate\msf{-compliant}}

\newcommand{\slot}{r}
\newcommand{\round}{\slot}

\newcommand{\utility}{U}
\newcommand{\reward}{R}

\newcommand{\welfare}{\msf{Welf}}
\newcommand{\bribe}{\beta}
\newcommand{\budget}{\msf{B}}
\newcommand{\deposit}{G}
\newcommand{\stake}{\mc{S}}

\newcommand{\exchangeRate}{X}
\newcommand{\utilityBoost}{B}

\newcommand{\electionFunc}{f_\msf{elect}}

\newcommand{\pubkey}{\msf{pk}}

\newcommand{\priceStab}{\msf{PSt}}
\newcommand{\priceAnar}{\msf{PAn}}

\newcommand{\rateMax}{x_\msf{max}}
\newcommand{\rateMin}{x_\msf{min}}
\newcommand{\rateDiff}{x}

\usepackage{tikz}
\usepackage{mdframed}

\title{
    Blockchain Bribing Attacks and the Efficacy of Counterincentives
}

\ifanon
    \author{Anonymous}
\else
    \author{
        Dimitris Karakostas \\ University of Edinburgh \\ d.karakostas@ed.ac.uk
        \and
        Aggelos Kiayias \\ University of Edinburgh and IOG \\ akiayias@inf.ed.ac.uk
        \and
        Thomas Zacharias \\ University of Glasgow \\ thomas.zacharias@glasgow.ac.uk
    }
\fi

\begin{document}
\date{}
\maketitle

\begin{abstract}
    We analyze bribing attacks \revision{in Proof-of-Stake} distributed ledgers from a
    game theoretic perspective. In bribing attacks, an adversary offers
    participants a reward in exchange for instructing them how to
    behave, with the goal of attacking the protocol's properties.
    \revision{Specifically, our work focuses on adversaries that target blockchain safety.}
    We consider two types of bribing, depending on how the bribes are awarded:
    i) \emph{guided bribing}, where the bribe is given as long as the bribed party
    behaves as instructed;
    ii) \emph{effective bribing}, where bribes are conditional on the attack's
    success, \wrt well-defined metrics. 
    We analyze each type of attack in a game theoretic setting and identify
    relevant equilibria. 
    In guided bribing, we show that the protocol is not an
    equilibrium and then describe good equilibria, where the attack is
    unsuccessful, and a negative one, where all parties are bribed such that
    the attack succeeds. 
    In effective bribing, we show that both the protocol and the ``all bribed''
    setting are equilibria.
    Using the identified equilibria, we then compute bounds on the Prices of
    Stability and Anarchy.
    Our results indicate that additional mitigations are needed for guided bribing,
    so our analysis concludes with incentive-based mitigation
    techniques, namely slashing and dilution.
    Here, we present two positive results, that both render the protocol an
    equilibrium and achieve maximal welfare for all parties, and a negative
    result, wherein an attack becomes more plausible if it severely affects the
    ledger's token's market price.
\end{abstract}

\section{Introduction}\label{sec:introduction}

Bitcoin's~\cite{nakamoto2008bitcoin} paradigm for creating
decentralized databases, which combined a hash chain and a
Proof-of-Work (PoW) mechanism, enabled a solution to the consensus problem in a
setting with open participation. In exchange for maintaining the database,
participants are awarded tokens that can be sold on the open market.

Bitcoin's protocol has been explored and expanded with a plethora of research
ideas and implementations, forming an ecosystem of distributed ledger systems.
\revision{Among them, Proof-of-Stake (PoS) systems feature prominently, mainly due to
their low environmental cost~\cite{wendl2023environmental}. As opposed to PoW,
where participation requires consuming energy to solve moderately hard
puzzles, PoS relies on the amount of tokens that a party has. Therefore, to
participate in a PoS ledger, a party needs only to acquire and hold the
system's tokens for a time period.}

Typically, in order to argue about a ledger's security, researchers take two
approaches.

The ``cryptographic'' approach assumes that a fraction of parties
is \emph{honest} and follows the prescribed protocol faithfully, while the rest
are \emph{Byzantine} and can deviate arbitrarily. These works show
that Bitcoin-like ledgers are secure under honest
majority~\cite{C:GarKiaLeo17,EC:PasSeeShe17}.

The ``game-theoretic'' approach assumes that all parties are \emph{rational},
aiming to maximize a well-defined utility function. Because distributed ledgers
typically operate in a completely open manner, without any restriction or
validation of the participants' identities, game theoretic guarantees are
particularly useful. As such, an active line of research tries to prove that
ledgers, such as Bitcoin, are game theoretically compatible, \eg being Nash
Equilibria~\cite{FC:EyaSir14,FC:SBBR16,DBLP:conf/sigecom/KiayiasKKT16}.

A well-known, incentive-oriented, threat to blockchain systems is
\emph{bribery} attacks~\cite{FCW:Bonneau16a}. Here, an adversary $\adversary$
wants to disrupt the distributed ledger
via a bribing budget, which is allocated at will as bribes to protocol participants. When a party
$\party$ accepts a bribe, $\adversary$ instructs $\party$ to deviate from the
ledger protocol in a well-defined manner, \ie perform an
infraction.\footnote{The details of the deviation are known to $\party$
before choosing to get bribed.} If enough parties
accept the bribe, $\adversary$ can successfully violate one of the ledger's
properties. \revision{Moreover, if the attack is visible and the market treats it as severe,
then it can result in a reduction of the market price
(exchange rate) of the ledger's native token.} So, whether $\party$ accepts the
bribe depends on its rationality: if $\party$'s utility \revision{taking everything into account} increases via bribing,
compared to acting honestly, the rational choice for $\party$ is to accept the
bribe.

\revision{
Our work focuses on attacks against blockchain safety
(cf.~\cref{subsec:ledger-security-properties} for definitions). Such attacks are particularly
devastating, since they violate one of the two core ledger security properties
and enable real world attacks such as double spending, that has resulted in millions
in losses~\cite{etc-attack,zencash-attack}.
Safety is also notably vulnerable during a system's bootstrapping
period~\cite{rana2022optimal}, when it has not yet attracted enough parties to
safeguard it against malicious takeovers.  During this period,
the participants of another ledger may attempt to attack it in order to
discredit it and claim its market share, as was observed during the infamous
Bitcoin Cash split~\cite{bch-hash-war}.
Therefore, an adversary could be motivated to break safety, and use bribing as
a means to succeed, for various  reasons.
}

Several works demonstrate bribing
mechanisms~\cite{judmayer2019pay,FCW:LiaKat17,USENIX:LVTS17,FCW:McCHicMei18,FCW:CXGSLLS18,EPRINT:KhaNadWat20,FCW:VelTeuLuu17,FC:TeuJaiSax16}.
In practice, identifying whether a bribery occurred is
difficult, especially if payouts occur off-chain. Therefore, showcasing that an
incentive mechanism is resilient to bribing is a powerful tool for designing
robust protocols. \revision{This becomes more pressing in the case of PoS, where
the system does not rely on external hardware assumptions, but instead depends
on how its tokens are distributed among parties.}
 
Because some protocol deviations are identifiable, a vulnerable protocol could be enhanced with defenses that
disincentivize bribing that leads to such identifiable attacks. One such
proposed mitigation \revision{in PoS ledgers} is to hold parties \emph{financially accountable}.
Specifically, parties are required to lock assets in
escrow before participating in the ledger protocol. If a party
misbehaves in an accountable manner, then its deposit gets \emph{slashed},
meaning it is confiscated. Otherwise, the party can withdraw its deposit at the
end of its participation. An alternative approach is to \emph{dilute} a misbehaving
party's assets by excluding it from receiving newly-issued tokens, thus
avoiding the (technical and legal) complications of locking assets. The latter approach is
taken by Cardano~\cite{cardano-dilution} to disincentivize abstaining, whereas
Cosmos employs a combination of slashing and
dilution~\cite{cosmos-dilution,cosmos-slashing}.
Nonetheless, both techniques have drawbacks, as slashing requires
large amounts of assets to be locked, while dilution reduces
the asset value of regular, non-participating users.
Therefore, identifying the exact cases when introducing a defense
and analyzing how effective, and under which conditions, such defenses are, is
vital for balancing the various tradeoffs.

\subsection*{Analysis Outline and Contributions}

Our work explores the efficiency of bribing attacks and defense mechanisms \revision{in PoS systems} from
a game-theoretic perspective. Note that balancing between the benefits of
accepting a bribe and the damage caused by suffering a penalty inherently
presumes rationality of the protocol participant, which renders a
game-theoretic model a natural choice. Our analysis identifies both good and
bad equilibria, depending on whether bribing leads to an attack or not. As a
result, we identify when a protocol can handle bribing on its own and when a
defense mechanism is needed, and what guarantees such defense provides.

In more detail, our approach is as follows.

First, we define the formal model of our setting (Section~\ref{sec:bribing}).
We consider the relevant infraction predicate,
which captures the behavior of the parties that get bribed
and participate in the attack, \eg by double
signing to create conflicting ledgers.
We also define the exchange rate's behavior \wrt a bribing
attack. We assume that the exchange rate remains constant, if no attack is
successful, but is reduced if a successful attack is observed.
This behavior aims to capture the market's response to such an attack.
Specifically, a bribing attack can violate one of the system's core properties,
resulting in users losing their trust in the system, or even the system
becoming unusable.  Consequently, it is expected that the market price of the
system's token would be decreased, as its usability evaporates.

Next, we define and analyze two modes of bribing.
In the \emph{guided} mode (Section~\ref{sec:bribing_guided}), the adversary pays
the bribes only as long as a party performs an infraction, regardless if the attack is
successful.
In the \emph{effective} mode (Section~\ref{sec:bribing_effective}), the bribes
are paid if the attack is successful, \ie if the exchange rate is reduced.
In both cases, the adversary instructs the bribed parties to act in a specific
manner.

First, we explore whether the extreme scenarios, where all parties either
follow the protocol or accept the bribes, are Nash Equilibria or compliant
strategies (cf. Section~\ref{sec:preliminaries}). In guided bribing, the
protocol is not an equilibrium, whereas the ``all bribed'' setting is. In
effective bribing, both of these extreme scenarios are equilibria.

Second, we explore mixed scenarios, where some parties are bribed and others
behave honestly. In guided bribing, we show a family of equilibria that behave
as such and are all positive, in the sense that an attack is
unsuccessful. In effective bribing, no such mixed scenarios were identified.
 
Next, we analyze the aggregate utility of all
parties (``welfare''). For each setting, we identify bounds of welfare across
equilibria and then evaluate the Prices of Stability and 
Anarchy (cf. Section~\ref{subsec:welfare}).

Finally, Section~\ref{sec:accountable} explores counterincentives for
guided bribing. We identify two positive equilibria, each applicable to a
different set of realistic assumptions, and show that both yield
maximal welfare to parties. This result suggests that countermeasures (slashing or dilution)
can help deterring  guided bribing attacks and help a
protocol become an equilibrium. Nonetheless, we also present a negative
equilibrium, where all parties are incentivized to accept a bribe. Notably, the
ability of an attacker to maintain this equilibrium depends on the attack's
effect on the system's exchange rate: the more destructive an
attack on the exchange rate, the lower the cost for the attacker to maintain the negative equilibrium becomes.

\ifconference
Due to space restrictions, we refer to the paper's full version for all
proofs~\cite{karakostas2024blockchain}.
\else
To make reading more effortless, we refer to Appendix~\ref{app:proofs} for all
proofs.
\fi

\section{Related work}\label{sec:related}

\subsubsection*{Bribing}

Bribing attacks were first described in a peer-reviewed work by
Bonneau~\cite{FCW:Bonneau16a}, who offered various bribing mechanisms
and discussed potential implications. At the same time, it was shown
that such attacks can be exacerbated when miners are paid only via transaction
fees and not fixed block rewards~\cite{CCS:CKWN16}.  
The first practical example was implemented on
Ethereum~\cite{FCW:McCHicMei18}, via smart contracts that enable
bribing both ``in-bound'', \ie denominated in the token of the attacked
system, and ``out-of-band'', \ie using a separate cryptocurrency system;
interestingly, this payout scheme enables both bribing modes explored in our work
(guided and effective).
Followup works increased the efficiency of bribing attacks, by
enabling double-spend collusion across different systems in a trustless
manner~\cite{judmayer2019pay}, or utilized smart contracts for
payouts~\cite{FCW:CXGSLLS18}.
Such works are complementary to ours, since we analyze player dynamics assuming
the existence of a bribing mechanism.

For a detailed research overview of PoW bribing attacks
we refer to~\cite{EPRINT:JSZTEGMW20}. Although we
focus on Proof-of-Stake (PoS) systems, \cite{EPRINT:JSZTEGMW20} offers a broad understanding of
incentive manipulation and presents motivations for bribing attacks, \eg
double-spending~\cite{FCW:Bonneau16a,liao2017incentivizing,ebrahimpour2021analysis},
breaking consensus~\cite{FCW:Bonneau18,kroll2013economics,FCW:LiaKat17},
censorship~\cite{EPRINT:WinHerFau19}, frontrunning~\cite{SP:DGKLZBBJ20}, or breaking Hashed Time Lock Contracts~\cite{EPRINT:KhaNadWat20}.

\revision{
We note that our work focuses on bribing as a means to perform safety
attacks. Although the literature on blockchain bribing primarily considers
arguably less destructive attacks, like transaction ordering manipulation or
temporary censorship, there do exist various works that explore bribing for
safety
violations~\cite{FC:TeuJaiSax16,FCW:LiaKat17,FCW:McCHicMei18}.}

More recently, a model for bribery attacks was proposed
in~\cite{ESORICS:SunRuaSu20}. This work explores attacks against longest-chain
blockchains, like Bitcoin, and assumes some miners behave honestly, whereas
others act rationally, accepting the bribe if it increases their (short-term)
utility. In addition, this work assumes in-bound bribes, \ie which are paid via
on-chain transactions in the attacked system's native tokens. In contrast, our
model is more general, since we assume that:
\begin{inparaenum}[i)]
    \item all parties act rationally, without requiring some to always act honestly;
    \item bribes are paid out-of-band, \eg in USD, thus are not affected by
        fluctuations on the blockchain's token price (which might occur as a
        result of the bribe).
\end{inparaenum}

\subsubsection*{Accountability and Slashing}

The existence of protocol violations and attacks, such as bribing, has
highlighted the importance of \emph{accountability}: the ability to identify,
and possibly penalize, offending parties. Accountability has been a topic of
discussion in distributed systems for many
years~\cite{haeberlen2007peerreview,haeberlen2009fault}. 

In the context of BFT protocols and blockchains, a seminal
work~\cite{CCS:SWNKV21} formalized the guarantees of various protocols in
terms of forensic support, \ie identifying malicious parties in a distributed
manner. Previously, accountability was notably discussed in Casper
FFG~\cite{buterin2017casper}, Ethereum's PoS proposed protocol.
Gasper~\cite{buterin2020combining}, a Casper variant, arguably achieved
accountable safety under a $\frac{1}{3}$ adversarial
power~\cite{casper-incentives}. However, a liveness attack against Gasper was
later identified~\cite{SP:NeuTasTse21}. That work highlighted
the availability-accountability dilemma~\cite{EPRINT:NeuTasTse21} where, in an
environment with dynamic participation, no protocol can be both accountably
safe and live.
Other works explored accountability for alternative
blockchain systems, such as rollups~\cite{tas2022accountable} and permissioned
protocols~\cite{graf2020accountability}.

Finally, Kannan and Deb~\cite{kannan23slashing} outline a slashing analysis that
considers bribing. They argue that relying on
the market's response and a reduction of the token's price is insufficient
and, in the absence of such response, all parties are incentivized to accept
bribes. Our work expands this idea, by defining a formal model and
conducting a rigorous game theoretic analysis. Interestingly, our results
agree partially to the intuitive arguments of that work. Specifically, we show
that if all parties are bribed, then indeed a party is not incentivized to
unilaterally be honest. However, we also show that, under effective bribing or
under slashing, the protocol is an equilibrium, so no party is unilaterally
incentivized to get bribed, if all others are honest. Finally, the results of
Section~\ref{sec:accountable} also give support to the idea that slashing is a
helpful mitigation to (guided) bribing attacks.

\section{Execution Framework and Preliminaries}\label{sec:preliminaries}

We assume a distributed protocol $\proto$, which is executed by a set of
parties $\partySet$ over a number of time slots. We consider that all parties are rational, i.e., they try to maximize their
utility.
Every party $\party \in \partySet$ is activated on each time slot, following a
schedule set by an environment $\env$, which also provides the 
inputs.
Each party $\party$ is associated with a number
$\power_{\party} \in [0, 1]$. $\power_{\party}$ identifies
$\party$'s \emph{percentage} of protocol participation power, \eg
hashing or staking power.
Note that we focus on PoS systems, so the
distribution of participation power (stake) is 
public.\footnote{Privacy-preserving PoS protocols, like Ouroboros Crypsinous~\cite{SP:KKKZ19}, assume a private stake distribution, but remain theoretical, whereas our focus is on real-world systems.}
We use the following notation:
\begin{inparaenum}[i)]
    \item $\secparam$: $\proto$'s security parameter;
    \item $\mathsf{negl}(\cdot)$: a negligible function (asymptotically smaller than the inverse of any polynomial);
    \item $[n]$: the set $\{1, \ldots, n\}$;
    \item $E[X]$: expectation of a random variable $X$.
\end{inparaenum}

\subsection{Execution}\label{subsec:basic}

The execution is divided in time slots of a specific length. Before the
execution starts, each party selects a strategy from the strategy space
$\strategySet$, that consists of all possible efficient
algorithms.\footnote{Similar static models are considered
in~\cite{KiayiasS21,karakostas2022compliance,schwarzschilling2023time}.}

We assume a
peer-to-peer network, where parties communicate using a variant of
a \emph{diffuse} functionality~\cite{EC:GarKiaLeo15}. We model the network via a special party, the \emph{router} $\router$.
$\router$ receives every message when it is created and is responsible for
delivering it to its recipient. The order of delivery can vary,
depending on the real-world network's implementation.\footnote{For instance, a
realistic router could deliver messages from distant geographic regions or
sources with low bandwidth after messages from nearby well-connected sources.} 
In our setting, we assume that the adversary $\adversary$, \ie
the party that offers bribes to participants, controls $\router$.

The information that a party $\party$ possesses about the state and action of
another party $\party'$ is determined by the data obtained from the
diffuse functionality throughout the execution.

A protocol's execution $\execution_{\env, \router, \profile, \slot}$ until a given time slot
$\slot$ is probabilistic and parameterized by:
\begin{inparaenum}[i)]
    \item the environment $\env$;
    \item a router $\router$;
    \item the strategy profile $\profile$ of the participating parties.
\end{inparaenum}
As discussed, $\env$
provides the parties with inputs and schedules their activation. For notation
simplicity, when $\slot$ is omitted, $\execution_{\env, \router, \profile}$ refers to the
end of the execution, which occurs after polynomially many time slots.

An \emph{execution trace} $\trace_{\env, \router, \profile, \slot}$ until a time slot
$\slot$ is the value that the random variable $\execution_{\env, \router, \profile, \slot}$ takes for a fixed:
\begin{inparaenum}[i)]
    \item environment $\env$;
    \item router $\router$;
    \item strategy profile $\profile$;
    \item random coins of $\env$, each party $\party \in \partySet$, and every protocol-specific oracle (see below).
\end{inparaenum}
A party $\party$'s view of an
execution trace $\trace_{\env, \router, \profile, \slot}^{\party}$ comprises
the messages $\party$ has sent and received until slot $\slot$. For
notation simplicity, we omit the subscripts $\{ \env, \router,
\profile, \slot \}$ from $\execution$ and $\trace$, unless
needed for clarity.

\subsection{Rewards and Utilities}\label{subsec:preliminaries_utility}

Let $\execution$ be an execution scheduled by the environment $\env$
with router $\router$. Parties follow a strategy profile $\profile$. Let
the following random variables over execution traces \wrt $\env$,
$\router$, $\profile$:
\begin{itemize}
    \setlength\itemsep{1pt}
    \item \revision{$\stake_{\party, \execution_{\env, \router, \profile}}$: \emph{stake} of party $\party$}
    \item $\reward_{\party, \execution_{\env, \router, \profile}}$: \emph{ledger rewards} of party $\party$
    \item $\exchangeRate_{\execution_{\env, \router, \profile}}$: \emph{exchange rate} between ledger's token and USD
    \item $\utilityBoost_{\party, \execution_{\env, \router, \profile}}$: \emph{external rewards} of party $\party$
    \item $\deposit_{\party, \execution_{\env, \router, \profile}}$: \emph{compliance payout} of party $\party$
\end{itemize}

In our setting, the ledger rewards $\reward_{\party, \execution_{\env, \router, \profile}}$ are the tokens that a party
receives for participation in the protocol. The exchange rate $\exchangeRate_{\execution_{\env, \router, \profile}}$
reflects the (USD) market price of the ledger's token.
The external rewards $\utilityBoost_{\party, \execution_{\env, \router, \profile}}$ are bribes, given by the adversary
to parties under the condition of performing certain infractions.
The compliance payout $\deposit_{\party, \execution_{\env, \router, \profile}}$ corresponds to the return of
$\party$'s deposit, if the system requires the locking of tokens to enable
participation, or the issuing of dilution rewards to $\party$.
\revision{Finally, stake $\stake_{\party, \execution_{\env, \router, \profile}}$ is the amount of tokens that a party owns
throughout the execution, in order to use the distributed ledger's
applications. For simplicity, in the rest of this work we assume that each
party's stake is proportional to their power, \st $\stake_{\party, \execution_{\env, \router, \profile}} =
\power_\party \cdot \stake$, where $\stake$ is the total stake in the
system.\footnote{This assumption presents two implications. First, it implies
that no stake shifts occur throughout the execution. This can be relaxed by
exploring typical user practices and considering their expected amount of stake
at the end of the execution. Second, it implies that each party's power is
proportional to their ``skin in the game''. This assumption approximates
systems without stake delegation~\cite{SCN:KarKiaLar20}, but it can also be
relaxed by considering stake distributions which are alternative functions of
the power. Such work would possibly not affect the qualitative results of this
paper, but could result in different bounds in some of the results.}
We note that stake concerns tokens that are free to use, that is
they are not locked, as is the case for $\deposit$.}

Throughout this work, we consider two utility functions:
\begin{enumerate}
    \item \emph{Reward}: $\utility_{\party}(\profile) = \revision{ E[\stake_{\party, \execution_{\env, \router, \profile}} \cdot \exchangeRate_{\execution_{\env, \router, \profile}}] } + E[\reward_{\party, \execution_{\env, \router, \profile}} \cdot \exchangeRate_{\execution_{\env, \router, \profile}}]+E[\utilityBoost_{\party, \execution_{\env, \router, \profile}}]$
    \item \emph{Accountable Reward}: $\utility_{\party}(\profile) = \revision{E[\stake_{\party, \execution_{\env, \router, \profile}} \cdot \exchangeRate_{\execution_{\env, \router, \profile}}] } + $\\
        $E[\reward_{\party, \execution_{\env, \router, \profile}} \cdot \exchangeRate_{\execution_{\env, \router, \profile}}]+E[\utilityBoost_{\party, \execution_{\env, \router, \profile}}]+E[\deposit_{\party, \execution_{\env, \router, \profile}}\cdot\exchangeRate_{\execution_{\env, \router, \profile}}]$
\end{enumerate}
\noindent For simplicity, we will use the notation \revision{$\stake_\party$}, $\reward_{\party,
\profile}$, $\exchangeRate_\profile$,  $\utilityBoost_{\party,\profile}$,
$\deposit_{\party, \profile}$, when $\execution$, $\env$, and $\router$ are
implicit.


\subsection{Distributed Ledger}

Next we describe an abstraction of a distributed ledger protocol $\proto$,
with the goal of capturing the core elements of real-world
protocols like Algorand~\cite{EPRINT:CGMV18}, HotStuff~\cite{PODC:YMRGA19},
and Tendermint~\cite{buchman2016tendermint}.

\emph{Blocks and chains}
The protocol-related messages are blocks. A block $\block$ contains
transactions with 
application-specific information and a hash-based pointer $\hashPointer$ to another block.
Therefore, blocks form a tree, where each block defines a single parent and
many blocks might define the same parent, \ie creating a fork. A single branch
of the tree is a chain $\chain$, with all chains starting from the same block,
$\genesis$. A ledger $\ledger$ is an array of transactions $[\tx_1, \tx_2,
\dots]$; the length of $\ledger$ (denoted by $|\ledger|$) is the number of
transactions in it. $\ledger$ can be extracted from a blockchain $\chain$ by
appending the transactions in $\chain$'s blocks, in the order they appear in
the blocks.

\emph{Validity predicate} 
The predicate $\messageValidityPredicate$ defines if a block can be
appended to a chain. Given chain $\chain$ and block $\block$,
$\messageValidityPredicate$ outputs $1$ if $\chain || \block$ 
(the concatenation of $\chain$ and $\block$) is acceptable \wrt the protocol's
application rules. An implementation of $\messageValidityPredicate$ should
ensure \eg that no double-spending is possible, no assets can be forged, etc.

\emph{Parties} 
Each party $\party$ that participates in $\proto$ is identifiable by a public
key $\pubkey_\party$ and has some protocol power $\power_\party$.\footnote{Depending on the implementation, $\power$ can represent
stake, in Proof-of-Stake systems, or hashing power, in Proof-of-Work.}
Each party keeps a local chain of blocks $\chain_\party$, which they update as
described below, and a set of (valid) transactions which are
not in $\chain_\party$ (``mempool'').

\emph{Leader and committee} 
All parties have access to function $\electionFunc(\cdot, \cdot)$ which,
given a chain $\chain$ and a time slot $\slot$, outputs:
\begin{inparaenum}[i)]
    \item $\leader$: the leader-party for $\slot$;
    \item $\committee$: a set of parties that form the committee for $\slot$.
\end{inparaenum}
For a given slot $\slot$, $\leader$ is eligible to construct a block
$\block$ and diffuse it to the network. Each committee member $\party \in
\committee$ retrieves the block and validates it \wrt their local chain using
$\messageValidityPredicate(\chain_\party, \block)$. If the predicate's output is
$1$, $\party$ signs $\block$ and diffuses it to the network (\ie sends it to
the router $\router$). If a party $\party'$ receives $\block$ and enough
committee signatures,\footnote{Specifically, the aggregate power of the parties
that issued the signatures should be above a protocol-defined threshold, which
is typically $\frac{2}{3}$ of the aggregate stake of all committee members.}
then $\party'$ appends $\block$ to its local chain.

\noindent\emph{Remark.}
We assume that
a party participates in consensus at a frequency which is a linear function
of its (proportional) power. 
In practice, for the real world systems explored in this work, it holds that
$E(\msgNumber_\party) = \power_\party \cdot \rounds$, 
where $\msgNumber_\party$ is the number of protocol messages that $\party$ outputs and $\rounds$ is the total number of rounds.
This assignment method is widely used, since it satisfies
two useful properties: 
i) sybil proofness;
ii) collusion proofness~\cite{chen2019axiomatic}.
In particular, parties cannot participate at a higher rate by splitting their
power across multiple, seemingly independent entities (sybil behavior), or by forming
collective entities.

\ifsubmission
\else
\ifsubmission
\section{Distributed Ledger Security Properties}\label{subsec:ledger-security-properties}
\else
\paragraph{Security properties}\label{subsec:ledger-security-properties}
\fi
A distributed ledger should guarantee two core properties, safety and
liveness (cf. Definition~\ref{def:ledger-properties}). We denote by $\prefix$
the prefix operation, \ie if $\ledger \prefix \ledger'$ then $\forall
i \in [1, |\ledger|]: \tx_i = \tx'_i$.

\begin{definition}\label{def:ledger-properties}
    A secure distributed ledger protocol $\proto$, run by $\totalParties$ parties,
    should satisfy the following properties:
    \begin{itemize}
        \item \emph{Safety}: Let $\ledger_{\party_1, \round_1},
            \ledger_{\party_2, \round_2}$ be the ledgers output by parties
            $\party_1, \party_2$ on rounds $\round_1, \round_2$ respectively.
            The following condition must hold: $\forall \round_1, \round_2\; \forall
            \party_1, \party_2: \ledger_{\party_1, \round_1} \prefix
            \ledger_{\party_2, \round_2} \cup \ledger_{\party_2, \round_2}
            \prefix \ledger_{\party_1, \round_1}$.
        
        \item \emph{Liveness}: If a transaction $\tx$ is provided as input to
            an honest party on slot $\slot$, all honest parties output a ledger
            containing $\tx$ at slot $\slot + \livenessParam$, $\livenessParam$ being the ledger's liveness parameter.
    \end{itemize}
\end{definition}

Ledger protocols considered in this work typically satisfy safety as long
as $\frac{2}{3}$ of the participating power is controlled by parties that
follow $\proto$; similarly, for liveness the threshold is $\frac{1}{3}$. In the
rest of the paper we will denote the security threshold by $\threshold$ and set
$\threshold = \frac{1}{3}$, unless specified otherwise.

\fi

\emph{Ledger Attacks}
Our work focuses on safety infractions\ifsubmission~(cf. Appendix~\ref{subsec:ledger-security-properties})\fi, in
particular \emph{double signing}.
A party double signs if it proposes (when being a leader) or signs (when being
a committee member) two blocks, $\block_1, \block_2$, \st for
every chain $\chain$ that contains $\block_1$,
$\messageValidityPredicate(\chain, \block_2) = 0$ and vice versa, \ie every
chain that contains both $\block_1$ and $\block_2$ is invalid.
A double signing attack occurs if, and 
for a set of parties $\infractionPartySet \subseteq \partySet$ where
$\sum_{\party \in \infractionPartySet} \power_\party \geq \threshold$,
every party in $\partySet$ double signs on a slot $\slot$.
Note that double-signing might take different meanings, depending on the system's
internals. For example, in Ouroboros~\cite{C:KRDO17} double-signing occurs when
a slot leader creates two conflicting blocks that correspond to the same slot
and the same height in the chain. Instead, in Algorand~\cite{EPRINT:GHMVZ17},
double-signing means that a party, when elected as member of the committee,
votes for two conflicting blocks 
for the same slot.


\subsection{Block-proportional Rewards}\label{subsec:preliminaries_proportion}

We consider block-proportional rewards, with a fixed block reward $\reward$
as follows. At the end of the execution, a special party $\observer$,
called the \emph{observer}, outputs a chain $\chain_{\observer,\trace}$, where
$\trace$ is the execution trace. The observer models a passive user that is
always online, \eg a full node. Let $\msgNumber_{\party, \trace}$ be the number
of blocks created by party $\party$ in $\chain_{\observer, \trace}$. Then,
the ledger rewards of $\party$ are:
$\reward_{\party, \profile}(\trace) = \msgNumber_{\party, \trace} \cdot \reward$.\footnote{We
slightly abuse notation here, where $\reward$ with subscript $\party, \profile$
denotes the protocol-related reward of $\party$ under profile $\profile$,
whereas $\reward$ without any subscript denotes the fixed reward that each block yields, which is a protocol parameter.}

\subsection{Approximate Nash Equilibrium}\label{subsec:nash_equilibrium}

An approximate Nash equilibrium is a common tool for expressing a solution to a
non-cooperative game of $\totalParties$ parties $\party_1,\ldots,\party_n$. Each party $\party_i$
employs a strategy $\strategy_i$, a set of rules and actions
the party makes depending on what has happened in the game,
\ie it defines the part of the entire distributed protocol $\proto$ performed
by $\party_i$. There exists an ``honest'' strategy, defined by $\proto$, which
parties may employ; for ease of notation,
$\proto$ denotes both the distributed protocol and the honest strategy.
A \emph{strategy profile} is a vector of all players' strategies.
Each party $\party_i$ has a game \emph{utility} $\utility_i$, which is a real
function that takes as input a strategy profile. A strategy profile is an
$\epsilon$-Nash equilibrium when no party can increase its utility more than
$\epsilon$ by \emph{unilaterally} changing its strategy
(Definition~\ref{def:equilibrium}).

\begin{definition}\label{def:equilibrium}
    Let:
    \begin{inparaenum}[i)]
        \item $\epsilon$ be a non-negative real number;
        \item $\strategySet$ be the set of strategies a party may employ;
        \item $\profile^* = (\strategy^*_i, \strategy^*_{-i})$ be a strategy
            profile of $\partySet$, where $\strategy^*_i$ is the strategy
            followed by $\party_i$;
        \item $\strategy^*_{-i}$ denote the $\totalParties - 1$ strategies
            employed by all parties except $\party_i$. 
    \end{inparaenum}
    We say that $\profile^*$ is an \emph{$\epsilon$-Nash equilibrium} \wrt a
    utility vector $\bar{\utility} = \langle \utility_1, \ldots,
    \utility_\totalParties \rangle$ if:
    $\forall \party_i \in \partySet \; \forall \strategy_i \in \strategySet \setminus \{ \strategy^*_i \} : \utility_i(\strategy^*_i, \strategy^*_{-i}) \geq \utility_i(\strategy_i, \strategy^*_{-i}) - \epsilon$.
\end{definition}

For simplicity, when all parties have the same utility $U$, $\profile^*$ is an
$\epsilon$-Nash equilibrium \wrt $U$. Also, $\proto$ is an $\epsilon$-Nash equilibrium \wrt $U$ when $\profile_\proto=\langle\proto,\ldots,\proto\rangle$, \st all parties follow the honest strategy.

\subsection{Welfare and Prices of \{ Stability, Anarchy \}}\label{subsec:welfare}

An important consideration in our analysis is a system's \emph{welfare},
a measure of efficiency of a strategy profile.
We define the welfare of a strategy profile as the sum of (expected) utilities
of all parties under said profile:
$$
\welfare(\profile)=\sum_{\party\in\partySet}\utility_{\party}(\profile)\;,
$$
where $\utility_\party(\profile)$
is the utility of party $\party$ under strategy profile $\profile$.

Two relevant notions are \emph{Price of Stability} and \emph{Price of Anarchy}~\cite{koutsoupias1999worst}, denoted by $\priceStab$ and $\priceAnar$ respectively. 
$\priceStab$ expresses how
close an equilibrium's welfare is to the maximum achievable welfare. For approximate equilibria,
$\priceStab$ is defined as the ratio of the maximum welfare, across all
strategy profiles, over the maximum welfare achieved across all $\negl(\secparam)$-Nash equilibria:
$$
\priceStab = \frac{\max_{\profile \in \strategySet^\totalParties} \welfare(\profile)}{\max_{\profile \in \equilibriaSet} \welfare(\profile)}\;,
$$
where $\equilibriaSet$ is the set of all $\negl(\secparam)$-Nash equilibria.

$\priceAnar$ expresses  the ratio between the maximum achievable welfare and the 'worst' equilibrium (i.e., the one with the minimum welfare), as formalized below: 
$$
\priceAnar = \frac{\max_{\profile \in \strategySet^\totalParties} \welfare(\profile)}{\min_{\profile \in \equilibriaSet} \welfare(\profile)}\;.
$$

In this work, we consider a relaxation of the Price of Anarchy notion, by searching across equilibria where the parties's strategies are in some subset $\mathbb{T}\subset\strategySet$. In what we call \emph{Price of Anarchy \wrt $\mathbb{T}$}, denoted by $\mathbb{T}\text{-}\priceAnar$, we assume that there is some level of coordination among the parties, in the sense that they agree to follow only strategies that are included in $\mathbb{T}$. Formally,
$$
\mathbb{T}\text{-}\priceAnar = \frac{\max_{\profile \in \strategySet^\totalParties} \welfare(\profile)}{\min_{\profile \in \equilibriaSet\cap\mathbb{T}^\totalParties} \welfare(\profile)}\;.
$$
Naturally, unless $\equilibriaSet\cap\mathbb{T}^\totalParties=\emptyset$, $1\leq\mathbb{T}\text{-}\priceAnar\leq\priceAnar $.

\subsection{Compliance}\label{subsec:compliance}

Compliance~\cite{karakostas2022compliance} is another tool based on Nash
dynamics. The core notion in compliance is the \emph{infraction predicate}
$\infractionPredicate$, which abstracts a deviant behavior that the analysis
aims to capture. Conversely, $\infractionPredicate$ identifies the
\emph{compliant} strategies. Specifically, a strategy $\strategy$ is compliant
if for all execution traces when a party follows $\strategy$, $\infractionPredicate$
is $0$.

Using the infraction predicate, the notion of an
$(\epsilon, \infractionPredicate)$-compliant protocol $\proto$ is defined.
$\epsilon$ specifies the gain threshold after which a party
switches strategies. In particular, $\epsilon$
specifies when a strategy profile $\profile'$ is \emph{directly reachable}
from profile $\profile$, where $\profile'$ results from the
unilateral deviation of party $\party_i$ from $\profile$ and, 
by this deviation, the $\party_i$'s utility increases by more than $\epsilon$,
while $\profile'$ sets a \emph{best response} for $\party_i$.
Generally, $\profile'$ is \emph{reachable} from $\profile$, if $\profile'$ results
from a ``path'' of strategy profiles, starting from $\profile$, which are
sequentially related via direct reachability. Finally, the \emph{cone}
of a profile $\profile$ is defined as the set of all strategies that are
reachable from $\profile$, including $\profile$ itself.

Following these definitions, $\proto$ is $(\epsilon,
\infractionPredicate)$-compliant if the cone of the ``all honest''
profile $\profile_\proto$ contains only profiles consisting of $\compliant$
strategies. In a compliant protocol, parties may
(unilaterally) deviate from the honest strategy only in a compliant manner,
without violating the predicate $\infractionPredicate$.

\section{Formal Model of Bribing Attacks}\label{sec:bribing}

In this section, we describe the model, under which the following sections will
explore bribing attacks. Briefly, we assume that an adversary offers bribes to
parties in order to perform infractions that lead to particularly harmful
attacks. Following, we first define the network setting and then discuss the
relation between infractions and attacks and offer examples of both. Next, we
describe the behavior of the exchange rate, that is the ledger's token price,
in the presence of such attacks. Finally, we put restrictions on the adversary,
in terms of the amount of funds it can allocate for bribes.

Before we proceed, let us outline a list of assumptions that hold for
the protocols that we analyze. Assuming that all parties act honestly, \ie
follow the prescribed protocol, the following conditions should hold:
\begin{inparaenum}[i)]
    \item the block rewards that are given by the system as a whole are
        maximized;
    \item exactly one block is created for every execution slot;
    \item the random variable $\msgNumber_{\party,\profile_\proto}$, of ledger-related messages
        that a party $\party$ creates during the execution, \revision{respects the Chernoff concentration bounds. Namely, for any $\theta\in(0,1)$:}
\end{inparaenum}
\begin{align*}
\revision{\Pr\big[\msgNumber_{\party, \profile_\proto}\geq(1+\theta) \cdot E[\msgNumber_{\party, \profile_\proto}] \big]}&\revision{\leq e^{-\frac{\theta^2}{3}\cdot E[\msgNumber_{\party, \profile_\proto}] }}\\
\revision{\Pr\big[\msgNumber_{\party, \profile_\proto}\leq(1-\theta)\cdot E[\msgNumber_{\party, \profile_\proto}] \big]}&\revision{\leq e^{-\frac{\theta^2}{2}\cdot E[\msgNumber_{\party, \profile_\proto}] }}
\end{align*}

\subsection{Network}

In this work, we assume a \emph{synchronous} network. Specifically, the router
$\router_\text{sync}$ delivers all messages at the end of the round during
which they were created. 

We note that various protocols have been analyzed and shown secure in
the partially synchronous setting. Here, ``Global Stabilization
Time'' (GST) exists, unknown to the protocol's designers and participants.
Before GST, the router $\router_\text{psync}$ can introduce arbitrary delays to
all messages.  After GST, $\router_\text{psync}$ has to behave as the
synchronous router. In our setting, since $\adversary$ controls the router, it
can arbitrarily delay all messages before GST. Consequently, it can violate
liveness, by preventing all blocks before GST from being finalized. However,
this means that $\adversary$ can prevent parties from receiving rewards for the
slots before GST. Therefore, the expected rewards that a party receives, and
hence their utility, depends on the GST and the adversarial behavior. In
addition, because the GST is unknown to participants, it is impossible to
estimate even a lower bound on utility. As a result, a game theoretic analysis,
where parties compute and compare the expected utility of each available
strategy, is not possible assuming a partially synchronous network.

\subsection{Infraction Predicates}\label{subsec:bribe_IPs}

As described in Section~\ref{subsec:compliance}, we follow the model
of~\cite{karakostas2022compliance}. Briefly, a
deviant behavior is captured by an \emph{infraction predicate}
$\infractionPredicate$. For each party $\party$, this predicate is $0$ if, for
every execution, $\party$ follows the honest protocol.
$\infractionPredicate$ is $1$ if, at some execution trace, $\party$'s strategy
deviates from the protocol in a well-defined manner.

We define the strategy $\strategy_\infractionPredicate$ as the one where
$\party$ performs the infraction defined by $\infractionPredicate$, whenever
possible, and behaves honestly otherwise. The strategy profile where all
parties perform $\infractionPredicate$ is denoted by
$\profile_\infractionPredicate = \langle \strategy_\infractionPredicate, \ldots, \strategy_\infractionPredicate \rangle$.
Also $\infractionPredicate(\trace, \party)$ denotes the value of infraction predicate $\infractionPredicate$ \wrt $\party$ for an execution trace $\trace$.

In this work, each predicate is parameterized by a time frame $\Delta$. This
time frame defines when the party is expected to perform an infraction. We
assume that $\Delta$ is sufficiently large, \st if a party wants to perform the
infraction, it can do so with overwhelming probability. In other words, the
probability that a party cannot perform an infraction even though its strategy
instructs it to do so, is negligible in the security parameter $\secparam$.

Each predicate is associated with a threshold $\threshold$. This threshold
denotes the percentage of power that makes an attack against the ledger
feasible. In practice, $\threshold$ sets the power needed for a subset of parties
$\infractionPartySet$, \st if everyone in
$\infractionPartySet$ performs the infraction, an attack can occur with
overwhelming probability.

\ifsubmission
\else
\ifsubmission
\section{Predicate Examples}\label{subsec:predicate_examples}
\else
\subsubsection*{Predicate Examples}\label{subsec:predicate_examples}
\fi

The compliance model allows us to capture and analyze well-defined deviations.
For instance,
predicates \ref{eq:infraction_ds} and \ref{eq:infraction_censor} capture the
behavior that results in violations of safety (double signing) or
liveness (censorship) (cf. Section~\ref{sec:preliminaries}).

\begin{equation}\label{eq:infraction_ds}
    \infractionPredicate_\mathsf{ds}(\trace,\party) = 
    \left\{\begin{array}{ll} 
        1, & \mbox{if in $\trace$, } \exists \slot \in \Delta: \party \mbox{ double-signs in } \slot\\
        0, &\mbox{otherwise} 
    \end{array}\right.\;
\end{equation}

\begin{equation}\label{eq:infraction_censor}
    \infractionPredicate_\mathsf{censor}(\trace, \party) = 
    \left\{\begin{array}{ll} 
        1, & \mbox{if in $\trace$, } \forall \slot \in \Delta: \party \mbox{ censors } \\
        & \mbox{transaction } \tx\\
        0, &\mbox{otherwise} 
    \end{array}\right.\;
\end{equation}


In both cases, the predicate is parameterized by a value $\Delta$. This is the
time window within which the infraction takes place. In one case (double
signing), the party needs to act at least once in this window. In the other
(censorship), it needs to act for the whole duration of $\Delta$.


\fi

\subsubsection*{Attack success probability}
As stated above and shown in the examples\ifsubmission~of Appendix~\ref{subsec:predicate_examples}\fi, predicates are parameterized by a
time frame $\Delta$, within which the attack should conclude.
An attack occurs if, within $\Delta$ and for a duration of
rounds (which is an attack parameter), the aggregate power of parties that
perform an infraction is above the threshold $\threshold$
(Definition~\ref{def:attack-failure-event}).

\begin{definition}\label{def:attack-failure-event}
    Let $\infractionPredicate$ be an infraction predicate with threshold
    $\threshold$.
    Also let $\hat{\round}$ be a time slot and $\lambda',\lambda,\Delta$ be integers such that $\lambda' \leq \lambda \leq \Delta$.
    An \emph{attack associated with $\infractionPredicate$ occurs}, if the following event occurs:
    \begin{align*}
        \exists \infractionPartySet \subseteq \partySet \mbox{ such that } \sum_{\party \in \infractionPartySet} \power_\party \geq \threshold: \\
        \big[ \exists \round^* \geq \hat{\round}:
        \forall \round \in [\round^*, \round^* + \lambda): \sum_{\party \in \infractionPartySet} \power_{\committee, \party} \geq \threshold \big] \land \\
        \big[ \exists \round' \in [\round^*, \round^* + \lambda - \lambda'):
        \forall \round \in [\round', \round' + \lambda'): \leader_{\round} \in \infractionPartySet \big]
    \end{align*}
    where $\infractionPartySet$ is the set of parties that violate $\infractionPredicate$,
    $\power_{\committee, \party}$ is the power percentage of $\party$ within the elected committee $\committee$,
    and $\leader_\round$ is the leader of time slot $\round$.
\end{definition}

Note that $\lambda$ is the parameter of the security property which is
violated when the infraction is performed. For example, for safety attacks it holds
$\lambda = 1$, since it suffices to perform a successful double signing for one
round. For liveness attacks, $\lambda$ is the liveness parameter, \ie the
number of rounds that a party waits until finalizing a published transaction.

$\lambda'$ is the number of rounds for which the leader should be
adversarial. For example, for safety attacks $\lambda' = 1$. For liveness
attacks censorship is successful if
enough committee members refuse to sign blocks that contain the censored
transaction $\tx$, regardless of the leader's actions, so $\lambda' = 0$.

In the rest of the paper, we assume $\hat{\round}, \Delta, \lambda',
\lambda$ are chosen in such a manner that the event of
Definition~\ref{def:attack-failure-event} occurs with overwhelming probability.
Therefore, an attack is successful as long as enough parties, \ie with
aggregate power above the given threshold, are willing to perform the
infraction.

\subsection{Exchange Rate}

We now describe how the exchange rate behaves in the presence of attacks.
In detail, assume that the parties in $\partySet$ follow a strategy profile
$\profile$. Let $\infractionPredicate$ be an infraction predicate.
$\adversary$ instructs all parties that accept a bribe to violate
$\infractionPredicate$. Also let $\threshold \in (0,1)$ be the security
threshold, \ie the percentage of aggregate power that bribed parties need to
control, in order to perform an attack based on $\infractionPredicate$.

The exchange rate is defined as follows:
\begin{equation}\label{eq:exchange}
    \exchangeRate_\profile(\trace) = \left\{\begin{array}{ll} 
        \rateMax, &\mbox{if in $\trace$} : \sum_{\party \in \partySet: \infractionPredicate(\trace, \party) = 1} \power_\party < \threshold \\
        \rateMin, &\mbox{if in $\trace$} : \sum_{\party \in \partySet: \infractionPredicate(\trace, \party) = 1} \power_\party \geq \threshold 
    \end{array}\right.\;
\end{equation}
 
The values $\rateMax, \rateMin$ express the ``toxicity'' effect of the attack.
In particular, the difference $\rateDiff = \rateMax - \rateMin$ states the
level by which the exchange rate drops, in case of a successful attack.  In the
extreme case, when the attack is catastrophic for the system, the ledger's
token price goes to $0$ (\ie $\rateMin = 0$).

\revision{
We note that assuming that the market responds to an attack, \st the exchange
rate drops, is not always realistic. For example, if the attack is not visible
and provable, \eg in the case of censorship attacks~\cite{SP:TYME21}, the
exchange rate may remain unaffected. Similarly, even if the attack is visible, but the market does not deem it
as severe, $\rateDiff$ might be small.\footnote{Historical examples of
such attacks are described in~\cite[Section 7.3]{karakostas2022compliance}.}
In our work, such behavior of the exchange rate affects the results where the
bound depends on $\rateMin$. Typically, if $\rateMin$ is close to $\rateMax$,
then the attack may require smaller budget, since the attacker would not need
to compensate the bribed parties for potential losses due to the price drop.}

\noindent\emph{Remark.} Our work follows the approach where the
exchange rate takes one of two constant values, depending on the attackers' power
and $\infractionPredicate$. This enables a concrete
analysis of the studied protocol and provides evidence of its behavior against
said attack. In reality, the exchange rate may be influenced by various factors. However, to
argue concretely about an attack's efficacy in the general setting, some
level of simplification is necessary. We capture this by bounding the
impact on the exchange rate drop across all traces. Intuitively, 
$\rateMax$ and $\rateMin$ act as the
bounds of the exchange rate when the attack is infeasible ($\rateMax$) and when it is successful ($\rateMin$).

\subsection{Bribing Budget}

It is easy to see that, if the adversary $\adversary$ has unlimited funds for
bribes, then it can break the security of any system of rational
participants. Specifically, $\adversary$ would offer every party 
$\party$\footnote{More precisely, to enough parties that have an aggregate power above $\threshold$.}
a bribe which is $\epsilon$ larger than the expected rewards that the system
offers $\party$, when $\party$ acts honestly. In that case, $\party$ has an
incentive to accept the bribe and increase its utility by $\epsilon$.
Therefore, it is necessary to restrict the total amount of funds that $\adversary$
can offer in bribes.

We assume that the adversary $\adversary$ has a budget $\budget$
for bribes. In particular, let $\{ \bribe_\party \}_{\party \in \partySet}$
denote an allocation of bribes to all parties. It should hold that
$\sum_{\party \in \partySet} \bribe_\party = \budget$ and $\forall \party \in \partySet: \bribe_\party \geq 0$.
$\adversary$ allocates this budget arbitrarily,
without restriction on the number and the level of bribes 
offered to each party.

We note that bribes are paid in the base unit of account, \eg USD.
If a party accepts a bribe, the receivable amount does not depend on
exchange rate fluctuations, which might occur due to the attack's success.
The exact details of how bribing takes place and how the bribes
are distributed are outside the scope of this paper.

There are various candidate bounds for the bribing budget. One could be the
aggregate amount of rewards that the system gives to all parties.
Another could be the aggregate amount of rewards for
parties that collectively control more than $\threshold$ of power. In both
cases though, in various scenarios $\adversary$ can trivially allocate bribes
in a way that incentivizes enough parties to deviate.\footnote{$\adversary$
would give bribes equal to each party's expected rewards. If $\rateMin > 0$,
then each party would be incentivized to accept the bribe and also collect the
(reduced) rewards.}

Our work assumes a tighter bribing budget. For an attack which
results from an infraction $\infractionPredicate$, consider the following:
\begin{itemize}
    \item $\threshold \in [0, 1]$ is the threshold of power proportion that parties need to control, in order to
        perform $\infractionPredicate$ \st the attack succeeds;
    \item $\rateDiff > 0$ is the difference of the exchange rate, as a result
        of a successful attack;
    \item $\rounds$ is the total number of rounds;
    \item $\reward$ is the (average) reward given per round;
    \item $\stake$ is the total stake in the system.
\end{itemize}
$\adversary$'s bribing budget for this attack is:
$
\budget_\infractionPredicate \leq \threshold \cdot \rateDiff \cdot (\rounds \cdot \reward + \revision{\stake}).
$

Intuitively, the budget cannot exceed a percentage $\threshold$ of the value of rewards \revision{and stake}
that is lost as a result of the attack. Observe that, in order to offer
positive results for all reward allocations, this bound is tight. If we allowed
a larger budget, $\adversary$ could trivially perform an attack in the case
when rewards are allocated to each party exactly proportionally to their power.
Specifically, $\adversary$ would give to $\threshold$\% of parties a bribe
equal to their expected lost rewards \revision{and stake}, plus some $\epsilon$ value.

We note that this restriction is useful for particularly devastating attacks.
If the bribing budget is high, a positive result shows that the system can
withstand attacks by richer adversaries. However, if the budget is low, a
positive result is weaker, since it holds only for small (less wealthy)
adversaries. Therefore, 
if an attack is not particularly
destructive to the exchange rate, then real-world attackers can violate the
budget restriction assumption more easily.
This bound applies to the next two sections; in Section~\ref{sec:accountable}
we will revisit it to take into account counterincentives, such as slashing.

\section{Guided Bribing}\label{sec:bribing_guided}

In this section, we assume that a party receives its external rewards, \ie a
bribe, as long as it performs a specific infraction. The bribe
is issued regardless if the attack, which depends on the infraction, is
successful (\st the exchange rate drops to $\rateMin$).
Formally, for each party $\party \in \partySet$, the bribe $\bribe_\party$ is a
non-negative real value. For some infraction predicate $\infractionPredicate$,
the external rewards are defined as follows:

\begin{equation}\label{eq:external_guided}
    \utilityBoost_{\party,\profile}(\trace) = 
    \left\{
        \begin{array}{ll}
            \bribe_\party, & \mbox{if } \infractionPredicate(\trace,\party) = 1 \\
            0, & \mbox{otherwise}
        \end{array}
    \right.
\end{equation}

We call the scenario of Eq.~\eqref{eq:external_guided} \emph{guided bribing} \wrt $\infractionPredicate$.

\subsection{Equilibria and Compliance}\label{subsec:bribing_guided_equilibria}

The first outcome of our analysis is Theorem~\ref{thm:guided}\ifconference\else~(proof in Appendix~\ref{app:eq_comp})\fi, which offers a
variety of results. In particular, the theorem assumes guided bribing \wrt
$\infractionPredicate$, a synchronous network,
block-proportional rewards with $\reward$ rewards per block, and the
\emph{Reward} utility function (cf. Section~\ref{sec:preliminaries}).

First, we show that $\profile_\infractionPredicate$ is a
$\negl(\secparam)$-Nash equilibrium, if no party holds more than
$1-\threshold$ of the participation power. The main idea is
that, if all parties deviate, no single party can prevent the attack, so
each party's utility only increases by accepting the bribe.

Second, $\profile_\infractionPredicate$ is a $\negl(\secparam)$-Nash
equilibrium if a party $\party$ holds at least $\threshold$ of
the participation power and its bribe is sufficiently large. The idea is
similar to the first result. If $\party$ can unilaterally perform the attack
and is incentivized to do so, the other parties can only increase
their utility by getting bribed. In order for this condition to hold,
$\party$'s bribe should substantially exceed its expected reward loss, that is
the number of blocks that it creates multiplied by $\reward \cdot x$.

Third, the protocol $\proto$ is not $\infractionPredicate$-compliant, if there
exists a party $\party$ whose participation power is less than $\threshold$ and
its bribe is a positive value. The main idea here is that $\party$ cannot
perform the attack on its own. Therefore, assuming all other parties follow
$\proto$, $\party$ is incentivized to unilaterally deviate, in order to get
both the bribe and the rewards (multiplied by the unaffected rate $\rateMax$).



\begin{theorem}\label{thm:guided}
    Assume the following:
    \begin{itemize}
        \item $\proto$: a protocol with block-proportional rewards (cf.
            Section~\ref{subsec:preliminaries_proportion}), with $\reward$
            rewards per block;
        \item \revision{each party's stake is $\stake_\party = \power_\party \cdot \stake$;}
        \item $\profile_\proto$: the all-honest strategy profile;
        \item $\threshold$: the security threshold \wrt an infraction
            predicate $\infractionPredicate$;
        \item for every strategy profile $\profile$: 
            \begin{itemize}
                \item the exchange rate $\exchangeRate_\profile$ \wrt
                    $\infractionPredicate, \threshold$ follows
                    Eq.~\eqref{eq:exchange};
                \item the external rewards $\utilityBoost_{\party,\profile}$,
                    due to guided bribing \wrt $\infractionPredicate$,
                    follow Eq.~\eqref{eq:external_guided}.
            \end{itemize}
    \end{itemize}

    Then, the following hold:

    \begin{enumerate}
        \item\label{thm:guided_nash1} If for every party $\party \in
            \partySet$, it holds that $\power_\party \leq 1 - \threshold$, then
            exists $\epsilon$ negligible in $\secparam$ \st the strategy
            profile $\profile_\infractionPredicate$ is an $\epsilon$-Nash
            equilibrium \wrt Reward under $\router_\text{sync}$.
         
        \item\label{thm:guided_nash2} Assume there exists a party
            $\party \in \partySet$ \st $\power_\party\geq\threshold$. Let
            $\theta > 0$ be a real value \st the probability
            $\Pr \big[ \msgNumber_{\party, \profile_\proto}>(1+\theta) \cdot E[\msgNumber_{\party, \profile_\proto}] \big]$
            is negligible in $\secparam$, where $\msgNumber_{\party,
            \profile_\proto}$ is the number of blocks created by $\party$ in
            $\profile_\proto$. 
            If $\bribe_\party \geq \big(\revision{\stake_\party} + (1 + \theta) \cdot E[\msgNumber_{\party, \profile_\proto}] \cdot \reward \big) \cdot \rateMax$, 
            then exists $\epsilon$ negligible in $\secparam$ \st the strategy
            profile $\profile_\infractionPredicate$ is an $\epsilon$-Nash
            equilibrium \wrt Reward under $\router_\text{sync}$.
         
        \item\label{thm:guided_compliance} Assume there exists a party
            $\party \in \partySet$ \st $\power_\party < \threshold$ and
            $\bribe_\party > 0$. Then, for every constant $\gamma \in (0, 1)$,
            $\proto$ is not
            $(\gamma \cdot \bribe_\party, \infractionPredicate)$-compliant \wrt
            Reward under $\router_\text{sync}$.
        %
    \end{enumerate}
\end{theorem}


Theorem~\ref{thm:guided} showcases the effectiveness of bribes to distorting
the participants' behavior. In effect, a protocol cannot incentivize on its
own all parties to behave honestly. In Section~\ref{sec:accountable} we will
revisit this result and show how a protocol can become an equilibrium with the
introduction of counterincentives.

\subsection{Maximal Sets and Promising Parties}\label{subsubsec:bribing_guided_equilibria_maximal}

We now explore the different equilibria that exist under guided bribing. This
exploration particularly helps to understand the conditions that
form an equilibrium and the dynamics that can result in one. In addition, it
will be helpful in the following section, which will evaluate the
Prices of Stability and Anarchy.

We first define the notion of \emph{maximal subsets} of parties
(Definition~\ref{def:maximal_set}). Intuitively, a maximal subset includes
parties that perform an infraction and have less power than the
security threshold $\threshold$. Importantly,
by adding any additional party to a maximal subset, the aggregate power gets
above $\threshold$.

\begin{definition}\label{def:maximal_set}
    Assume the following:
    \begin{itemize}
        \item $\proto$ is a protocol run by parties in $\partySet$ in $\rounds$
            rounds, with block-proportional rewards
            (Section~\ref{subsec:preliminaries_proportion}) with $\reward$
            rewards per block; 
        \item $\threshold$ is a security threshold \wrt $\infractionPredicate$;
        \item for every strategy profile $\profile$:
            \begin{itemize}
                \item the exchange rate $\exchangeRate_\profile$ \wrt
                    $\infractionPredicate, \threshold$ follows
                    Eq.~\eqref{eq:exchange}; 
                \item the external rewards $\utilityBoost_{\party,\profile}$
                    due to guided bribing \wrt $\infractionPredicate$ follow
                    Eq.~\eqref{eq:external_guided};
            \end{itemize}
        \item $\msgNumber_{\party, \profile_\proto}$ denotes the number of
            blocks created by $\party$ in all-honest strategy profile
            $\profile_\proto$.
    \end{itemize}

    Let a subset $\infractionPartySet \subset \partySet$ of parties. We say that 
    \emph{$\infractionPartySet$ is maximal \wrt $\threshold, \{\bribe_\party\}_{\party \in \partySet}$}, 
    if the following conditions hold:
    \begin{enumerate}
        \item\label{item:maximal_condition1}
            $\sum_{\party \in \infractionPartySet} \power_\party < \threshold$.
        \item\label{item:maximal_condition2} 
            $\forall \party^* \in \partySet \setminus \infractionPartySet$: 
            $\sum_{\party \in \infractionPartySet \cup \{ \party^* \}} \power_\party \geq \threshold$ and 
            $\bribe_{\party^*} \leq (E[\msgNumber_{\party^*, \profile_\proto}] \cdot \reward + \revision{\stake_{\party^*}}) \cdot \rateDiff$.
    \end{enumerate}

    In addition, we say that a party $\party \in \partySet$ is 
    \emph{promising \wrt $\{\bribe_\party\}_{\party \in \partySet}$}, if it holds that
    $\bribe_{\party} > (E[\msgNumber_{\party, \profile_\proto}] \cdot \reward + \revision{\stake_\party}) \cdot \rateDiff$. 
    We denote the set of parties that are promising \wrt $\{\bribe_\party\}_{\party \in \partySet}$ by
    $\partySet_\text{prom}^{\{\bribe_\party\}}$.
\end{definition} 

By definition, a maximal subset contains all promising parties. Thus, in order
for a maximal subset to exist, it is necessary that the aggregate stake of the
promising parties is less than $\threshold$. Interestingly, as we prove in 
Lemma~\ref{lem:maximal_existence}\ifconference\else~(proof in Appendix~\ref{app:guided_max_prom})\fi, this condition is also sufficient.

Lemma~\ref{lem:maximal_existence} provides two sets of conditions, any of which
implies that the aggregate stake of the promising parties is less than
$\threshold$, \st a maximal subset exists. The first set assumes that the
expected number of blocks, that each party produces, is a monotonically
increasing function of its power. The second set assumes that the expected
number of blocks is a linear function of the party's power. For the linear function:
i) $\hat{\rounds}$ expresses the multiplicative factor;
ii) the additive factor $\frac{\rounds - \hat{\rounds}}{\totalParties}$ is derived from the restriction that the aggregate
number of blocks (across all parties) is $\rounds$.

We define both sets to make the lemma as generic as possible.
Nonetheless, most real-world protocols, including the ones considered in this
work, satisfy the second set.

\begin{lemma}\label{lem:maximal_existence}
    Assume the following:
    \begin{itemize}
        \item $\proto$ is a protocol run by parties in $\partySet$ in $\rounds$ rounds, with block-proportional rewards (Section~\ref{subsec:preliminaries_proportion}) with $\reward$
            rewards per block; 
        \item \revision{each party's stake is $\stake_\party = \power_\party \cdot \stake$;}
        \item $\threshold$ is a security threshold \wrt $\infractionPredicate$;
        \item for every strategy profile $\profile$:
            \begin{itemize}
                \item the exchange rate $\exchangeRate_\profile$ \wrt
                    $\infractionPredicate, \threshold$ follows
                    Eq.~\eqref{eq:exchange}; 
                \item the external rewards $\utilityBoost_{\party,\profile}$
                    due to guided bribing \wrt $\infractionPredicate$ follow
                    Eq.~\eqref{eq:external_guided};
            \end{itemize}
        \item $\msgNumber_{\party, \profile_\proto}$ denotes the number of
            blocks created by $\party$ in all-honest strategy profile
            $\profile_\proto$.
    \end{itemize}

    Consider the following two sets of conditions:
    \begin{enumerate}
        \item 
            (i) $\threshold \geq \frac{1}{2}$; \\
            (ii) $\forall \infractionPartySet, \mathbb{B} \subseteq \partySet: \sum_{\party\in\infractionPartySet} \power_\party \leq \sum_{\party \in \mathbb{B}} \power_\party 
                \Rightarrow \sum_{\party \in \infractionPartySet} E[\msgNumber_{\party, \profile_\proto}] \leq \sum_{\party \in \mathbb{B}} E[\msgNumber_{\party, \profile_\proto}]$; \\
            (iii) $\sum_{\party \in \partySet} \bribe_\party \leq \frac{1}{2} \cdot (\rounds \cdot \reward + \revision{\stake}) \cdot \rateDiff$.
         
        \item 
            (i) $\forall \party \in \partySet: E[\msgNumber_{\party, \profile_\proto}] = \power_\party \cdot \hat{\rounds} + \frac{\rounds - \hat{\rounds}}{\totalParties}$, where $\hat{\rounds} \in [0, \rounds]$;\\ 
            (ii)  $\sum_{\party \in \partySet} \bribe_\party \leq \Big( \threshold \cdot \hat{\rounds} + \frac{\rounds - \hat{\rounds}}{\totalParties} \Big) \cdot \reward \cdot \rateDiff + \revision{\threshold \cdot \stake \cdot \rateDiff}$, where $\hat{\rounds}$ as above.
    \end{enumerate}
    If either set of conditions holds, there exists a subset
    of parties $\infractionPartySet_m$ that is maximal \wrt
    $\threshold, \{\bribe_\party\}_{\party\in\partySet}$ (cf. Definition~\ref{def:maximal_set}).
\end{lemma}


So far we defined maximal sets and showed the conditions for the existence of such
set. Now, Theorem~\ref{thm:maximal_sufficient}\ifconference\else~(proof in Appendix~\ref{app:guided_max_prom})\fi~shows that, if a
maximal set is formed and parties outside this set follow the honest
protocol, the strategy profile that is formed is an equilibrium. In
essence, we show that the existence of a maximal set is sufficient
for the formation of an equilibrium.

Regarding notation, assume a subset of parties $\infractionPartySet \subseteq \partySet$
and some infraction predicate $\infractionPredicate$.
$\profile_\infractionPredicate^\infractionPartySet$ is the
profile where all parties in $\infractionPartySet$ follow
$\strategy_\infractionPredicate$ and all parties in
$\partySet\setminus\infractionPartySet$ behave honestly.

\begin{theorem}\label{thm:maximal_sufficient}
    Assume the following:
    \begin{itemize}
        \item $\proto$: a protocol run by parties in $\partySet$ in $\rounds$ rounds, with block-proportional rewards (Section~\ref{subsec:preliminaries_proportion}) with $\reward$
            rewards per block; 
        \item \revision{each party's stake is $\stake_\party = \power_\party \cdot \stake$;}
        \item $\threshold$: the security threshold \wrt an infraction
            predicate $\infractionPredicate$;
        \item for every strategy profile $\profile$: 
            \begin{itemize}
                \item the exchange rate $\exchangeRate_\profile$ \wrt
                    $\infractionPredicate, \threshold$ follows
                    Eq.~\eqref{eq:exchange};
                \item the external rewards $\utilityBoost_{\party,\profile}$,
                    due to guided bribing \wrt $\infractionPredicate$,
                    follow Eq.~\eqref{eq:external_guided}.
            \end{itemize}
    \end{itemize}

    If $\infractionPartySet \subset \partySet$ is maximal \wrt $\threshold,
    \{\bribe_\party\}_{\party\in\partySet}$ (cf.
    Definition~\ref{def:maximal_set}), then there exists an $\epsilon$
    negligible in $\secparam$ \st
    $\profile_\infractionPredicate^\infractionPartySet$ is an $\epsilon$-Nash
    equilibrium \wrt utility Reward under $\router_\text{sync}$.

\end{theorem}


\noindent\emph{Remark.}
    A subset
    $\infractionPartySet$'s maximality is a sufficient condition for
    $\profile_\infractionPredicate^\infractionPartySet$ to be an equilibrium (Theorem~\ref{thm:maximal_sufficient}).
    However, as Theorem~\ref{thm:guided} implies, this condition is not
    necessary. Although $\partySet$ is not maximal (since
    $\sum_{\party \in \partySet}\power_\party = 1 > \threshold$), the strategy
    profile $\profile_\infractionPredicate^\partySet= \profile_\infractionPredicate$
    is an $\epsilon$-Nash equilibrium for some negligible $\epsilon$, when the
    participation power of all parties is no more than $1 - \threshold$.

\subsection{Price of \{ Stability, Anarchy \}}\label{subsec:bribing_guided_PSt}

We now explore the Price of Stability
($\priceStab$) and the Price of Anarchy ($\priceAnar$).
These notions consider welfare loss for all parties in
two settings. First, a decentralized setting, where each party acts following their own
incentives. Second, a centrally-planned setting, where an authority chooses
each party's strategy, aiming to maximize welfare. Intuitively,
$\priceStab$ expresses the minimum such loss, which results from the best
equilibrium \wrt welfare, whereas $\priceAnar$ expresses the maximum
loss, \ie for the worst equilibrium. Our results for guided
bribing are formalized in two theorems. 

First, Theorem~\ref{thm:guided_pst}\ifconference\else~(proof in Appendix~\ref{app:guided_price})\fi~offers bounds for $\priceStab$.
Intuitively, consider the case when there exists some initial coordination
between all parties. For example, the parties may initially agree to follow the
protocol and then, after the system is deployed, act rationally.\footnote{We
remind that the all-honest strategy profile is not an equilibrium, so
some parties are expected to deviate after the system's
deployment.}
Assuming some such initial coordination, the upper bound on $\priceStab$
expresses the minimum opportunity cost (in terms of welfare), that the parties
will suffer, \ie if they end up in the most favorable equilibrium.
Similarly, the lower bound shows the opportunity cost that the parties will
necessarily suffer, at all times, even if they end up in the best equilibrium.
Note that both bounds are proportional to $\threshold$ and
$1-\frac{\rateMin}{\rateMax}$, so that the larger these values are, the higher both
bounds become (\ie the Price of Stability increases).

\begin{theorem}\label{thm:guided_pst}
    Assume the following:
    \begin{itemize}
        \item $\proto$: a protocol run by parties in $\partySet$ in $\rounds$
            rounds, with block-proportional rewards (Section~\ref{subsec:preliminaries_proportion}) with $\reward$
            rewards per block;
        \item $\threshold$: a security threshold \wrt $\infractionPredicate$;
        \item for every strategy profile $\profile$:
            \begin{itemize}
                \item the exchange rate $\exchangeRate_\profile$ \wrt
                    $\infractionPredicate, \threshold$ follows
                    Eq.~\eqref{eq:exchange}; 
                \item the external rewards $\utilityBoost_{\party, \profile}$
                    due to guided bribing \wrt $\infractionPredicate$
                    follow Eq.~\eqref{eq:external_guided};
            \end{itemize}
        \item one of the sets of conditions in
            Lemma~\ref{lem:maximal_existence} holds, so there exists a subset
            of $\partySet$ that is maximal \wrt
            $\threshold,\{\bribe_\party\}_{\party\in\partySet}$. 
    \end{itemize}

   Then, the following hold:
    \begin{enumerate}
        \item\label{item:guided_pst1}
            $\priceStab \leq 1 + \revision{ \dfrac{\Gamma}{\stake + \rounds \cdot \reward} \cdot\Big(1-\dfrac{\rateMin}{\rateMax}\Big) }$, 
            where \revision{ $\Gamma \in \Big\{ \frac{1}{2} \cdot (\rounds \cdot \reward + \stake),   (\threshold\cdot\hat{\rounds} + \frac{\rounds - \hat{\rounds}}{\totalParties}) \cdot \reward + \threshold \cdot\stake  \Big\} $ } 
            for some $\hat{\rounds} \leq \rounds$.
        \item\label{item:guided_pst2}
            Assume that $\frac{5}{2}\cdot\rateMin\leq\rateMax\revision{\leq\frac{\rounds\cdot\reward}{6}\cdot\rateMin}$ and that for some
            $\hat{\rounds} \geq \frac{\threshold - \frac{3}{\totalParties}}{\threshold - \frac{2}{\totalParties}} \cdot \rounds$,
            the second set of conditions in Lemma~\ref{lem:maximal_existence}
            holds. Then, there exist a participation power allocation
            $\{\power_\party\}_{\party \in \partySet}$ and a bribe allocation
            $\{\bribe_\party\}_{\party \in \partySet}$ \st for any constant
            $\gamma \in (0,1)$ and $\totalParties \geq 8^{\frac{1}{1 - \gamma}}$, 
            it holds that:
        \[ \priceStab \geq 1 + \Big( \threshold - \frac{1}{\totalParties^\gamma} \Big) \cdot \Big( 1 - \frac{\rateMin}{\rateMax} \Big) - \negl(\secparam)\;. \]
    \end{enumerate}
\end{theorem}


Second, Theorem~\ref{thm:guided_pan}\ifconference\else~(proof in Appendix~\ref{app:guided_price})\fi~offers bounds for $\priceAnar$.
Intuitively, consider the case of no initial coordination among
the parties. Therefore, they might end up in the worst possible equilibrium. In
this case, an upper bound on $\priceAnar$ expresses the maximum opportunity cost (in
terms of welfare) that the parties can face. 
In addition, a lower bound on $\priceAnar$ expresses the opportunity cost that
they will necessarily have at all cases, if they end up in the worst-case
equilibrium.

Our results show that $\priceAnar$ is not always upper bounded. Instead, there exist
extremely adverse settings, when the welfare of the worst equilibrium tends to
$0$ \revision{as $\rateMin$ tends to $0$}. In addition, $\priceAnar$ is lower bounded, such that the worst
equilibrium, in terms of welfare, results in a Price of Anarchy at least as high
as the bound.

\begin{theorem}\label{thm:guided_pan}
    Assume the following:
    \begin{itemize}
        \item $\proto$ is a protocol run by parties in $\partySet$ in $\rounds$
            rounds, with block-proportional rewards (cf.
            Subsection~\ref{subsec:preliminaries_proportion}) with $\reward$
            rewards per block;
        \item \revision{each party's stake is $\stake_\party = \power_\party \cdot \stake$;}
        \item $\threshold$ is a security threshold \wrt $\infractionPredicate$;
        \item for every strategy profile $\profile$:
            \begin{itemize}
                \item the exchange rate $\exchangeRate_\profile$ \wrt
                    $\infractionPredicate, \threshold$ is defined as in
                    Eq.~\eqref{eq:exchange}; 
                \item the external rewards $\utilityBoost_{\party,\profile}$
                    due to guided bribing \wrt $\infractionPredicate$ are
                    defined as in Eq.~\eqref{eq:external_guided};
            \end{itemize}
    \end{itemize}
    Then, the following hold:
    \begin{enumerate}
        \item\label{item:guided_pan1} 
            If for every party $\party \in \partySet$, it holds that
            $\power_\party \leq 1 - \threshold$, and
            $\sum_{\party \in \partySet} \bribe_\party$ is a non-negligible value
            upper bounded by
            $\threshold \cdot (\rounds \cdot \reward + \revision{\stake}) \cdot \rateDiff$, then the Price
            of Anarchy is bounded by:
            \[ \frac{1}{\threshold + (1 - \threshold) \cdot \frac{\rateMin}{\rateMax}} - \negl(\secparam) \leq \priceAnar < \infty\;. \]
        \item\label{item:guided_pan2} 
            If for every party $\party \in \partySet$, it holds that
            $\power_\party < \threshold$, and
            $\sum_{\party \in \partySet} \bribe_\party = \negl(\secparam)$, then
            \revision{$\priceAnar \geq \Big( 1 + \frac{\rounds \cdot \reward}{\stake} \Big) \cdot \frac{\rateMax}{\rateMin}$}.
    \end{enumerate}
\end{theorem}


\section{Effective Bribing}\label{sec:bribing_effective}

We now study a stricter bribing setting, which we call
``effective''. As with guided bribing, a party needs to perform a specific
infraction, in order to receive its external rewards (bribe). In guided bribing
this was the only condition, so the bribe was paid regardless of the
infraction's outcome. In effective bribing, the bribe is paid as long as the
infraction contributes to a successful attack. In particular, the party
receives the bribe if the exchange rate drops to $\rateMin$.

Under \emph{effective bribing}, the external rewards are formally
defined for some infraction predicate $\infractionPredicate$ as follows:
\begin{equation}\label{eq:external_effective}
    \utilityBoost_{\party,\profile}(\trace) = \left\{ 
        \begin{array}{ll}
            \bribe_\party, & \mbox{if } \infractionPredicate(\trace, \party) = 1 \mbox{ and } \exchangeRate_\profile(\trace) = \rateMin \\
            0, & \mbox{otherwise}
        \end{array}
    \right.\;
\end{equation}
where $\exchangeRate_\profile(\trace)$ is defined in Eq.~\eqref{eq:exchange}.

Following, we explore the equilibria formed under effective bribing, as well
as the Price of Stability and Price of Anarchy.

\subsection{Equilibria}\label{subsec:bribing_effective_equilibria}

When exploring equilibria under effective bribing, our first result, Theorem~\ref{thm:effective}\ifconference\else~(proof in Appendix~\ref{app:eff_eq})\fi, offers three interesting cases.
First, we show that, as was the case of guided bribing, the strategy profile
where all parties attempt the infraction is an equilibrium. 
Second, the all-honest strategy profile is also an equilibrium. This is in
opposition to the guided setting, where the protocol strategy was not an
equilibrium.
Third, the strategy profile where all parties fully abstain from participating in
the execution, denoted by $\profile_\text{abs}$, is an equilibrium. This result
will be particularly useful in the following subsection, where we show that the
Price of Anarchy is sometimes unbounded.

\begin{theorem}\label{thm:effective}
    Assume the following:
    \begin{itemize}
        \item $\proto$: a protocol run by parties in $\partySet$ in $\rounds$
            rounds, with block-proportional rewards (Section~\ref{subsec:preliminaries_proportion}) with $\reward$
            rewards per block;
            \item \revision{each party's stake is $\stake_\party = \power_\party \cdot \stake$;}
        \item $\threshold$: a security threshold \wrt $\infractionPredicate$;
        \item $\profile_\proto$: the all-honest strategy profile;
        \item $\profile_\text{abs}$: the all-abstain strategy profile;
        \item for every strategy profile $\profile$:
            \begin{itemize}
                \item the exchange rate $\exchangeRate_\profile$ \wrt
                    $\infractionPredicate, \threshold$ follows
                    Eq.~\eqref{eq:exchange}; 
                \item the external rewards $\utilityBoost_{\party, \profile}$
                    due to effective bribing \wrt $\infractionPredicate$
                    follow Eq.~\eqref{eq:external_effective};
            \end{itemize}
    \end{itemize}

    Then, the following hold:

    \begin{enumerate}
        \item\label{thm:effective_nash1} 
            If for every party $\party \in \partySet$, it holds that
            $\power_\party \leq 1 - \threshold$, then there is an $\epsilon$
            negligible in $\secparam$ \st the strategy profile
            $\profile_\infractionPredicate$ is an $\epsilon$-Nash equilibrium
            \wrt Reward under $\router_\text{sync}$.
         
        \item\label{thm:effective_nash2}
            Assume that there exists a party $\party \in \partySet$ \st
            $\power_\party \geq \threshold$. Let $\theta > 0$ be a real value \st
            the probability 
            $\Pr \big[ \msgNumber_{\party, \profile_\proto} > (1 + \theta) \cdot E[\msgNumber_{\party, \profile_\proto}] \big]$ 
            is negligible in $\secparam$, where
            $\msgNumber_{\party, \profile_\proto}$ is the number of blocks
            created by $\party$ in $\profile_\proto$. If 
            $\bribe_\party \geq \big( (1 + \theta) \cdot E[\msgNumber_{\party, \profile_\proto}] \cdot \reward + \revision{\stake_\party} \big) \cdot \rateMax$, 
            then there is an $\epsilon$ negligible in $\secparam$ such that the
            strategy profile $\profile_\infractionPredicate$ is an
            $\epsilon$-Nash equilibrium \wrt Reward under
            $\router_\text{sync}$.
         
        \item\label{thm:effective_nash3}
            If for every party $\party \in \partySet$, it holds that
            $\power_\party < \threshold$, then $\proto$ is a Nash equilibrium
            \wrt Reward under $\router_\text{sync}$.
         
        \item\label{thm:effective_nash4}
            If for every party $\party \in \partySet$, it holds that
            $\power_\party < \threshold$, then $\profile_\text{abs}$ is a Nash
            equilibrium \wrt Reward under $\router_\text{sync}$.
    \end{enumerate}
\end{theorem}


\subsection{Price of \{ Stability, Anarchy \}}\label{subsec:effective_Pst}

As with guided bribing, we now explore the Price of Stability and Price of
Anarchy in the case of effective bribing.

Theorem~\ref{thm:effective_pst}\ifconference\else~(proof in Appendix~\ref{app:eff_price})\fi~assumes that no
party can unilaterally perform an attack, \ie all parties control less than
$\threshold$ power. In this setting, there exist two extreme equilibria, in
terms of welfare. The best equilibrium achieves the same welfare as the
centralized setting, so $\priceStab = 1$. For the worst
equilibrium though welfare is $0$, so $\priceAnar = \infty$ (as $\rateMin \rightarrow 0$). 
 
\begin{theorem}\label{thm:effective_pst}
    Assume the following:
    \begin{itemize}
        \item $\proto$ is a protocol run by parties in $\partySet$ in $\rounds$
            rounds, with block-proportional rewards (cf.
            Subsection~\ref{subsec:preliminaries_proportion}) with $\reward$
            rewards per block;
        \item $\threshold$ is a security threshold \wrt $\infractionPredicate$;
        \item for every strategy profile $\profile$:
            \begin{itemize}
                \item the exchange rate $\exchangeRate_\profile$ \wrt
                    $\infractionPredicate, \threshold$ is defined as in
                    Eq.~\eqref{eq:exchange}; 
                \item the external rewards $\utilityBoost_{\party, \profile}$
                    due to effective bribing \wrt $\infractionPredicate$ are
                    defined as in Eq.~\eqref{eq:external_effective};
            \end{itemize}
        \item for every party $\party \in \partySet$, it holds that $\power_\party < \threshold$ \revision{and $\stake_\party = \power_\party \cdot \stake$};
        \item $\sum_{\party \in \partySet} \bribe_\party \leq (\rounds \cdot \reward + \revision{\stake}) \cdot \rateDiff$.\footnote{Note that this is a tighter bound than the one already applied by the bribing budget.}
    \end{itemize}

    Then:
    \begin{inparaenum}[1)]
        \item $\priceStab=1$,
        \item \revision{$\priceAnar \geq (1 + \frac{\rounds \cdot \reward}{\stake}) \cdot \frac{\rateMax}{\rateMin}$}
    \end{inparaenum}
    \wrt Reward under $\router_\text{sync}$.
\end{theorem}


Theorem~\ref{thm:effective_pst} shows that the all-honest and
all-abstain equilibria determine the best ($\priceStab = 1$) and worst
($\priceAnar = \infty$, as $\rateMin \rightarrow 0$) scenarios one can expect in a world without any
coordination. 

Subsequently, Theorem~\ref{thm:effective_T-pan}\ifconference\else~(proof in Appendix~\ref{app:eff_price})\fi~explores the worst-case
scenario, in a setting where the parties coordinate by either (i) being honest
or (ii) follow $\strategy_\infractionPredicate$. It is meaningful to exclude
all other strategies, since under the Reward utility function, rational parties
have no other reason to deviate from the protocol guidelines, apart from
performing the infraction that allows them to get their external rewards.
Observe that the lower bound is in inverse relation with $\threshold$ and both bounds
are in inverse relation with $\rateMin$. Intuitively, if an attack
is easy to conduct (\ie low threshold $\threshold$) and its effects are
particularly adverse (\ie results in a bid drop of the exchange rate), both
bounds of the Price of Anarchy \wrt $\{\proto,
\strategy_\infractionPredicate\}$ increase.

\begin{theorem}\label{thm:effective_T-pan}
    Assume the following:
    \begin{itemize}
        \item $\proto$ is a protocol run by parties in $\partySet$ in $\rounds$ rounds, with block-proportional rewards (cf.
            Subsection~\ref{subsec:preliminaries_proportion}) with $\reward$
            rewards per block; 
        \item $\threshold$ is a security threshold \wrt $\infractionPredicate$;
        \item for every strategy profile $\profile$:
            \begin{itemize}
                \item the exchange rate $\exchangeRate_\profile$ \wrt
                    $\infractionPredicate, \threshold$ is defined as in
                    Eq.~\eqref{eq:exchange}; 
                \item the external rewards $\utilityBoost_{\party, \profile}$
                    due to effective bribing \wrt $\infractionPredicate$ are
                    defined as in Eq.~\eqref{eq:external_effective};
            \end{itemize}
        \item $\msgNumber_{\party, \profile_\proto}$ is the number of blocks
            created by $\party$ in $\profile_\proto$;
        \item for every party $\party \in \partySet$, it holds that
            $\power_\party \leq 1 - \threshold$, \revision{$\stake_\party = \power_\party \cdot \stake$}, and
            $\sum_{\party \in \partySet} \bribe_\party \leq \threshold \cdot (\rounds \cdot \reward + \revision{\stake}) \cdot \rateDiff$.
    \end{itemize}

    Then, the following hold:

    \begin{enumerate}
        \item\label{item:effective_T-pan1} 
            Price of Anarchy \wrt $\{\proto, \strategy_\infractionPredicate\}$
            is lower bounded by:
        \[ \{\proto, \strategy_\infractionPredicate\} \text{-} \priceAnar \geq \frac{1}{\threshold + (1 - \threshold) \cdot \frac{\rateMin}{\rateMax}} - \negl(\secparam)\;. \]

        \item\label{item:effective_T-pan2} 
            Price of Anarchy \wrt $\{\proto, \strategy_\infractionPredicate\}$ 
            is upper bounded by:
            \[ \{\proto, \strategy_\infractionPredicate\} \text{-} \priceAnar \leq \frac{(\revision{\stake} + \rounds \cdot \reward) \cdot \rateMax}{(\revision{\stake} + \rounds \cdot \reward) \cdot \rateMin + \Phi} + \negl(\secparam)\;, \]
        where $\Phi = \min_{\mathbb{A} \subseteq \partySet: \sum_{\party \in \mathbb{A}} \power_\party \geq \threshold} \sum_{\party \in \mathbb{A}} \bribe_\party$.

        \ignore{
        \item\label{item:effective_T-pan2} The Price of Anarchy \wrt $\{\proto,\strategy_\infractionPredicate\}$ is upper bounded by
        \[\{\proto,\strategy_\infractionPredicate\}\text{-}\priceAnar\leq\frac{\rounds}{\rounds\cdot\frac{\rateMin}{\rateMax}+\Gamma\cdot\big(1-\frac{\rateMin}{\rateMax}\big)}+\negl(\secparam)\;,\]
        where  $\Gamma=\min_{\mathbb{A}\subseteq\partySet:\sum_{\party\in\mathbb{A}}\power_\party\geq\threshold}\sum_{\party\in\mathbb{A}}E[\msgNumber_{\party, \profile_\proto}]$.
        \item\label{item:effective_T-pan3} Assume that for some $\hat{\rounds}\in[0,\rounds]$ and for every party $\party\in\partySet$, it holds that $E[\msgNumber_{\party, \profile_\proto}]=\power_\party\cdot\hat{\rounds}+\frac{\rounds-\hat{\rounds}}{\totalParties}$. Then,
        \[\{\proto,\strategy_\infractionPredicate\}\text{-}\priceAnar\leq\frac{\rounds}{\rounds\cdot\frac{\rateMin}{\rateMax}+\Big(\threshold\cdot\hat{\rounds}+\frac{\rounds-\hat{\rounds}}{\totalParties}\Big)\cdot\big(1-\frac{\rateMin}{\rateMax}\big)}+\negl(\secparam)\;.\]
        }
    \end{enumerate}
\end{theorem}


We note that the two theorems make different assumptions about each party's
power. Theorem~\ref{thm:effective_pst} assumes that no party has power above
$\threshold$. Therefore, no party can unilaterally perform an attack,
so it is implied that the all-honest strategy is an equilibrium.
Theorem~\ref{thm:effective_T-pan} assumes that no party has power above
$1 - \threshold$. Intuitively, this means that no party can unilaterally
prevent an attack from happening, so if all other parties perform the
infraction, each party would also do so (\ie the all-infraction strategy is an
equilibrium).

Finally, Corollary~\ref{cor:ffective_T-pan} is a direct consequence of
Theorem~\ref{thm:effective_T-pan}. Specifically, it sets an upper bound on
Price of Anarchy \wrt $\{\proto, \strategy_\infractionPredicate\}$, when a subset of parties with
enough power to perform an attack (\ie above $\threshold$) are offered
negligible bribes.

\begin{corollary}\label{cor:ffective_T-pan}
    Under the conditions of Theorem~\ref{thm:effective_T-pan}, if there is a subset
    $\mathbb{A} \subseteq \partySet$ \st
    $\sum_{\party \in \mathbb{A}} \power_\party \geq \threshold$ and
    $\sum_{\party \in \mathbb{A}} \bribe_\party = \negl(\secparam)$, then
    $\{\proto, \strategy_\infractionPredicate\} \text{-} \priceAnar \leq \frac{\rateMax}{\rateMin} + \negl(\secparam)$ \;.
\end{corollary}

\section{Accountable Rewards Under Guided Bribing}\label{sec:accountable}

Our final contribution is to evaluate possible mitigations against bribing
attacks. \revision{In particular, we evaluate mechanisms that aim to protect
against visible and provable attacks, such as double signing, \st the protocol
can penalize a party when it observes such attack. Notably, these defenses
cannot protect against attacks when a non-interactive proof of misbehavior
cannot be produced, \eg censorship attacks.}

There are two mechanisms that we consider: 
\begin{itemize} 
    \item \emph{slashing}: In this setting, parties are required to lock a certain
        amount of assets at the beginning of the execution. If a party
        misbehaves in a publicly verifiable manner, that is if a
        non-interactive proof of misbehavior can be created, then it forfeits
        its deposit.\footnote{A non-interactive proof is needed due to the open
        nature of such systems. In essence, if a party joins the protocol at a
        later point in time, it should be able to validate that past acts of
        misbehavior.} Otherwise, if no such proof is shown, the party gets back
        its deposit at the end of the execution. 
    \item \emph{dilution}. In this setting, the system mints and distributes
        rewards on a regular basis. If a party misbehaves in a publicly
        verifiable manner, it is barred from receiving any newly minted
        rewards. This has a dilution effect on the party's cumulative stake in
        the system (its stake percentage drops). 
\end{itemize}

Under slashing, we assume that each party's deposit is proportional to its participation power.
Formally, given an aggregate deposit amount
$\deposit > 0$, the mitigation value 
of party $\party$
\wrt $\infractionPredicate$ is:\footnote{Again here (as with $\reward$ above) we slightly abuse notation for $\deposit$. When the subscripts $\party, \profile$ are used $\deposit_{\party, \profile}$ denotes the random variable of $\party$'s deposit under strategy profile $\profile$. When no subscript is used, $\deposit$ denotes the aggregate deposit among all parties.}
\begin{equation}\label{eq:deposit}
    \deposit_{\party, \profile}(\trace) = \left\{ 
        \begin{array}{ll}
            0, & \mbox{if } \infractionPredicate(\trace, \party) = 1 \\
            \power_\party \cdot \deposit, & \mbox{otherwise}
        \end{array}
    \right.\;.
\end{equation}
In effect, $\party$ forfeits its deposit, only if $\party$ performs an infraction.\footnote{Again observe that the ``if'' direction does not necessarily hold. Specifically, a party might perform an infraction which is not provable. In that case, the deposit amount would not be ``slashed''.}
Otherwise, if the party does not perform the given infraction, then a deposit amount equal to $\power_\party \cdot \deposit$ is returned, i.e., $\party$ is not slashed.

This mechanism is expressed as an additional term (in USD) in the Accountable Reward utility function, equal to the expectation of the random variable $\deposit_{\party, \profile}\cdot\exchangeRate_{\profile}$.
By Eq.~\eqref{eq:exchange} and~\eqref{eq:deposit}, this random variable is defined as follows:
\begin{equation}\label{eq:compliance_rewards}
    \deposit_{\party, \profile}\cdot\exchangeRate_{\profile}(\trace) = \left\{ 
        \begin{array}{ll}
            0, & \mbox{if } \infractionPredicate(\trace, \party) = 1 \\
            \power_\party \cdot \deposit\cdot\rateMin, &\mbox{if } \begin{array}{ll}\infractionPredicate(\trace, \party) = 0  \mbox{ and } \\\sum_{\hat{\party} \in \partySet: \infractionPredicate(\trace, \hat{\party}) = 1} \power_{\hat{\party}} \geq \threshold \end{array} \\
            \power_\party \cdot \deposit\cdot\rateMax, & \mbox{otherwise} 
        \end{array}
    \right.\;.
\end{equation}

The situation is similar for the case of dilution. If the total
amount of dilution rewards is $\deposit$, then the mitigation value of party
$P$ is equal to an equation identical to Eq.~\eqref{eq:deposit}
(we omit it for brevity). 
This means that when the party is performing an infraction its
rewards are zeroed, so its effective stake is ``diluted''.

Following, we will refer to the value of Eq.~\eqref{eq:deposit}
as ``compliance payout'', in order to capture both types of mitigations.

The rest of this section explores the \emph{guided} bribing effects 
under accountable rewards. We show two ``positive'' results
(Theorems~\ref{thm:accountable_protocol_nash}
and~\ref{thm:acc_maximal_sufficient}), in the sense that they describe
equilibria where a bribing attack is not successful, and one negative. 

Note that effective bribing is not explored here, since the protocol was
already shown to be an equilibrium without the proposed mitigations
(Theorem~\ref{thm:effective}).

\subsection{Protocol Equilibrium and Negative Equilibrium}\label{subsec:accountable_protocol_eq}

Theorem~\ref{thm:accountable_protocol_nash}\ifconference\else~(proof in Appendix~\ref{app:acc_eq1})\fi~shows that the
protocol is an equilibrium, if each party's bribe is less than its compliance payout. In essence, if a
party's compliance payout is large enough, then it is not incentivized to accept the
bribe and perform the infraction.

This shows that compliance payouts is an effective defense mechanism in helping make
a protocol an equilibrium. Recall that in Theorem~\ref{thm:guided},
a party would accept a bribe if it expected the attack to not be
successful, \ie if its rewards would not be reduced. By introducing a utility
reduction \emph{regardless} of the attack's success, we introduce a negative
utility difference, if the compliance payout is larger than the bribe, thus
making the protocol an equilibrium.

We note that the bribing budget is upper-bounded by
the \emph{aggregate} amount of compliance payouts across all
parties ($\deposit$). Also the briber is restricted on how it can distribute this
budget, since we require that no single party's bribe exceeds its compliance payout. 

\begin{theorem}\label{thm:accountable_protocol_nash}
    Assume the following:
    \begin{itemize}
        \item $\proto$: a protocol with block-proportional rewards (cf.
            Section~\ref{subsec:preliminaries_proportion}), with $\reward$
            rewards per block;
        \item $\profile_\proto$: the all-honest strategy profile;
        \item $\threshold$: the security threshold \wrt an infraction
            predicate $\infractionPredicate$;
        \item for every strategy profile $\profile$: 
            \begin{itemize}
                \item the exchange rate $\exchangeRate_\profile$ \wrt
                    $\infractionPredicate, \threshold$ follows
                    Eq.~\eqref{eq:exchange};
                \item the external rewards $\utilityBoost_{\party, \profile}$,
                    due to guided bribing \wrt $\infractionPredicate$, 
                    follow Eq.~\eqref{eq:external_guided};
               \item the compliance payout $ \deposit_{\party, \profile}$ \wrt $\infractionPredicate$ follows Eq.~\eqref{eq:deposit};
            \end{itemize}
        \item For every party $\party \in \partySet$, it holds that $\power_\party < \threshold$ \revision{and $\stake_\party = \power_\party \cdot \stake$}.  
    \end{itemize}
    Then, it holds that there is an $\epsilon$ negligible in $\secparam$ \st
    $\proto$ is an $\epsilon$-Nash equilibrium \wrt Accountable Reward under
    $\router_\text{sync}$ if and only if
    for every $\party \in \partySet$: $\beta_\party - \power_\party \cdot \deposit \cdot \rateMax \leq \negl(\secparam)$.
\end{theorem}


Second, Theorem~\ref{thm:accountable_negative_nash}\ifconference\else~(proof in Appendix~\ref{app:acc_eq1})\fi~presents a
negative equilibrium, where enough parties are incentivized to get bribed \st
an attack is feasible. The proof idea follows similarly to
Theorem~\ref{thm:guided}, where a party cannot unilaterally prevent the attack
from happening, so it cannot unilaterally prevent the exchange rate from
becoming $\rateMin$. Here, we
require that the bribe of each party is larger than their compliance payout
value, denominated in USD after an attack occurs (\st the exchange rate is
$\rateMin$).

\begin{theorem}\label{thm:accountable_negative_nash}
    Assume the following:
    \begin{itemize}
        \item $\proto$: a protocol with block-proportional rewards (cf.
            Section~\ref{subsec:preliminaries_proportion}), with $\reward$
            rewards per block;
        \item $\threshold$: the security threshold \wrt an infraction predicate $\infractionPredicate$;
        \item $\profile_\infractionPredicate$: the strategy profile where all parties perform $\infractionPredicate$;
        \item for every strategy profile $\profile$: 
            \begin{itemize}
                \item the exchange rate $\exchangeRate_\profile$ \wrt
                    $\infractionPredicate, \threshold$ follows
                    Eq.~\eqref{eq:exchange};
                \item the external rewards $\utilityBoost_{\party, \profile}$,
                    due to guided bribing \wrt $\infractionPredicate$, 
                    follow Eq.~\eqref{eq:external_guided};
               \item the compliance payout $ \deposit_{\party, \profile}$ \wrt $\infractionPredicate$ follows Eq.~\eqref{eq:deposit};
            \end{itemize}
          \item For every party $\party \in \partySet$, it holds that $\power_\party < 1 - \threshold$ \revision{and $\stake_\party = \power_\party \cdot \stake$}.  
    \end{itemize}
    Then, it holds that there is an $\epsilon$ negligible in $\secparam$ \st
    $\profile_\infractionPredicate$ is an $\epsilon$-Nash equilibrium \wrt Accountable Reward under
    $\router_\text{sync}$ if
    for every $\party \in \partySet$: $\beta_\party - \power_\party \cdot \deposit \cdot \rateMin \geq \negl(\secparam)$.
\end{theorem}


Theorem~\ref{thm:accountable_negative_nash} offers a helpful intuition about
the feasibility of bribing attacks, in the presence of counterincentives.
Specifically, the more harmful an attack
is, the lower $\deposit \cdot \rateMin$ is expected to become. Consequently, a devastating
bribing attack is more easily sustainable, since the bribing budget needed to
maintain an attack equilibrium gets lower. In the extreme scenario, when
$\rateMin = 0$, Theorem~\ref{thm:accountable_negative_nash} becomes equivalent
to Theorem~\ref{thm:guided}, \ie compliance payouts have no effect on attack
prevention.

\subsection{Promising Parties}\label{subsec:accountable_equilibria_maximal}

We now perform an analysis akin to
Section~\ref{subsubsec:bribing_guided_equilibria_maximal}. We
will show an equilibrium where some parties are incentivized to accept a bribe,
but not enough \st the attack is successful.

First, we define an alternative notion of ``promising'' parties
(Definition~\ref{def:acc_maximal_set}). Here, a party is promising if
its bribe is higher than its compliance payout. This notion is close to
Section~\ref{subsubsec:bribing_guided_equilibria_maximal}, where the bribe was
assumed to be higher than the received rewards.

\begin{definition}\label{def:acc_maximal_set}
    Assume the following:
    \begin{itemize}
        \item $\proto$ is a protocol run by parties in $\partySet$;
        \item $\threshold$ is a security threshold \wrt $\infractionPredicate$;
        \item for every strategy profile $\profile$:
            \begin{itemize}
                \item the exchange rate $\exchangeRate_\profile$ \wrt
                    $\infractionPredicate, \threshold$ follows
                    Eq.~\eqref{eq:exchange};
                \item the external rewards $\utilityBoost_{\party, \profile}$
                    due to guided bribing \wrt $\infractionPredicate$
                    follow Eq.~\eqref{eq:external_guided};
               \item the compliance payout $ \deposit_{\party, \profile}$ \wrt
                   $\infractionPredicate$ follows
                    Eq.~\eqref{eq:deposit};
            \end{itemize}
    \end{itemize}

    We say that a party $\party \in \partySet$ is 
    \emph{promising \wrt $\{\bribe_\party\}_{\party \in \partySet}, \deposit$}, if it holds that
    $\bribe_{\party} > \power_\party \cdot \deposit\cdot\rateMax$. 
    We denote the set of parties that are promising \wrt $\{\bribe_\party\}_{\party \in \partySet}, \deposit$ by
    $\partySet_\text{prom}^{\{\bribe_\party\}, \deposit}$.
\end{definition} 

We note that, if a promising party exists, then the assumptions of
Theorem~\ref{thm:accountable_protocol_nash} no longer hold. 

The second step is to restrict the bribing budget
(Lemma~\ref{lem:acc_maximal_existence}\ifconference\else, proof in Appendix~\ref{app:acc_prom}\fi). In particular, we assume that the
budget is upper bounded by $\threshold \cdot \deposit\cdot\rateMax$, where $\deposit$ is the
total aggregate compliance payout and $\threshold$ is the power threshold needed for
performing an attack.

\begin{lemma}\label{lem:acc_maximal_existence}
    Assume the following:
    \begin{itemize}
        \item  $\proto$: a protocol run by parties in $\partySet$.
        \item $\threshold$: a security threshold \wrt $\infractionPredicate$;
        \item for every strategy profile $\profile$:
            \begin{itemize}
                \item the exchange rate $\exchangeRate_\profile$ \wrt
                    $\infractionPredicate, \threshold$ follows
                    Eq.~\eqref{eq:exchange};
                \item the external rewards $\utilityBoost_{\party, \profile}$
                    due to guided bribing \wrt $\infractionPredicate$
                    follow Eq.~\eqref{eq:external_guided};
                \item the compliance payout $ \deposit_{\party, \profile}$ \wrt $\infractionPredicate$ 
                    follows Eq.~\eqref{eq:deposit};
            \end{itemize}
        \item $\sum_{\party \in \partySet} \bribe_\party \leq \threshold \cdot \deposit\cdot\rateMax$.
    \end{itemize}

    Then, it holds that $\sum_{\party \in \partySet_\text{prom}^{\{ \bribe_\party \}, \deposit}} \power_\party < \threshold$.
\end{lemma}


The restriction on the bribing budget is an additional assumption to the
previous result. Whereas before we restricted how $\adversary$ can allocate its
budget, we now restrict the budget but allow $\adversary$ to
allocate it arbitrarily among bribes to parties.

Our second result is given in
Theorem~\ref{thm:acc_maximal_sufficient}\ifconference\else~(proof in Appendix~\ref{app:acc_prom})\fi.\footnote{Recall that by
$\profile_\infractionPredicate^\infractionPartySet$, we denote the strategy
profile where all parties in $\infractionPartySet$ follow
$\strategy_\infractionPredicate$ and all parties in
$\partySet\setminus\infractionPartySet$ behave honestly.}
This result shows that, if promising parties exist and the budget is restricted
as in Lemma~\ref{lem:acc_maximal_existence}, then there exists an equilibrium,
where some of the promising parties are bribed, but with not enough power to
successfully perform the attack.

\begin{theorem}\label{thm:acc_maximal_sufficient}
    Assume the following:
    \begin{itemize}
        \item $\proto$: a protocol run by parties in $\partySet$;
        \item \revision{each party's stake is $\stake_\party = \power_\party \cdot \stake$;}
        \item $\threshold$: the security threshold \wrt an infraction
            predicate $\infractionPredicate$;
        \item for every strategy profile $\profile$: 
            \begin{itemize}
                \item the exchange rate $\exchangeRate_\profile$ \wrt
                    $\infractionPredicate, \threshold$ follows
                    Eq.~\eqref{eq:exchange};
                \item the external rewards $\utilityBoost_{\party, \profile}$,
                    due to guided bribing \wrt $\infractionPredicate$, follow
                    Eq.~\eqref{eq:external_guided}.
                \item the compliance payout $ \deposit_{\party, \profile}$ \wrt $\infractionPredicate$ 
                    follows Eq.~\eqref{eq:deposit};
            \end{itemize}
    \end{itemize}

    If $\sum_{\party \in \partySet_\text{prom}^{\{\bribe_\party\} ,\deposit}} \power_\party < \threshold$, 
    then there exists an $\epsilon$ negligible in $\secparam$ \st
    $\profile_\infractionPredicate^{\partySet_\text{prom}^{\{\bribe_\party\}, \deposit}}$ is an $\epsilon$-Nash
    equilibrium \wrt utility Accountable Reward under $\router_\text{sync}$.

\end{theorem}


\subsubsection*{Bribing Budget Evaluation}

To assess Theorem~\ref{thm:acc_maximal_sufficient}'s assumptions, we turn to
real world systems. Slashing has been used in the context of
Proof-of-Stake (PoS) systems. Here, the protocol requires each public key to
deposit some tokens, in exchange for participating in the consensus protocol
(``staking'').  Notably, this is not possible in Proof-of-Work systems, where
the identity of the party that creates a block is not known by default.
We also note that some PoS systems, like Cardano, Algorand, and Avalanche don't
use slashing.\footnote{Cardano: \href{https://cardano.org}{cardano.org};
Algorand: \href{https://algorand.com}{algorand.com};
Avalanche: \href{https://www.avax.network/}{avax.network}}

In all analyzed systems, a valid block requires
$\frac{2}{3}$ of committee signatures. However, if a safety violation occurs,
then the number of parties that
double-sign, that is sign both blocks, can be as low as $\frac{1}{3}$. This is
the case because at most $\frac{1}{3}$ parties can abstain from signing either
of the blocks. Therefore, to estimate the bribing budget's bound \wrt slashing,
we set $\threshold = \frac{1}{3}$, which is the minimum number of parties that
can be penalized for a safety violation. Therefore, the bribing budget bound is
computed as $\frac{1}{3}$ of the deposited value.

Table~\ref{tab:slashing_deposits} outlines data for major PoS blockchains.
Ethereum\footnote{\url{https://ethereum.org/en/staking}} is by far the system
with the most value locked in staking deposits (\$$64.6$B). This is
somewhat expected, since Ethereum is the second most valued cryptocurrency,
after Bitcoin, and supports multiple applications with
its smart contract infrastructure. 
Next is 
Solana with \$$34.8$B locked value in staking, 
followed by Polygon (\$$2.7$B), 
Cosmos Hub (\$$2.38$B),
Tezos (\$$648$M),
and Polkadot (\$$90.4$M).\footnote{Solana: \href{https://solana.com}{solana.com};
Polygon: \href{https://staking.polygon.technology}{staking.polygon.technology};
Cosmos: \href{https://cosmos.network}{cosmos.network};
Tezos: \href{https://tezos.com/stats}{tezos.com};
Polkadot: \href{https://staking.polkadot.network}{staking.polkadot.network}.}

Table~\ref{tab:slashing_deposits} demonstrates that slashing can increase the
requirements for bribing attacks. Conversely,
Theorem~\ref{thm:acc_maximal_sufficient} applies for all adversaries that have
a smaller budget than those identified in the table. Naturally, the higher the
bribing budget bound, the more secure a system is, \ie it can defend against
adversaries with higher budget. 

\begin{table}
    \centering \def\arraystretch{1.5}

    \begin{center}
        \begin{tabular}{|c|c|c|c|}
            \hline

              \textbf{System}
            & \begin{tabular}[c]{@{}c@{}} \textbf{Deposits} \\ \textbf{(Token)} \\ \end{tabular}
            & \begin{tabular}[c]{@{}c@{}} \textbf{Deposits} \\ \textbf{(USD)} \\ \end{tabular}
            & \begin{tabular}[c]{@{}c@{}} \textbf{Bribing Budget} \\ \textbf{Bound (USD)} \\ \end{tabular} \\
            \hline

            Ethereum (ETH) & $28.8$M & \$$64.6$B & \$$21.5$B \\
            \hline

            Solana (SOL) & $380.4$M & \$$34.8$B & \$$11.6$B \\
            \hline

            Polygon (MATIC) & $3.5$B & \$$2.7$B & \$$890$M \\
            \hline

            Cosmos (ATOM) & $249.5$M & \$$2.38$B & \$$793$M \\
            \hline

            Tezos (XTZ) & $675.3$M & \$$648$M & \$$216$M \\
            \hline

            Polkadot (DOT) & $13.7$M & \$$90.4$M & \$$30.13$M \\
            \hline

        \end{tabular}
    \end{center}
    \caption{
        Slashing data for various systems ($26$ Jan
        $2024$). Prices used to compute USD
        values were obtained from
        \href{https://coinmarketcap.com}{coinmarketcap.com}. The minimum
        percentage of parties that can be slashed for double signing
        is $\threshold = \frac{1}{3}$, so
        the bribing budget bound is $\threshold$ multiplied by the
        USD deposit value.
    }
    \label{tab:slashing_deposits}
\end{table}

\subsection{Price of \{ Stability, Anarchy \}}\label{subsec:accountable_stability}

We now compute upper bounds for the Price of Stability ($\priceStab$) and
Anarchy ($\priceAnar$) under accountable rewards. Recall that $\priceStab$ expresses how ``good'', in
terms of welfare, the best equilibrium is, and $\priceAnar$ expresses how
``bad'' the worst equilibrium is, both being at least $1$.

Theorem~\ref{thm:accountable_protocol_pst}\ifconference\else~(proof in Appendix~\ref{app:acc_pst})\fi~assumes that no
promising parties exist. In this setting, the protocol yields maximal welfare and,
since it is an equilibrium (cf.
Theorem~\ref{thm:accountable_protocol_nash}), $\priceStab$ is negligibly close
to $1$.

\begin{theorem}\label{thm:accountable_protocol_pst}
    Assume the following:
    \begin{itemize}
        \item $\proto$: a protocol with block-proportional rewards (cf.
            Section~\ref{subsec:preliminaries_proportion}), with $\reward$
            rewards per block;
        \item \revision{each party's stake is $\stake_\party = \power_\party \cdot \stake$;}
        \item $\profile_\proto$: the all-honest strategy profile;
        \item $\threshold$: the security threshold \wrt an infraction
            predicate $\infractionPredicate$;
        \item for every strategy profile $\profile$: 
            \begin{itemize}
                \item the exchange rate $\exchangeRate_\profile$ \wrt
                    $\infractionPredicate, \threshold$ follows
                    Eq.~\eqref{eq:exchange};
                \item the external rewards $\utilityBoost_{\party, \profile}$,
                    due to guided bribing \wrt $\infractionPredicate$,
                    follow Eq.~\eqref{eq:external_guided};
               \item the deposit $ \deposit_{\party,\profile}$ \wrt $\infractionPredicate$ 
                   follows Eq.~\eqref{eq:deposit};
            \end{itemize}
          \item For every party $\party \in \partySet$, it holds that
              (i) $\power_\party < \threshold$ and 
              (ii) $\beta_\party - \power_\party \cdot \deposit\cdot\rateMax \leq \negl(\secparam)$.  
    \end{itemize}
Then, it holds that $\priceStab\leq 1+\negl(\secparam)$.
\end{theorem}


Theorem~\ref{thm:accountable_maximal_pst}\ifconference\else~(proof in Appendix~\ref{app:acc_pst})\fi~uses the
assumptions of Theorem~\ref{thm:acc_maximal_sufficient}, setting an upper bound on the bribing
budget of $\threshold \cdot \deposit\cdot\rateMax$. We show that the
identified equilibrium also yields maximal welfare, so the Price of
Stability is negligibly close to $1$ in this setting as well.

\begin{theorem}\label{thm:accountable_maximal_pst}
    Assume the following:
    \begin{itemize}
        \item $\proto$: a protocol run by parties in $\partySet$ in $\rounds$
            rounds, with block-proportional rewards (Section~\ref{subsec:preliminaries_proportion}) with $\reward$
            rewards per block;
        \item \revision{each party's stake is $\stake_\party = \power_\party \cdot \stake$;}
        \item $\threshold$: a security threshold \wrt $\infractionPredicate$;
        \item for every strategy profile $\profile$:
            \begin{itemize}
                \item the exchange rate $\exchangeRate_\profile$ \wrt
                    $\infractionPredicate, \threshold$ follows
                    Eq.~\eqref{eq:exchange}; 
                \item the external rewards $\utilityBoost_{\party, \profile}$
                    due to guided bribing \wrt $\infractionPredicate$
                    follow Eq.~\eqref{eq:external_guided};
                     \item the deposit $ \deposit_{\party,\profile}$ \wrt $\infractionPredicate$ follows Eq.~\eqref{eq:deposit};
            \end{itemize}
        \item the condition  $\sum_{\party \in \partySet} \bribe_\party \leq \threshold\cdot\deposit\cdot\rateMax$ in
            Lemma~\ref{lem:acc_maximal_existence} holds, so  $\sum_{\party \in \partySet_\text{prom}^{\{\bribe_\party\},\deposit}} \power_\party < \threshold$. 
    \end{itemize}
Then, it holds that $\priceStab \leq 1 +\negl(\secparam)$.
\ignore{
    Then, the following hold:
    \begin{enumerate}
        \item\label{item:accountable_maximal_pst1} 
            $\priceStab \leq 1 +\frac{\threshold\cdot\deposit}{\rounds\cdot\reward\cdot\rateMax}$.

        \item\label{item:accountable_maximal_pst2} 
            Assume that $\rateMax \geq 2 \cdot \rateMin$ and that for some
            $\hat{\rounds} \geq \frac{\threshold - \frac{3}{\totalParties}}{\threshold - \frac{2}{\totalParties}} \cdot \rounds$,
            the set of conditions in Lemma~\ref{lem:acc_maximal_existence}
            holds. Then, there exist a participation power allocation
            $\{\power_\party\}_{\party \in \partySet}$ and a bribe allocation
            $\{\bribe_\party\}_{\party \in \partySet}$ \st for any constant
            $\gamma \in (0,1)$ and $\totalParties \geq 8^{\frac{1}{1 - \gamma}}$, 
            it holds that:
        \[ \priceStab \geq 1 + \Big( \threshold - \frac{1}{\totalParties^\gamma} \Big) \cdot \Big( 1 - \frac{\rateMin}{\rateMax} \Big) - \negl(\secparam)\;. \]
    \end{enumerate}
   } 
\end{theorem}


In summary, we show that the two identified equilibria achieve the highest
welfare possible, under their own sets of assumptions. This is particularly
useful, because the bribing attack is unsuccessful in both cases.  Therefore,
accountable rewards disincentivize bribing attacks, while also yielding maximal
rewards.

Finally, Theorem~\ref{thm:accountable_pan}\ifconference\else~(proof in Appendix~\ref{app:acc_pst})\fi~uses the negative result of
Theorem~\ref{thm:accountable_negative_nash} to offer a lower bound on the Price
of Anarchy under accountable rewards. This bound helps with the intuition
outlined in Theorem~\ref{thm:accountable_negative_nash}, \ie if an attack is
more harmful (to the exchange rate) then the Price of Anarchy tends to increase.

\begin{theorem}\label{thm:accountable_pan}
    Assume the following:
    \begin{itemize}
        \item $\proto$: a protocol run by parties in $\partySet$ in $\rounds$ rounds with block-proportional rewards (cf.
            Section~\ref{subsec:preliminaries_proportion}), with $\reward$
            rewards per block;
        \item \revision{each party's stake is $\stake_\party = \power_\party \cdot \stake$;}
        \item $\threshold$: the security threshold \wrt an infraction predicate $\infractionPredicate$;
        \item $\profile_\infractionPredicate$: the strategy profile where all parties perform $\infractionPredicate$;
        \item for every strategy profile $\profile$: 
            \begin{itemize}
                \item the exchange rate $\exchangeRate_\profile$ \wrt
                    $\infractionPredicate, \threshold$ follows
                    Eq.~\eqref{eq:exchange};
                \item the external rewards $\utilityBoost_{\party, \profile}$,
                    due to guided bribing \wrt $\infractionPredicate$, 
                    follow Eq.~\eqref{eq:external_guided};
               \item the compliance payout $ \deposit_{\party, \profile}$ \wrt $\infractionPredicate$ follows Eq.~\eqref{eq:deposit};
            \end{itemize}
          \item For every party $\party \in \partySet$, it holds that:
              (i) $\power_\party < 1 - \threshold$;
              (ii) $\beta_\party - \power_\party \cdot \deposit \cdot \rateMin \geq \negl(\secparam)$;
              (iii)  $\beta_\party - \power_\party \cdot \deposit\cdot\rateMax \leq \negl(\secparam)$.
    \end{itemize}
    Then, it holds that $\priceAnar \geq \frac{(\revision{\stake} + \rounds\cdot\reward + \deposit)\cdot\rateMax}{(\revision{\stake} + \rounds\cdot\reward)\cdot\rateMin+\budget}- \negl(\secparam)$.
\end{theorem}


\section{Conclusion}

Our work explores if an adversary with
a limited budget can
attack a PoS blockchain system via bribes, \st the market price of the
ledger's native token is affected. We
investigate two types of bribing attacks.

In guided bribing (Section~\ref{sec:bribing_guided}), a party receives the bribe
only by acting as instructed.
We show that the protocol is not an equilibrium, as
some parties are
always incentivized to accept the bribe, whereas the setting where all
parties get bribed is an equilibrium (Theorem~\ref{thm:guided}).
We then identify sufficient conditions for an equilibrium's existence and bound the
Prices of Stability and Anarchy. Our results suggest that a protocol cannot
defend such attacks without special mechanisms
(which are explored later in Section~\ref{sec:accountable}).

In effective bribing (Section~\ref{sec:bribing_effective}), a party receives the
bribe only if the attack is successful. Here,
both extreme scenarios, where (i) all parties follow the protocol or (ii) all
parties get bribed, are Nash equilibria. 
Therefore, our results suggest that if a system reaches the positive
equilibrium, this can be maintained without any further defense mechanisms,
albeit a negative equilibrium is still possible.

Finally, we analyze the effect of financial penalization, like slashing or dilution, against
guided bribing (Section~\ref{sec:accountable}). We identify two sets of
conditions that lead to positive equilibria, where parties achieve
maximal welfare (Price of Stability is $1$). These conditions assume an upper
bound on the bribing budget which depends on the aggregate
deposited amount. As Table~\ref{tab:slashing_deposits} shows,
violating this assumption in many real world systems
requires a bribing budget of multiple millions or billions of USD.
However, we again identify a negative
equilibrium, which can be maintained with an attacking budget that is inversely
proportional to the attack's severity on the system's exchange rate.
Consequently, albeit defenses like slashing and dilution do help turn the
protocol into an equilibrium, they are not all-inclusive solutions since
negative equilibria can still manifest.

Our work also poses various open questions for future work. 

First, we assume the adversary only wants to break safety.
However, bribes could be used to target liveness. A bribed party
could try to censor a transaction by not proposing or signing blocks that
include it. In this case, abstaining from participation or failing to create a
block, which is a liveness violation, results in lost rewards, so the
expected block rewards, hence the utility, of a party is affected by the
nature of the infraction.  Analyzing such liveness attacks is particularly
complex and an interesting topic of further exploration.

Second, our results restrict the bribing budget. Although the
bounds are arguably high, they could be insufficient for the real world.
Future work can estimate bounds for deployed systems, \eg based on
historical market data (to estimate the difference $\rateMax -
\rateMin$) or deposits in systems with slashing (similar
to Section~\ref{subsec:accountable_equilibria_maximal}).
%

Third, we focus on mechanisms that can be naturally explored under the standard game-theoretic model. However, there could exist attack scenarios where the adversary externally incentivizes parties and controls others (e.g., by splitting its budget in bribes and acquiring stake). In this case, some parties can be considered rational while others are Byzantine, and it would be interesting to extend our results under alternative frameworks, such as BART~\cite{aiyer2005bar}.

Finally, we consider block-proportional rewards, but alternative
schemes, like resource-proportional
rewards~\cite{karakostas2022compliance}, have been used in practice
and are worth analyzing. Additionally, taking into account off-protocol incentives, \eg 
MEV~\cite{mazorra2022price}, could approximate the real world better.

\section*{Acknowledgements}
This work was supported by Input Output (iohk.io) through their funding of the Edinburgh Blockchain Technology Lab.
Part of the work of Thomas Zacharias was conducted while a member of the Blockchain Technology Lab at the University of Edinburgh.

\bibliographystyle{plain}
\bibliography{bibliography/additional,bibliography/cryptobib/abbrev0,bibliography/cryptobib/crypto_crossref}

\ifconference
\else
    \appendix
    \ifsubmission

\ifsubmission
\section{Distributed Ledger Security Properties}\label{subsec:ledger-security-properties}
\else
\paragraph{Security properties}\label{subsec:ledger-security-properties}
\fi
A distributed ledger should guarantee two core properties, safety and
liveness (cf. Definition~\ref{def:ledger-properties}). We denote by $\prefix$
the prefix operation, \ie if $\ledger \prefix \ledger'$ then $\forall
i \in [1, |\ledger|]: \tx_i = \tx'_i$.

\begin{definition}\label{def:ledger-properties}
    A secure distributed ledger protocol $\proto$, run by $\totalParties$ parties,
    should satisfy the following properties:
    \begin{itemize}
        \item \emph{Safety}: Let $\ledger_{\party_1, \round_1},
            \ledger_{\party_2, \round_2}$ be the ledgers output by parties
            $\party_1, \party_2$ on rounds $\round_1, \round_2$ respectively.
            The following condition must hold: $\forall \round_1, \round_2\; \forall
            \party_1, \party_2: \ledger_{\party_1, \round_1} \prefix
            \ledger_{\party_2, \round_2} \cup \ledger_{\party_2, \round_2}
            \prefix \ledger_{\party_1, \round_1}$.
        
        \item \emph{Liveness}: If a transaction $\tx$ is provided as input to
            an honest party on slot $\slot$, all honest parties output a ledger
            containing $\tx$ at slot $\slot + \livenessParam$, $\livenessParam$ being the ledger's liveness parameter.
    \end{itemize}
\end{definition}

Ledger protocols considered in this work typically satisfy safety as long
as $\frac{2}{3}$ of the participating power is controlled by parties that
follow $\proto$; similarly, for liveness the threshold is $\frac{1}{3}$. In the
rest of the paper we will denote the security threshold by $\threshold$ and set
$\threshold = \frac{1}{3}$, unless specified otherwise.

\fi

\section{Proofs}\label{app:proofs}

\subsection{Guided Bribing}

\subsubsection{Equilibria and Compliance}\label{app:eq_comp}

\paragraph{\textsc{Theorem}~\ref{thm:guided}}
 \emph{   Assume the following:
    \begin{itemize}
        \item $\proto$ is a protocol with block-proportional rewards (cf.
            Section~\ref{subsec:preliminaries_proportion}), with $\reward$
            rewards per block;
        \item \revision{each party's stake is $\stake_\party = \power_\party \cdot \stake$;}
        \item $\profile_\proto$ denotes the all-honest strategy profile;
        \item $\threshold$ is the security threshold \wrt an infraction
            predicate $\infractionPredicate$;
        \item for every strategy profile $\profile$: 
            \begin{itemize}
                \item the exchange rate $\exchangeRate_\profile$ \wrt
                    $\infractionPredicate, \threshold$ is defined as in
                    Eq.~\eqref{eq:exchange};
                \item the external rewards $\utilityBoost_{\party,\profile}$,
                    due to guided bribing \wrt $\infractionPredicate$, are
                    defined as in Eq.~\eqref{eq:external_guided}.
            \end{itemize}
    \end{itemize}
    Then, the following hold:
    \begin{enumerate}
        \item
            If for every party $\party \in
            \partySet$, it holds that $\power_\party \leq 1 - \threshold$, then
            exists $\epsilon$ negligible in $\secparam$ \st the strategy
            profile $\profile_\infractionPredicate$ is an $\epsilon$-Nash
            equilibrium \wrt Reward under $\router_\text{sync}$.
        \item
            Assume there exists a party
            $\party \in \partySet$ \st $\power_\party\geq\threshold$. Let
            $\theta > 0$ be a real value \st the probability
            $\Pr \big[ \msgNumber_{\party, \profile_\proto}>(1+\theta) \cdot E[\msgNumber_{\party, \profile_\proto}] \big]$
            is negligible in $\secparam$, where $\msgNumber_{\party,
            \profile_\proto}$ is the number of blocks created by $\party$ in
            $\profile_\proto$. 
            If $\bribe_\party \geq \big(\revision{\stake_\party} + (1 + \theta) \cdot E[\msgNumber_{\party, \profile_\proto}] \cdot \reward \big) \cdot \rateMax$, 
            then exists $\epsilon$ negligible in $\secparam$ \st the strategy
            profile $\profile_\infractionPredicate$ is an $\epsilon$-Nash
            equilibrium \wrt Reward under $\router_\text{sync}$.
        \item
            Assume there exists a party
            $\party \in \partySet$ \st $\power_\party < \threshold$ and
            $\bribe_\party > 0$. Then, for every constant $\gamma \in (0, 1)$,
            $\proto$ is not
            $(\gamma \cdot \bribe_\party, \infractionPredicate)$-compliant \wrt
            Reward under $\router_\text{sync}$.
    \end{enumerate}
    }
\begin{proof}
    \emph{(\ref{thm:guided_nash1}).}
    For some subset of parties $\infractionPartySet$, which follow
    $\strategy_\infractionPredicate$, let $F_\infractionPartySet$ be the event that the
    parties in $\infractionPartySet$ fail to perform the infraction. By the discussion on Definition~\ref{def:attack-failure-event}, if
    $\sum_{\party \in \infractionPartySet} \power_\party \geq \threshold$, 
    then $\Pr[F_\infractionPartySet] = \negl(\secparam)$. 

    Let $Q_\party$ be the event that $\party$ does not perform an infraction. By Section~\ref{subsec:bribe_IPs} , we have that $\Pr[Q_\party] = \negl(\secparam)$. Clearly, when $\neg Q_\party$ happens, the external rewards of $\party$ are $\bribe_\party$.

    Assume the strategy profile
    $\profile_\infractionPredicate = \langle \strategy_\infractionPredicate, \ldots, \strategy_\infractionPredicate \rangle$.
    It is straightforward that $\Pr[F_\partySet] = \negl(\secparam)$. 
    Therefore,
    $\neg F_\partySet$ happens with overwhelming probability, \ie all parties
    perform the infraction. Therefore, the exchange rate is $\rateMin$ and the parties
    receive their external rewards. By Eq.~\eqref{eq:exchange}
    and~\eqref{eq:external_guided}, and the fact that under block-proportional
    rewards, a party's maximum system's reward is polynomially bounded by
    $\rounds \cdot \reward$, where $\rounds$ is the number of rounds in the
    execution, the utility of some party $\party$ is:

    \begin{equation}\label{eq:utility_ds_all}
        \begin{split}
           & \utility_\party(\profile_\infractionPredicate)=\\
            =& \revision{\stake_\party \cdot E[\exchangeRate_{\profile_\infractionPredicate}] } + E[\reward_{\party, \profile_\infractionPredicate}\cdot\exchangeRate_{\profile_\infractionPredicate}]+E[\utilityBoost_{\party, \profile_\infractionPredicate}]=\\
            =& \revision{\stake_\party \cdot \sum_z z \cdot \Pr[\exchangeRate_{\profile_\infractionPredicate}=z]} + \sum_z z\cdot\Pr[\reward_{\party, \profile_\infractionPredicate}\cdot \exchangeRate_{\profile_\infractionPredicate}=z]+\\
            &+\sum_z z\cdot\Pr[\utilityBoost_{\party, \profile_\infractionPredicate}=z]\geq\\
            \geq& \revision{\stake_\party \cdot \sum_z z\cdot\Pr[\exchangeRate_{\profile_\infractionPredicate}=z\wedge (\neg F_\partySet)]} +\\
            &+\sum_z z\cdot\Pr[\reward_{\party, \profile_\infractionPredicate}\cdot \exchangeRate_{\profile_\infractionPredicate}=z\wedge (\neg F_\partySet)]+\\
            &+\sum_z z\cdot\Pr[\utilityBoost_{\party, \profile_\infractionPredicate}=z\wedge (\neg Q_\party)]=\\
            =& \revision{\stake_\party \cdot \sum_z z\cdot\Pr[\neg F_\partySet]\cdot\Pr[\exchangeRate_{\profile_\infractionPredicate}=z|\neg F_\partySet]} +\\
            &+\sum_z z\cdot\Pr[\neg F_\partySet]\cdot\Pr[\reward_{\party, \profile_\infractionPredicate}\cdot \exchangeRate_{\profile_\infractionPredicate}=z|\neg F_\partySet]+\\
            &+\sum_z z\cdot\Pr[\neg Q_\party]\cdot\Pr[\utilityBoost_{\party, \profile_\infractionPredicate}=z|\neg Q_\party]\geq\\
            \geq &\revision{\stake_\party \cdot \rateMin - \negl(\secparam)}+\\
            &+\sum_z z\cdot\Pr[\neg F_\partySet]\cdot\Pr[\reward_{\party, \profile_\infractionPredicate}\cdot \rateMin=z|\neg F_\partySet]+\\
            &+\sum_z z\cdot\Pr[\neg Q_\party]\cdot\Pr[\utilityBoost_{\party, \profile_\infractionPredicate}=z|\neg Q_\party]=\\
            =&\revision{\stake_\party \cdot \rateMin - \negl(\secparam)}+\\
            &+\sum_z z\cdot\Pr[\reward_{\party, \profile_\infractionPredicate}\cdot \rateMin=z\wedge\neg F_\partySet]+\\
            &+\sum_z z\cdot\Pr[\neg Q_\party]\cdot\Pr[\utilityBoost_{\party, \profile_\infractionPredicate}=z|\neg Q_\party]\geq\\
            \geq&\revision{\stake_\party \cdot \rateMin - \negl(\secparam)}+\\
            &+\sum_z z\cdot\big(\Pr[\reward_{\party, \profile_\infractionPredicate}\cdot \rateMin=z]-\negl(\secparam)\big)+\\
            &+\sum_z z\cdot\Pr[\neg Q_\party]\cdot\Pr[\utilityBoost_{\party, \profile_\infractionPredicate}=z|\neg Q_\party]\geq\\
            \geq&\revision{\stake_\party \cdot \rateMin - \negl(\secparam)}+\\
            &+\sum_z z\cdot\Pr[\reward_{\party, \profile_\infractionPredicate}\cdot \rateMin=z]-\sum_z z\cdot\negl(\secparam)+\\
            &+(1-\negl(\secparam))\cdot\sum_z z\cdot\Pr[\utilityBoost_{\party, \profile_\infractionPredicate}=z|\neg Q_\party]\geq\\
            \geq&\revision{\stake_\party \cdot \rateMin - \negl(\secparam)}+\\
            &+E[\reward_{\party, \profile_\infractionPredicate}]\cdot\rateMin-(\rounds \cdot \reward)^2\cdot\negl(\secparam)+(1-\negl(\secparam))\cdot\bribe_\party=\\
            =& \revision{\stake_\party \cdot \rateMin} + E[\reward_{\party, \profile_\infractionPredicate}]\cdot\rateMin+\bribe_\party-\negl(\secparam).
        \end{split}
    \end{equation}
    Assume now that some party $\party^*$ unilaterally deviates from
    $\profile_\infractionPredicate$ to some strategy profile $\profile$. 

    By assumption, $\power_{\party^*} \leq 1 - \threshold$. Observe that, under
    $\profile$, all other parties, which collectively hold at least
    $\threshold$ of participating power, perform an infraction whenever they
    can. So when $\neg F_{\partySet \setminus \{ \party^* \}}$ happens, the
    exchange rate is $\rateMin$ irrespective of the strategy of $\party^*$.
    Since $\sum_{\party \in \partySet \setminus \{ \party^* \}} \power_\party \geq \threshold$,
    we have that $\Pr \big[ F_{\party \in \partySet \setminus \{ \party^* \}} \big] = \negl(\secparam)$.
    In addition, a party's expected ledger rewards are maximized when it
    follows $\strategy_\infractionPredicate$. Given the fact that, under block-proportional
    rewards, a party's maximum system's reward is polynomially bounded by
    $\rounds \cdot \reward$, we get that:
     
    \begin{equation}\label{eq:utility_ds_all_deviate}
        \begin{split}
           &\utility_{\party^*}(\profile)=\\
            =&\revision{\stake_{\party^*} \cdot E[\exchangeRate_\profile]} + E[\reward_{\party^*, \profile}\cdot\exchangeRate_\profile]+E[\utilityBoost_{\party^*, \profile}]\leq\\
            \leq& \revision{\stake_{\party^*} \cdot E[\exchangeRate_\profile]} + E[\reward_{\party^*, \profile}\cdot\exchangeRate_\profile]+\bribe_{\party^*}=\\
            =&\revision{\stake_{\party^*} \cdot \sum_z z\cdot\Pr[\exchangeRate_\profile=z]} + \sum_z z\cdot\Pr[\reward_{\party^*, \profile}\cdot\exchangeRate_\profile=z]+\bribe_{\party^*}=\\
            =&\revision{\stake_{\party^*} \cdot \sum_z z\cdot\Pr\big[\exchangeRate_\profile=z\wedge\big(\neg F_{\party\in\partySet\setminus\{\party^*\}}\big)\big]}+\\
            &+\revision{\stake_{\party^*} \cdot \sum_z z\cdot\Pr\big[\exchangeRate_\profile=z\wedge  F_{\party\in\partySet\setminus\{\party^*\}}\big]}+\\
            &+\sum_z z\cdot\Pr\big[\reward_{\party^*, \profile}\cdot\exchangeRate_\profile=z\wedge\big(\neg F_{\party\in\partySet\setminus\{\party^*\}}\big)\big]+\\
            &+\sum_z z\cdot\Pr\big[\reward_{\party^*, \profile}\cdot\exchangeRate_\profile=z\wedge F_{\party\in\partySet\setminus\{\party^*\}}\big]+\bribe_{\party^*}\leq\\
            \leq&\revision{\stake_{\party^*} \cdot \sum_z \Big(z\cdot\Pr\big[\neg F_{\party\in\partySet\setminus\{\party^*\}}\big]\cdot\Pr\big[\exchangeRate_\profile=z\big|\neg F_{\party\in\partySet\setminus\{\party^*\}}\big]\Big)}+\\
            &+\revision{\stake_{\party^*} \cdot \sum_z z\cdot\Pr\big[ F_{\party\in\partySet\setminus\{\party^*\}}\big]}+\\
            &+\sum_z \Big(z\cdot\Pr\big[\neg F_{\party\in\partySet\setminus\{\party^*\}}\big]\cdot\\
            &\quad\cdot\Pr\big[\reward_{\party^*, \profile}\cdot \exchangeRate_\profile=z\big|\neg F_{\party\in\partySet\setminus\{\party^*\}}\big]\Big)+\\
            &+\sum_z z\cdot\Pr\big[ F_{\party\in\partySet\setminus\{\party^*\}}\big]+\bribe_{\party^*}=\\
            =&\revision{\stake_{\party^*} \cdot \rateMin \cdot \Pr\big[\neg F_{\party\in\partySet\setminus\{\party^*\}}\big]}+\revision{\stake_{\party^*} \cdot \sum_z z\cdot\negl(\secparam)}+\\
            &+\sum_z \Big(z\cdot\Pr\big[\neg F_{\party\in\partySet\setminus\{\party^*\}}\big]\cdot\\
            &\quad\cdot\Pr\big[\reward_{\party^*, \profile}\cdot \rateMin=z\big|\neg F_{\party\in\partySet\setminus\{\party^*\}}\big]\Big)+\\
            &+\sum_z z\cdot\Pr\big[ F_{\party\in\partySet\setminus\{\party^*\}}\big]+\bribe_{\party^*}\leq\\
            \leq&\revision{\stake_{\party^*} \cdot \rateMin \cdot (1 - \negl(\secparam))} + \revision{\stake_{\party^*} \cdot \rateMax^2 \cdot \negl(\secparam)}\\
            &+\sum_z z\cdot\Pr\big[\reward_{\party^*, \profile}\cdot \rateMin=z\wedge\neg F_{\party\in\partySet\setminus\{\party^*\}}\big]+\\
            &+\sum_z z\cdot\negl(\secparam)+\bribe_{\party^*}\leq\\
            \leq&\revision{\stake_{\party^*} \cdot \rateMin + \negl(\secparam)} + \\
            &+\sum_z z\cdot\Pr\big[\reward_{\party^*, \profile}\cdot \rateMin=z\big]+\\
            &+\sum_z z\cdot\negl(\secparam)+\bribe_{\party^*}\leq\\
            \leq& \revision{\stake_{\party^*} \cdot \rateMin} + \revision{\negl(\secparam)}+E[\reward_{\party^*, \profile}]\cdot\rateMin + (\rounds\cdot\reward)^2  \cdot\negl(\secparam)+\bribe_{\party^*}\leq\\
            \leq& \revision{\stake_{\party^*} \cdot \rateMin} + E[\reward_{\party^*, \profile}]\cdot\rateMin+\bribe_{\party^*}+\negl(\secparam)\leq\\
            \leq& \revision{\stake_{\party^*} \cdot \rateMin} + E[\reward_{\party^*, \profile_\infractionPredicate}]\cdot\rateMin+\bribe_{\party^*}+\negl(\secparam).
        \end{split}
    \end{equation}
    Therefore, by Equations~\eqref{eq:utility_ds_all}
    and~\eqref{eq:utility_ds_all_deviate}, we have that:
    \[ \utility_{\party^*}(\profile) \leq \utility_{\party^*}(\profile_\infractionPredicate) + \negl(\secparam) .\]

    So, $\profile_\infractionPredicate$ is an $\epsilon$-Nash equilibrium \wrt
    Reward under $\router_\text{sync}$, for some $\epsilon = \negl(\secparam)$.

\medskip
\emph{(\ref{thm:guided_nash2})}. For every party $\party^* \neq \party$, it holds that 
    $\power_{\party^*} \leq 1 - \threshold$. By following the same steps
    as above, we show that for any strategy profile $\profile^*$ that results
    as a unilateral deviation of $\party^*$ from $\profile_\infractionPredicate$,
    it holds that:
    \[ \utility_{\party^*}(\profile^*) \leq \utility_{\party^*}(\profile_\infractionPredicate) + \negl(\secparam) \]

    Therefore, it suffices to examine the unilateral deviations of $\party$.

    By assumption, we have that $\power_\party \geq \threshold$ and
    $\bribe_\party \geq \big(\revision{\stake_\party} + (1 + \theta) \cdot E[\msgNumber_{\party, \profile_\proto}] \cdot \reward \big) \cdot \rateMax$, 
    for $\theta > 0$.
    Assume that $\party$ unilaterally deviates from
    $\profile_\infractionPredicate$ by following a strategy profile $\profile$.
    Without loss of generality, we assume that in terms of block creation,
    $\party$ behaves honestly. Indeed, any deviation from honest block creation
    (\eg abstaining) would only reduce $\party$'s utility.

    Let $D$ be the event that $\party$ performs an infraction \wrt
    $\infractionPredicate$ in the execution. If $D$ occurs then, since
    $\power_\party \geq \threshold$, the exchange rate will drop to $\rateMin$
    and $\party$ will receive its external rewards. On the contrary, if $D$
    does not occur, then the exchange rate will be $\rateMax$ and $\party$ will
    receive no external rewards. Since block creation is the same as in an
    honest manner, the values of the random variable $\reward_{\party, \profile}$ 
    are independent of the occurrence of $D$.

    Let $\msgNumber_{\party, \profile}$ be the number of blocks created by
    $\party$ in $\profile$. By the assumption for $\theta$ and given that
    $\reward_{\party, \profile} \leq \msgNumber_{\party, \profile} \cdot \reward$,
    we have that:

    \begin{equation*}
        \begin{split}
            &\utility_\party(\profile)=\\
            =&\revision{\stake_\party \cdot E[\exchangeRate_{\profile}]} + E[\reward_{\party, \profile}\cdot\exchangeRate_{\profile}]+E[\utilityBoost_{\party, \profile}]=\\
            =&\sum_z z\cdot\Pr[(\revision{\stake_\party} + \reward_{\party, \profile})\cdot\exchangeRate_{\profile}+\utilityBoost_{\party, \profile}=z]=\\
            =&\sum_z z\cdot\Pr[(\revision{\stake_\party} + \reward_{\party, \profile})\cdot\exchangeRate_{\profile}+\utilityBoost_{\party, \profile}=z\wedge D]+\\
            &+\sum_z z\cdot\Pr\big[(\revision{\stake_\party} + \reward_{\party, \profile})\cdot\exchangeRate_{\profile}+\utilityBoost_{\party, \profile}=z\wedge \neg D\wedge\\
            &\quad \wedge\big(\msgNumber_{\party, \profile}>(1+\theta)\cdot E[\msgNumber_{\party, \profile_\proto}]\big)\big]+\\
            &+\sum_z z\cdot\Pr\big[(\revision{\stake_\party} + \reward_{\party, \profile})\cdot\exchangeRate_{\profile}+\utilityBoost_{\party, \profile}=z\wedge \neg D\wedge\\
            &\quad \wedge\big(\msgNumber_{\party, \profile}\leq(1+\theta)\cdot E[\msgNumber_{\party, \profile_\proto}]\big)\big]\leq\\
            \leq&\Pr[D]\cdot\sum_z z\cdot\Pr[(\revision{\stake_\party} + \reward_{\party, \profile})\cdot\exchangeRate_{\profile}+\utilityBoost_{\party, \profile}=z| D]+\\
            &+\sum_z z\cdot\Pr\big[\msgNumber_{\party, \profile}>(1+\theta)\cdot E[\msgNumber_{\party, \profile_\proto}]\big]+\\
            &+\sum_z z\cdot\Pr\big[(\revision{\stake_\party} + \reward_{\party, \profile})\cdot\exchangeRate_{\profile}+\utilityBoost_{\party, \profile}=z\wedge \neg D\wedge\\
            &\quad \wedge\big(\revision{\stake_\party} + \reward_{\party, \profile} \leq \revision{\stake_\party} + (1+\theta)\cdot E[\msgNumber_{\party, \profile_\proto}]\cdot\reward\big)\big]=\\
            =&\Pr[D]\cdot\sum_z z\cdot\Pr[(\revision{\stake_\party} + \reward_{\party, \profile})\cdot\rateMin+\bribe_\party=z| D]+\\
            &+\sum_z z\cdot\Pr\big[\msgNumber_{\party, \profile}>(1+\theta)\cdot E[\msgNumber_{\party, \profile_\proto}]\big]+\\
            &+\sum_z z\cdot\Pr\big[(\revision{\stake_\party} + \reward_{\party, \profile})\cdot\exchangeRate_{\profile}+\utilityBoost_{\party, \profile}=z\wedge \neg D\wedge\\
            &\quad \wedge\big(\revision{\stake_\party} +\reward_{\party, \profile} \leq\revision{\stake_\party} +(1+\theta)\cdot E[\msgNumber_{\party, \profile_\proto}]\cdot\reward\big)\big]\leq\\
            \leq&\Pr[D]\cdot\sum_z z\cdot\Pr[(\revision{\stake_\party} + \reward_{\party, \profile})\cdot\rateMin+\bribe_\party=z]+\\
            &+\sum_z z\cdot\Pr\big[\msgNumber_{\party, \profile_\proto}>(1+\theta)\cdot E[\msgNumber_{\party, \profile_\proto}]\big]+\\
            &+\sum_z z\cdot\Pr\big[(\revision{\stake_\party} + \reward_{\party, \profile})\cdot\exchangeRate_{\profile}+\utilityBoost_{\party, \profile}=z\wedge \neg D\wedge\\
            &\quad \wedge\big((\revision{\stake_\party} +\reward_{\party, \profile})\cdot \rateMax\leq\\
            &\quad\quad\leq\big(\revision{\stake_\party} +(1+\theta)\cdot E[\msgNumber_{\party, \profile_\proto}]\cdot\reward\big)\cdot \rateMax\big)\big]\leq\\
                    \end{split}
    \end{equation*}
                \begin{equation}\label{eq:utility_ds_P}
        \begin{split}
            \leq&\Pr[D]\cdot ((\revision{\stake_\party} + E[\reward_{\party, \profile}])\cdot\rateMin+\bribe_\party)+\\
            &+\big(\revision{\stake} + (\rounds\cdot\reward)^2\big)\cdot\negl(\secparam)+\\
            &+\sum_z z\cdot\Pr\big[(\revision{\stake_\party} + \reward_{\party, \profile})\cdot\exchangeRate_{\profile}+\utilityBoost_{\party, \profile}=z\wedge \neg D\wedge\\
            &\quad\wedge(\revision{\stake_\party} +\reward_{\party, \profile})\cdot \rateMax\leq\bribe_\party\big]\leq\\
            \leq&\Pr[D]\cdot ((\revision{\stake_\party} + E[\reward_{\party, \profile}])\cdot\rateMin+\bribe_\party)+\negl(\secparam)+\\
            &+\sum_z z\cdot\Pr\big[(\revision{\stake_\party} + \reward_{\party, \profile})\cdot \rateMax+0=z\wedge \neg D\wedge\\
            &\quad\wedge(\revision{\stake_\party} +\reward_{\party, \profile})\cdot \rateMax\leq\bribe_\party\big]\leq\\
            \leq& \revision{\stake_\party \cdot \rateMin} + E[\reward_{\party, \profile}]\cdot\rateMin+\Pr[D]\cdot\bribe_\party+\negl(\secparam)+\\
            &+\Pr[\neg D]\cdot\sum_z z\cdot\Pr\big[(\revision{\stake_\party} + \reward_{\party, \profile}) \cdot \rateMax=z\big|\neg D\wedge\\
            &\quad\wedge(\revision{\stake_\party} +\reward_{\party, \profile})\cdot \rateMax\leq\bribe_\party\big]\leq\\
            \leq& \revision{\stake_\party \cdot \rateMin} + E[\reward_{\party, \profile_\infractionPredicate}]\cdot\rateMin+\Pr[D]\cdot\bribe_\party+\negl(\secparam)+\\
            &+\Pr[\neg D]\cdot\sum_{z\leq\bribe_\party} z\cdot\Pr\big[(\revision{\stake_\party} + \reward_{\party, \profile})\cdot \rateMax=z\big|\neg D\wedge\\
            &\quad\wedge(\revision{\stake_\party} +\reward_{\party, \profile})\cdot \rateMax\leq\bribe_\party\big]\leq\\
            \leq& \revision{\stake_\party \cdot \rateMin} + E[\reward_{\party, \profile_\infractionPredicate}]\cdot\rateMin+\Pr[D]\cdot\bribe_\party+\negl(\secparam)+\\
            &+\Pr[\neg D]\cdot\sum_{z\leq\bribe_\party} \bribe_\party\cdot\Pr\big[(\revision{\stake_\party} + \reward_{\party, \profile})\cdot \rateMax=z\big|\neg D\wedge\\
            &\quad\wedge(\revision{\stake_\party} +\reward_{\party, \profile})\cdot \rateMax\leq\bribe_\party\big]=\\
            =&\revision{\stake_\party \cdot \rateMin} + E[\reward_{\party, \profile_\infractionPredicate}]\cdot\rateMin+\Pr[D]\cdot\bribe_\party+\\
            &+\Pr[\neg D]\cdot\bribe_\party\cdot 1+\negl(\secparam)=\\
            =&\revision{\stake_\party \cdot \rateMin} + E[\reward_{\party, \profile_\infractionPredicate}]\cdot\rateMin+\bribe_\party+\negl(\secparam).
        \end{split}
    \end{equation}

    Following the steps in Eq.~\eqref{eq:utility_ds_all}, we can show that 
    $\utility_\party(\profile_\infractionPredicate) = \revision{\stake_\party \cdot \rateMin} + E[\reward_{\party, \profile_\infractionPredicate}] \cdot \rateMin + \bribe_\party - \negl(\secparam)$.
    The latter, along with Eq.~\eqref{eq:utility_ds_P}, implies that
    \[ \utility_{\party}(\profile) \leq \utility_{\party}(\profile_\infractionPredicate) + \negl(\secparam) .\]

    Therefore, $\profile_\infractionPredicate$ is an $\epsilon$-Nash
    equilibrium \wrt Reward under $\router_\text{sync}$, for some
    $\epsilon = \negl(\secparam)$.

\medskip
\emph{(\ref{thm:guided_compliance})}. Let $\party$ be a party with $\power_\party < \threshold$ and
    $\bribe_\party > 0$. For every unilateral deviation of $\party$ from
    $\profile_\proto$, since $\power_\party < \threshold$ and the other parties
    follow the protocol $\proto$ honestly, it holds that the exchange rate
    remains always $\rateMax$. \revision{Therefore, $\party$'s stake is $\stake_\party
    \cdot \rateMax$.}

    Assume that $\party$ unilaterally deviates from $\profile_\proto$ by
    following $\strategy_\infractionPredicate$. Let $\profile^*$ be the
    resulting strategy profile, where all parties are honest except from
    $\party$ that follows $\strategy_\infractionPredicate$. 

    In any execution trace $\trace$, where $\party$ follows $\profile^*$, all
    the blocks created by $\party$ will be included in the chain of the
    observer just as if $\party$ was honest. Additionally, if $\party$
    performs an infraction \wrt $\infractionPredicate$, $\party$ will receive its 
    external rewards. Recall that the event $D$ where
    $\party$ performs an infraction when following
    $\strategy_\infractionPredicate$ happens with $1 - \negl(\secparam)$
    probability. Thus, given that $\bribe_\party > 0$, we have that, for every
    constant $\gamma \in (0,1)$:

    \begin{equation}\label{eq:utility_ds_comp_2}
        \begin{split}
            &\utility_{\party}(\profile^*)=\\
            =& \revision{\stake_\party \cdot E[\exchangeRate_{\profile^*}]} + E[\reward_{\party, \profile^*}\cdot\exchangeRate_{\profile^*}]+E[\utilityBoost_{\party, \profile^*}]=\\
            =& \revision{\stake_\party \cdot \rateMax} + E[\reward_{\party, \profile_\proto}\cdot \rateMax]+\sum_z z\cdot\Pr[\utilityBoost_{\party, \profile^*}=z]\geq\\
            \geq& \revision{\stake_\party \cdot \rateMax} + E[\reward_{\party, \profile_\proto}\cdot \rateMax]+\sum_z z\cdot\Pr[\neg D]\cdot\Pr[\utilityBoost_{\party, \profile^*}=z|\neg D]\geq\\
            \geq& \revision{\stake_\party \cdot E[\exchangeRate_{\profile_\proto}]} + E[\reward_{\party, \profile_\proto}\cdot\exchangeRate_{\profile_\proto}]+(1-\negl(\secparam))\cdot\bribe_\party>\\
            >&\utility_{\party}(\profile_\proto)+\gamma\cdot\bribe_\party.
        \end{split}
    \end{equation}
     
    Next, we show that for any unilateral deviation $\profile$ of $\party$ from
    the all-honest profile, if $\profile$ is $\infractionPredicate$-compliant
    (\ie $\party$ deviates from $\profile_\proto$ but never performs an
    infraction \wrt $\infractionPredicate$), it holds that
    $\utility_{\party}(\profile) \leq \utility_{\party}(\profile_\proto)$. 
    The latter holds because:
    \begin{inparaenum}[(i)]
        \revision{\item stake is the same in both profiles, since the exchange rate is not affected;}
        \item the expected ledger rewards are maximized in $\profile_\proto$,
            \ie $E[\reward_{\party, \profile}] \leq E[\reward_{\party,
            \profile_\proto}]$;
        \item since $\infractionPredicate$-compliant, the external rewards of
            $\party$ in $\profile$ are always $0$.
    \end{inparaenum}
    Consequently, if there is a best response for $\party$ with strictly
    greater utility than $\profile_\proto$, then it must be not
    $\infractionPredicate$-compliant.

    By the above, the fact that $\profile^*$ is not
    $\infractionPredicate$-compliant, and Eq.~\eqref{eq:utility_ds_comp_2}, we
    conclude that there is a strategy profile $\profile_\mathsf{best}$ (not
    necessarily $\profile^*$) that:
    \begin{inparaenum}[(i)]
        \item is a unilateral deviation from $\profile_\proto$,
        \item for any constant $\gamma\in (0,1)$,
            $\utility_{\party}(\profile_\mathsf{best}) > \utility_{\party}(\profile_\proto) + \gamma \cdot \bribe_\party$, 
        \item sets a best response for $\party$, and
        \item not $\infractionPredicate$-compliant.
    \end{inparaenum}

    So, for every constant $\gamma \in (0,1)$, $\profile_\mathsf{best}$ is
    directly $(\gamma \cdot \bribe_\party)$-reachable from $\profile_\proto$,
    which implies that $\proto$ is not 
    $(\gamma \cdot \bribe_\party, \infractionPredicate)$-compliant 
    \wrt Reward under $\router_\text{sync}$.



\end{proof}

\subsubsection{Maximal Sets and Promising Parties}\label{app:guided_max_prom}

\paragraph{\textsc{Lemma}~\ref{lem:maximal_existence}}
\emph{    Assume the following:
    \begin{itemize}
        \item $\proto$ is a protocol run by parties in $\partySet$ in $\rounds$ rounds, with block-proportional rewards (cf.
            Subsection~\ref{subsec:preliminaries_proportion}) with $\reward$
            rewards per block; 
        \item \revision{each party's stake is $\stake_\party = \power_\party \cdot \stake$;}
        \item $\threshold$ is a security threshold \wrt $\infractionPredicate$;
        \item for every strategy profile $\profile$:
            \begin{itemize}
                \item the exchange rate $\exchangeRate_\profile$ \wrt
                    $\infractionPredicate, \threshold$ is defined as in
                    Eq.~\eqref{eq:exchange}; 
                \item the external rewards $\utilityBoost_{\party,\profile}$
                    due to guided bribing \wrt $\infractionPredicate$ are
                    defined as in Eq.~\eqref{eq:external_guided};
            \end{itemize}
        \item $\msgNumber_{\party, \profile_\proto}$ denotes the number of
            blocks created by $\party$ in all-honest strategy profile
            $\profile_\proto$.
    \end{itemize}
    Consider the following two sets of conditions:
    \begin{enumerate}
        \item 
            (i) $\threshold \geq \frac{1}{2}$; \\
            (ii) $\forall \infractionPartySet, \mathbb{B} \subseteq \partySet: \sum_{\party\in\infractionPartySet} \power_\party \leq \sum_{\party \in \mathbb{B}} \power_\party 
                \Rightarrow \sum_{\party \in \infractionPartySet} E[\msgNumber_{\party, \profile_\proto}] \leq \sum_{\party \in \mathbb{B}} E[\msgNumber_{\party, \profile_\proto}]$; \\
            (iii) $\sum_{\party \in \partySet} \bribe_\party \leq \frac{1}{2} \cdot (\rounds \cdot \reward + \revision{\stake}) \cdot \rateDiff$.
        \item 
            (i) $\forall \party \in \partySet: E[\msgNumber_{\party, \profile_\proto}] = \power_\party \cdot \hat{\rounds} + \frac{\rounds - \hat{\rounds}}{\totalParties}$, where $\hat{\rounds} \in [0, \rounds]$;\\ 
            (ii)  $\sum_{\party \in \partySet} \bribe_\party \leq \Big( \threshold \cdot \hat{\rounds} + \frac{\rounds - \hat{\rounds}}{\totalParties} \Big) \cdot \reward \cdot \rateDiff + \revision{\threshold \cdot \stake \cdot \rateDiff}$, where $\hat{\rounds}$ as above.
    \end{enumerate}
    If either the two sets of conditions holds, then there exists a subset
    of parties $\infractionPartySet_m$ that is maximal \wrt
    $\threshold, \{\bribe_\party\}_{\party\in\partySet}$ (cf. Definition~\ref{def:maximal_set}).
    }
\begin{proof}
    We first show that if
    $\sum_{\party \in \partySet_\text{prom}^{\{\bribe_\party\}}} \power_\party < \threshold$,
    then there exists a subset of parties $\infractionPartySet_m$ that is
    maximal \wrt $\threshold, \{\bribe_\party\}_{\party\in\partySet}$. 
    In particular, the following algorithm outputs such a maximal subset:

    \begin{enumerate}
        \item[] On input $\big( \threshold, \reward, \rateDiff, \{\revision{\stake_\party,} \power_\party, \bribe_\party, E[\msgNumber_{\party, \profile_\proto}]\}_{\party \in \partySet} \big)$, 
            execute the following steps:
        \item Set $\infractionPartySet_m \leftarrow \partySet_\text{prom}^{\{\bribe_\party\}} := \{\party \in \partySet: \bribe_\party > (E[\msgNumber_{\party, \profile_\proto}] \cdot \reward + \revision{\stake_\party}) \cdot \rateDiff\}$;
        \item Set $\mathbf{L}$ as the list that contains the elements of the multiset $\{\power_\party: \party \in \partySet \setminus \partySet_\text{prom}^{\{\bribe_\party\}}\}$ sorted in increasing order;
        \item Set $\party^*$ as the party that corresponds to the first element of $\mathbf{L}$;
        \item If $\sum_{\party \in \infractionPartySet_m \cup \{\party^*\}} \power_\party < \threshold$, then
            (i) set $\infractionPartySet_m \leftarrow \infractionPartySet_m \cup \{\party^*\}$, 
            (ii) delete $\power_{\party^*}$ from $\mathbf{L}$, and 
            (iii) go to Step 3. 
            Else, output $\infractionPartySet_m$.
    \end{enumerate}

    Clearly, by the check in Step 4, the output of the algorithm satisfies Condition~\ref{item:maximal_condition1} of
    Definition~\ref{def:maximal_set}. We will prove that it also satisfies Condition~\ref{item:maximal_condition2} of
    Definition~\ref{def:maximal_set}. Namely, by the initialization of
    $\infractionPartySet_m$ and $\mathbf{L}$ in Steps 2 and 3, we have that for
    every party $\party$ that is not included in the output of the algorithm,
    it holds that $\bribe_\party \leq (E[\msgNumber_{\party, \profile_\proto}] \cdot \reward + \revision{\stake_\party}) \cdot \rateDiff$. 
    In addition, if there is a party
    $\party' \in \partySet \setminus \infractionPartySet_m$ \st
    $\sum_{\party \in \infractionPartySet_m \cup \{\party'\}} \power_\party < \threshold$,
    then for the party $\party'' \in \partySet \setminus \infractionPartySet_m$
    with the minimum participation power among all parties outside
    $\infractionPartySet_m$, it holds that
    $\sum_{\party \in \infractionPartySet_m \cup \{\party''\}} \power_\party < \threshold$.
    By definition $\party''$ should be included in $\infractionPartySet_m$ at
    Step 4, which leads to contradiction. Thus, for every party $\party^*$
    that is not included in the output of the algorithm, it also holds that
    $\sum_{\party \in \infractionPartySet_m \cup \{\party^*\}} \power_\party \geq \threshold$.
    So, the output of the algorithm is maximal \wrt
    $\threshold, \{\bribe_\party\}_{\party \in \partySet}$.\smallskip

    Now assume that the first set of conditions in the lemma statement holds.
    We will show that $\sum_{\party \in \partySet_\text{prom}^{\{\bribe_\party\}}} \power_\party < \threshold$.
    For the sake of contradiction, assume that
    $\sum_{\party \in \partySet_\text{prom}^{\{\bribe_\party\}}} \power_\party \geq \threshold$.
    Since $\threshold \geq \frac{1}{2}$ (condition 1.(i)), we have that
    \[ \sum_{\party \in \partySet_\text{prom}^{\{\bribe_\party\}}} \power_\party \geq \threshold \geq 1 - \threshold \]
    and
    \[ \sum_{\party \in \partySet \setminus \partySet_\text{prom}^{\{\bribe_\party\}}} \power_\party = 1 - \sum_{\party \in \partySet_\text{prom}^{\{\bribe_\party\}}} \power_\party \leq 1 - \threshold \]
    By condition 1.(ii), the latter implies that 
    \[ \sum_{\party \in \partySet \setminus \partySet_\text{prom}^{\{\bribe_\party\}}} E[\msgNumber_{\party, \profile_\proto}] \leq \sum_{\party \in \partySet_\text{prom}^{\{\bribe_\party\}}} E[\msgNumber_{\party, \profile_\proto}]. \]
    and
      \[ \revision{\sum_{\party \in \partySet \setminus \partySet_\text{prom}^{\{\bribe_\party\}}} \stake_\party \leq \sum_{\party \in \partySet_\text{prom}^{\{\bribe_\party\}}} \stake_\party}. \]

    Therefore, we have that
    \begin{equation*}
        \begin{split}
            &(\rounds \cdot \reward + \revision{\stake}) \cdot \rateDiff=\\
            =&\Big(\sum_{\party\in\partySet} E[\msgNumber_{\party, \profile_\proto}]\Big)\cdot\reward\cdot \rateDiff + \revision{\sum_{\party\in\partySet} \stake_\party \cdot \rateDiff} =\\
            =&\Big(\sum_{\party\in\partySet\setminus\partySet_\text{prom}^{\{\bribe_\party\}}}E[\msgNumber_{\party, \profile_\proto}]+ \sum_{\party\in\partySet_\text{prom}^{\{\bribe_\party\}}}E[\msgNumber_{\party, \profile_\proto}]\Big)\cdot\reward\cdot \rateDiff + \\
            &+\revision{\Big(\sum_{\party\in\partySet\setminus\partySet_\text{prom}^{\{\bribe_\party\}}} \stake_\party + \sum_{\party\in\partySet_\text{prom}^{\{\bribe_\party\}}} \stake_\party \Big) \cdot \rateDiff} \leq\\
            \leq& 2 \cdot \Big(\sum_{\party\in\partySet_\text{prom}^{\{\bribe_\party\}}}E[\msgNumber_{\party, \profile_\proto}]\cdot\reward + \revision{\sum_{\party\in\partySet_\text{prom}^{\{\bribe_\party\}}} \stake_\party} \Big) \cdot \rateDiff =\\
            =&2 \cdot \sum_{\party\in\partySet_\text{prom}^{\{\bribe_\party\}}}\Big(E[\msgNumber_{\party, \profile_\proto}]\cdot\reward + \revision{\sum_{\party\in\partySet_\text{prom}^{\{\bribe_\party\}}}\stake_\party} \Big) \cdot \rateDiff <\\
            <&2\cdot \sum_{\party\in\partySet_\text{prom}^{\{\bribe_\party\}}}\bribe_\party\leq\\
            \leq&2\cdot \sum_{\party\in\partySet}\bribe_\party\;,
        \end{split}
    \end{equation*}
    which contradicts condition 1.(iii).\smallskip

    Now assume that the second set of conditions in the lemma statement holds.
    Again, for the sake of contradiction, assume that
    $\sum_{\party \in \partySet_\text{prom}^{\{\bribe_\party\}}} \power_\party \geq \threshold$.
    By condition 2.(i), we have that
    \begin{equation*}
        \begin{split}
            &\sum_{\party\in\partySet}\bribe_\party>\\
            >&\Big(\sum_{\party\in\partySet_\text{prom}^{\{\bribe_\party\}}}E[\msgNumber_{\party, \profile_\proto}]\Big) \cdot \reward \cdot \rateDiff + \revision{\Big( \sum_{\party\in\partySet_\text{prom}^{\{\bribe_\party\}}} \stake_\party \Big) \cdot \rateDiff} =\\
            =&\Big(\sum_{\party\in\partySet_\text{prom}^{\{\bribe_\party\}}}\Big(\power_\party\cdot\hat{\rounds}+\dfrac{\rounds-\hat{\rounds}}{\totalParties}\Big)\Big)\cdot \reward \cdot \rateDiff + \revision{ \Big( \sum_{\party\in\partySet_\text{prom}^{\{\bribe_\party\}}} \power_\party \cdot \stake \Big) \cdot \rateDiff} \geq\\
            \geq&\Big(\Big(\sum_{\party\in\partySet_\text{prom}^{\{\bribe_\party\}}}\power_\party\cdot\hat{\rounds}\Big)+\dfrac{\rounds-\hat{\rounds}}{\totalParties}\Big)\cdot\reward\cdot \rateDiff + \revision{ \Big( \sum_{\party\in\partySet_\text{prom}^{\{\bribe_\party\}}} \power_\party \cdot \stake \Big) \cdot \rateDiff} \geq\\
            \geq&\Big(\threshold\cdot\hat{\rounds}+\dfrac{\rounds-\hat{\rounds}}{\totalParties}\Big)\cdot\reward\cdot \rateDiff + \revision{\threshold \cdot \stake \cdot \rateDiff} \;.
        \end{split}
    \end{equation*}
    which contradicts condition 2.(ii).\smallskip

    Therefore, if either of the sets of conditions holds, we have that
    $\sum_{\party \in \partySet_\text{prom}^{\{\bribe_\party\}}} \power_\party < \threshold$
    which implies the existence of a maximal subset \wrt
    $\threshold, \{\bribe_\party\}_{\party\in\partySet}$.
\end{proof}


\paragraph{\textsc{Theorem}~\ref{thm:maximal_sufficient}}
   \emph{ Assume the following:
    \begin{itemize}
        \item $\proto$ is a protocol run by parties in $\partySet$ in $\rounds$ rounds, with block-proportional rewards (cf.
            Subsection~\ref{subsec:preliminaries_proportion}) with $\reward$
            rewards per block; 
        \item \revision{each party's stake is $\stake_\party = \power_\party \cdot \stake$;}
        \item $\threshold$ is the security threshold \wrt an infraction
            predicate $\infractionPredicate$;
        \item for every strategy profile $\profile$: 
            \begin{itemize}
                \item the exchange rate $\exchangeRate_\profile$ \wrt
                    $\infractionPredicate, \threshold$ is defined as in
                    Eq.~\eqref{eq:exchange};
                \item the external rewards $\utilityBoost_{\party,\profile}$,
                    due to guided bribing \wrt $\infractionPredicate$, are
                    defined as in Eq.~\eqref{eq:external_guided}.
            \end{itemize}
    \end{itemize}
    If $\infractionPartySet \subset \partySet$ is maximal \wrt $\threshold,
    \{\bribe_\party\}_{\party\in\partySet}$ (cf.
    Definition~\ref{def:maximal_set}), then there exists an $\epsilon$
    negligible in $\secparam$ \st
    $\profile_\infractionPredicate^\infractionPartySet$ is an $\epsilon$-Nash
    equilibrium \wrt utility Reward under $\router_\text{sync}$.
    }
\begin{proof}
    Let $\party^*$ be a party that unilaterally deviates from
    $\profile_\infractionPredicate^\infractionPartySet$ to strategy profile
    $\profile$. By the maximality of $\infractionPartySet$, we have that
    $\sum_{\party \in \infractionPartySet}\power_\party < \threshold$, so the
    exchange rate is always $\rateMax$ in
    $\profile_\infractionPredicate^\infractionPartySet$. 

    We distinguish the following cases.

    \underline{Case 1. $\party^*\in\partySet\setminus\infractionPartySet$:} 
    By the maximality of $\infractionPartySet$, we have that
    \[\sum_{\party\in\infractionPartySet\cup\{\party^*\}}\power_\party \geq \threshold
    \mbox{  and }\bribe_{\party^*}\leq (E[\msgNumber_{\party^*, \profile_\proto}] \cdot \reward + \revision{\stake_{\party^*}}) \cdot \rateDiff.\]
    So, for the infraction failure event
    $F_{\infractionPartySet \cup \{\party^*\}}$ it holds that
    $\Pr[F_{\infractionPartySet\cup\{\party^*\}}] = \negl(\secparam)$.

    Let $D$ be the event that $\party^*$ performs an infraction \wrt
    $\infractionPredicate$ by following $\profile$. Since $\profile$ is a
    unilateral deviation from
    $\profile_\infractionPredicate^\infractionPartySet$, all the parties in
    $\infractionPartySet$ will perform an infraction when they are able to. For
    every $\party \in \infractionPartySet$, let $F_\party$ be the event that
    $\party$ (following $\profile$) fails to perform an infraction. Since
    $\party \in \infractionPartySet$ behaves according to
    $\strategy_\infractionPredicate$, we have that
    $\Pr[F_\party] = \negl(\secparam)$. By Section~\ref{sec:bribing}, when
    $D \wedge (\bigwedge_{\party \in \infractionPartySet} \neg F_\party)$ happens,
    the exchange rate drops with $1 - \negl(\secparam)$ probability.

    Thus, we have that:

    \begin{equation*}
        \begin{split}
            &\utility_{\party^*}(\profile)=\\
            =&\revision{\stake_{\party^*} \cdot E[\exchangeRate_{\profile}]} + E[\reward_{\party^*, \profile}\cdot\exchangeRate_{\profile}]+E[\utilityBoost_{\party^*, \profile}]=\\
            =&\sum_z z\cdot\Pr[(\revision{\stake_{\party^*}} + \reward_{\party^*, \profile})\cdot\exchangeRate_{\profile}+\utilityBoost_{\party^*, \profile}=z]=\\
            =&\sum_z z\cdot\Pr\Big[(\revision{\stake_{\party^*}} + \reward_{\party^*, \profile})\cdot\exchangeRate_{\profile}+\utilityBoost_{\party^*, \profile}=z\wedge D\wedge(\bigwedge_{\party\in\infractionPartySet}\neg F_\party)\Big]+\\
            &+\sum_z z\cdot\Pr\Big[(\revision{\stake_{\party^*}} + \reward_{\party^*, \profile})\cdot\exchangeRate_{\profile}+\utilityBoost_{\party^*, \profile}=z\wedge D\wedge(\bigvee_{\party\in\infractionPartySet} F_\party))\Big]+\\
            &+\sum_z z\cdot\Pr\big[(\revision{\stake_{\party^*}} + \reward_{\party^*, \profile})\cdot\exchangeRate_{\profile}+\utilityBoost_{\party^*, \profile}=z\wedge \neg D\big]\leq\\
            \leq&\sum_z z\cdot\Pr\Big[D\wedge(\bigwedge_{\party\in\infractionPartySet}\neg F_\party)\Big]\cdot\\
            &\quad \cdot \Pr\Big[(\revision{\stake_{\party^*}} + \reward_{\party^*, \profile})\cdot\exchangeRate_{\profile}+\utilityBoost_{\party^*, \profile}=z\Big| D\wedge(\bigwedge_{\party\in\infractionPartySet}\neg F_\party)\Big]+\\
            &+\sum_z z\cdot\Pr\Big[\bigvee_{\party\in\infractionPartySet} F_\party\Big]+\\
            &+\sum_z z\cdot\Pr\big[\neg D\big]\cdot\Pr\big[(\revision{\stake_{\party^*}} + \reward_{\party^*, \profile})\cdot\exchangeRate_{\profile}+\utilityBoost_{\party^*, \profile}=z\big| \neg D\big]\leq\\
            \leq&\sum_z z\cdot\Big(\Pr[D]-\Pr\Big[D\wedge(\bigvee_{\party\in\infractionPartySet} F_\party)\Big]\Big)\cdot\\
            &\quad\cdot\Big(\Pr\Big[(\revision{\stake_{\party^*}} + \reward_{\party^*, \profile})\cdot\rateMin+\bribe_{\party^*}=z\Big| D\wedge(\bigwedge_{\party\in\infractionPartySet}\neg F_\party)\Big]+\\
            &\quad\quad+\negl(\secparam)\Big)+\\
            &+\sum_z z\cdot\negl(\secparam)+\\
            &+\sum_z z\cdot\Pr[\neg D]\cdot\Pr\big[(\revision{\stake_{\party^*}} + \reward_{\party^*, \profile})\cdot \rateMax+0=z\big|\neg D\big]\leq\\
            \leq&\sum_z z\cdot\big(\Pr[D]-\negl(\secparam)\big)\cdot\big(\Pr\big[(\revision{\stake_{\party^*}} + \reward_{\party^*, \profile})\cdot\rateMin+\bribe_{\party^*}=z\big]+\\
            &\quad\quad+\negl(\secparam)\big)+\\
            &+(\rounds\cdot\reward)^2\cdot\negl(\secparam)+\\
            &+\sum_z z\cdot\Pr[\neg D]\cdot\Pr\big[(\revision{\stake_{\party^*}} + \reward_{\party^*, \profile})\cdot \rateMax=z\big]\leq\\
               \end{split}
    \end{equation*}     
               \begin{equation*}
        \begin{split}
            \leq&\negl(\secparam)+\sum_z z\cdot\Pr[D]\cdot\Pr\big[(\revision{\stake_{\party^*}} + \reward_{\party^*, \profile})\cdot\rateMin+\bribe_{\party^*}=z\big]+\\
            &+(\rounds\cdot\reward)^2\cdot\negl(\secparam)+\\
            &+\sum_z z\cdot\Pr[\neg D]\cdot\Pr\big[(\revision{\stake_{\party^*}} + \reward_{\party^*, \profile})\cdot \rateMax=z\big]\leq\\
            \leq&\negl(\secparam)+\Pr[D]\cdot\big(E[\revision{\stake_{\party^*}} + \reward_{\party^*, \profile}]\cdot\rateMin+\bribe_{\party^*}\big)+\\
            &+(\rounds\cdot\reward)^2\cdot\negl(\secparam)+\\
            &+(1-\Pr[ D])\cdot\big((\revision{\stake_{\party^*}} + E[\reward_{\party^*, \profile}])\cdot \rateMax\big)\leq\\
            \leq&\negl(\secparam)+\Pr[D]\cdot\big((\revision{\stake_{\party^*}} + E[\reward_{\party^*, \profile_\proto}])\cdot\rateMin+\bribe_{\party^*}\big)+\\
            &+(\rounds\cdot\reward)^2\cdot\negl(\secparam)+\\
            &+(1-\Pr[ D])\cdot\big((\revision{\stake_{\party^*}} + E[\reward_{\party^*, \profile_\proto}])\cdot \rateMax\big)\leq\\
            \leq& (\revision{\stake_{\party^*}} + E[\reward_{\party^*, \profile_\proto}])\cdot \rateMax+\\
            &+\Pr[D]\cdot\big(\bribe_{\party^*}-(\revision{\stake_{\party^*}} + E[\reward_{\party^*, \profile_\proto}])\cdot(\rateMax-\rateMin)\big)+\negl(\secparam)=\\
            =& (\revision{\stake_{\party^*}}+E[\reward_{\party^*, \profile_\proto}])\cdot \rateMax+\\
            &+\Pr[D]\cdot\big(\bribe_{\party^*}-(\revision{\stake_{\party^*}} + E[\msgNumber_{\party^*, \profile_\proto}]\cdot\reward)\cdot\rateDiff\big)+ \negl(\secparam)\leq\\
            \leq& \revision{\stake_{\party^*} \cdot \rateMax} + E[\reward_{\party^*, \profile_\infractionPredicate^\infractionPartySet}]\cdot \rateMax+\Pr[D]\cdot 0+\negl(\secparam)\leq\\
            =&\utility_{\party^*}(\profile_\infractionPredicate^\infractionPartySet)+\negl(\secparam).
        \end{split}
    \end{equation*}

    \underline{Case 2. $\party^*\in\infractionPartySet$:} 
    In this case, the probability that $\party^*$ performs an infraction when
    following $\profile$ is no more than when following
    $\profile_\infractionPredicate^\infractionPartySet$ (by the definition of
    $\strategy_\infractionPredicate$). Besides, the ledger rewards of
    $\party^*$ are maximized in
    $\profile_\infractionPredicate^\infractionPartySet$ and the exchange rate
    is always $\rateMax$. Therefore, we directly get that 
    $\utility_{\party^*}(\profile) \leq \utility_{\party^*}(\profile_\infractionPredicate^\infractionPartySet)$.\smallskip

    In any case, there is an $\epsilon$ negligible in $\secparam$ \st
    $\utility_{\party^*}(\profile) \leq \utility_{\party^*}(\profile_\infractionPredicate^\infractionPartySet) + \epsilon$,
    \ie $\profile_\infractionPredicate^\infractionPartySet$ is an
    $\epsilon$-Nash equilibrium \wrt Reward under $\router_\text{sync}$.

\end{proof}

\subsubsection{Price of \{ Stability, Anarchy \}}\label{app:guided_price}

\paragraph{\textsc{Theorem}~\ref{thm:guided_pst}}
 \emph{   Assume the following:
    \begin{itemize}
        \item $\proto$ is a protocol run by parties in $\partySet$ in $\rounds$
            rounds, with block-proportional rewards (cf.
            Subsection~\ref{subsec:preliminaries_proportion}) with $\reward$
            rewards per block;
        \item $\threshold$ is a security threshold \wrt $\infractionPredicate$;
        \item for every strategy profile $\profile$:
            \begin{itemize}
                \item the exchange rate $\exchangeRate_\profile$ \wrt
                    $\infractionPredicate, \threshold$ is defined as in
                    Eq.~\eqref{eq:exchange}; 
                \item the external rewards $\utilityBoost_{\party, \profile}$
                    due to guided bribing \wrt $\infractionPredicate$ are
                    defined as in Eq.~\eqref{eq:external_guided};
            \end{itemize}
        \item one of the sets of conditions in
            Lemma~\ref{lem:maximal_existence} holds, so there exists a subset
            of $\partySet$ that is maximal \wrt
            $\threshold,\{\bribe_\party\}_{\party\in\partySet}$. 
    \end{itemize}
    Then, the following hold:
    \begin{enumerate}
        \item
            $\priceStab \leq 1 + \revision{ \dfrac{\Gamma}{\stake + \rounds \cdot \reward} \cdot\Big(1-\dfrac{\rateMin}{\rateMax}\Big) }$, 
            where \revision{ $\Gamma \in \Big\{ \frac{1}{2} \cdot (\rounds \cdot \reward + \stake),   (\threshold\cdot\hat{\rounds} + \frac{\rounds - \hat{\rounds}}{\totalParties}) \cdot \reward + \threshold \cdot\stake  \Big\} $ } 
            for some $\hat{\rounds} \leq \rounds$.
        \item
            Assume that $\frac{5}{2}\cdot\rateMin\leq\rateMax\revision{\leq\frac{\rounds\cdot\reward}{6}\cdot\rateMin}$ and that for some
            $\hat{\rounds} \geq \frac{\threshold - \frac{3}{\totalParties}}{\threshold - \frac{2}{\totalParties}} \cdot \rounds$,
            the second set of conditions in Lemma~\ref{lem:maximal_existence}
            holds. Then, there exist a participation power allocation
            $\{\power_\party\}_{\party \in \partySet}$ and a bribe allocation
            $\{\bribe_\party\}_{\party \in \partySet}$ \st for any constant
            $\gamma \in (0,1)$ and $\totalParties \geq 8^{\frac{1}{1 - \gamma}}$, 
            it holds that:
        \[ \priceStab \geq 1 + \Big( \threshold - \frac{1}{\totalParties^\gamma} \Big) \cdot \Big( 1 - \frac{\rateMin}{\rateMax} \Big) - \negl(\secparam)\;. \]
    \end{enumerate}
    }
\begin{proof}
    \emph{(\ref{item:guided_pst1}).} 
    Since one of the sets of conditions in Lemma~\ref{lem:maximal_existence}
    holds, the total budget of the briber is bounded by
    $\Gamma\cdot\reward\cdot \rateDiff$, where \revision{$\Gamma=\frac{1}{2} \cdot (\rounds \cdot \reward + \stake)$}
    (resp.
    \revision{$\Gamma=(\threshold \cdot \hat{\rounds} + \frac{\rounds - \hat{\rounds}}{\totalParties}) \cdot \reward + \threshold \cdot \stake$},
    for some $\hat{\rounds}\leq\rounds$) if the first (resp. second) set of
    conditions in Lemma~\ref{lem:maximal_existence} holds. Therefore:
    %
    \begin{equation}\label{eq:guided_pst1}
        \begin{split}
            &\max_{\profile \in \strategySet^\totalParties} \welfare(\profile)=\\
            =& \revision{\sum_{\party\in\partySet} \stake_\party \cdot E[\exchangeRate_{\profile}]} + \sum_{\party\in\partySet}E[\reward_{\party, \profile}\cdot\exchangeRate_{\profile}]+\sum_{\party\in\partySet}E[\utilityBoost_{\party, \profile}]\leq\\
            \leq& \revision{\stake \cdot \rateMax} + \rounds \cdot \reward \cdot \rateMax + \Gamma \cdot \rateDiff\;.
        \end{split}
    \end{equation}
     
    In addition, the existence of a subset that is maximal \wrt
    $\threshold,$ $\{\bribe_\party\}_{\party\in\partySet}$ is guaranteed. Let
    $\infractionPartySet$ be a maximal subset. By Theorem~\ref{thm:maximal_sufficient},
    the strategy profile $\profile_\infractionPredicate^\infractionPartySet$ is a
    $\negl(\secparam)$-Nash equilibrium, and since
    $\sum_{\party\in\infractionPartySet}\power_\party<\threshold$, the exchange rate
    when the parties follow $\profile_\infractionPredicate^\infractionPartySet$ is
    always $\rateMax$. As a result,
    $\welfare(\profile_\infractionPredicate^\infractionPartySet)\geq
    \revision{\stake \cdot \rateMax} + \rounds\cdot\reward\cdot\rateMax$, which implies that:
    \begin{equation}\label{eq:guided_pst2}
        \max_{\profile \in \equilibriaSet} \welfare(\profile)\geq \revision{\stake \cdot \rateMax} + \rounds\cdot\reward\cdot\rateMax\;.
    \end{equation}
     
    By Eq.~\eqref{eq:guided_pst1} and~\eqref{eq:guided_pst2}, we get that:
    \begin{equation*}
        \begin{split}
            \priceStab=&\frac{\max_{\profile \in \strategySet^\totalParties} \welfare(\profile)}{\max_{\profile \in \equilibriaSet} \welfare(\profile)}\leq\\
            \leq&\dfrac{\revision{\stake \cdot \rateMax} + \rounds\cdot\reward\cdot\rateMax + \Gamma \cdot \rateDiff}{\revision{\stake \cdot \rateMax} + \rounds\cdot\reward\cdot\rateMax}=\\
            =&1+\revision{ \dfrac{\Gamma}{\stake + \rounds \cdot \reward} \cdot\Big(1-\dfrac{\rateMin}{\rateMax}\Big) }\;.
        \end{split}
    \end{equation*}

    \paragraph{(\ref{item:guided_pst2}).} 
    Let $\partySet=\{\party_1,\ldots,\party_\totalParties\}$. Below, we
    consider a stake and bribe allocation such that there is only one promising
    party and this party has small participation power.
    \begin{itemize}
        \item For the party $\party_1$, let $\power_{\party_1}=\frac{1}{\totalParties}$ 
            and $\bribe_{\party_1}=\frac{2}{\totalParties}\cdot\rounds\cdot\reward\cdot\rateDiff+\revision{\power_{\party_1}\cdot\stake\cdot\rateDiff}$.
        \item For the party $\party_2$, let $\power_{\party_2}=\threshold-\frac{1}{\totalParties}$ 
            and $\bribe_{\party_2}=(\threshold-\frac{2}{\totalParties})\cdot\rounds\cdot\reward\cdot\rateDiff+\revision{\power_{\party_2}\cdot\stake\cdot\rateDiff}$.
        \item For $j=3,\ldots,\totalParties$, let $\power_{\party_j}=\frac{1-\threshold}{\totalParties-2}$ 
            and $\bribe_{\party_1}=\revision{\power_{\party_j}\cdot\stake\cdot\rateDiff}$.
    \end{itemize}
    For the above allocation, the only party that is promising \wrt to
    $\{\bribe_\party\}_{\party\in\partySet}$ is $\party_1$. Indeed, it is clear
    that $\party_3,\ldots,\party_n$ are not promising. For $\party_1$, we have
    that:

    \[ E[\msgNumber_{\party_1, \profile_\proto}]=\power_{\party_1}\cdot\hat{\rounds}+\dfrac{\rounds-\hat{\rounds}}{\totalParties}=\dfrac{\rounds}{\totalParties}\;, \]
     
    hence, $\bribe_{\party_1}=2\cdot E[\msgNumber_{\party_1, \profile_\proto}]\cdot\reward\cdot\rateDiff+\revision{\power_{\party_1}\cdot\stake\cdot\rateDiff}>\big(E[\msgNumber_{\party_1, \profile_\proto}]\cdot\reward+\revision{\stake_{\party_1}}\big)\cdot\rateDiff$, i.e., $\party_1\in\partySet_\text{prom}^{\{\bribe_\party\}}$.

    For  $\party_2$, we have that for 
    $\hat{\rounds}\geq\frac{\threshold-\frac{3}{\totalParties}}{\threshold-\frac{2}{\totalParties}}\cdot\rounds$:
    \begin{equation*}
        \begin{split}
            E[\msgNumber_{\party_2, \profile_\proto}]&=\power_{\party_2}\cdot\hat{\rounds}+\dfrac{\rounds-\hat{\rounds}}{\totalParties}=\\
            &=\Big(\threshold-\frac{1}{\totalParties}\Big)\cdot\hat{\rounds}+\dfrac{\rounds-\hat{\rounds}}{\totalParties}=\\
            &=\Big(\threshold-\frac{2}{\totalParties}\Big)\cdot\hat{\rounds}+\dfrac{\rounds}{\totalParties}\geq\\
            &\geq\Big(\threshold-\frac{3}{\totalParties}\Big)\cdot\rounds+\dfrac{\rounds}{\totalParties}=\\
            &=\Big(\threshold-\frac{2}{\totalParties}\Big)\cdot\rounds\;,
        \end{split}
    \end{equation*}
     
    hence, $\bribe_{\party_2}=\big(\threshold-\frac{2}{\totalParties}\big)\cdot\rounds\cdot\reward\cdot\rateDiff+\revision{\power_{\party_2}\cdot\stake\cdot\rateDiff}\leq\big(E[\msgNumber_{\party_2, \profile_\proto}]\cdot\reward+\revision{\stake_{\party_2}}\big)\cdot\rateDiff$, 
    \ie  $\party_2\notin\partySet_\text{prom}^{\{\bribe_\party\}}$.

    Since $\partySet_\text{prom}^{\{\bribe_\party\}}=\{\party_1\}$ and
    $\power_{\party_1}+\power_{\party_2}=\threshold$ and by
    Definition~\ref{def:maximal_set}, we observe that any subset that is
    maximal \wrt $\threshold,\{\bribe_\party\}_{\party\in\partySet}$ includes
    $\party_1$ and a subset of
    $\partySet=\{\party_3,\ldots,\party_\totalParties\}$. Therefore, the
    welfare when the parties follow $\profile_\infractionPredicate^\infractionPartySet$,
    where $\infractionPartySet$ is any maximal subset, is
    $(\rounds\cdot\reward+\revision{\stake})\cdot\rateMax+\bribe_{\party_1}$, unless $\party_1$
    fails to perform an infraction, which happens with $\negl(\secparam)$
    probability. Therefore, by definition of $\bribe_{\party_1}$:
    \begin{equation}\label{eq:guided_pst3}
    \begin{split}
        &\mathsf{max}\big\{\welfare(\profile_\infractionPredicate^\infractionPartySet):\infractionPartySet\mbox{ is maximal \wrt }\threshold,\{\bribe_\party\}_{\party\in\partySet}\big\}\geq\\
        \geq& (\rounds\cdot\reward+\revision{\stake})\cdot\rateMax+\tfrac{1}{\totalParties}\cdot(2\cdot\rounds\cdot\reward+\revision{\stake})\cdot\rateDiff-\negl(\secparam)\;.
        \end{split}
    \end{equation}
    Next, let $\profile^*$ be a $\negl(\secparam)$-Nash equilibrium that maximizes the welfare across all such equilibria, \ie
    \[ \welfare(\profile^*)=\max_{\profile \in \equilibriaSet} \welfare(\profile)\;. \]

    By Theorem~\ref{thm:maximal_sufficient} and Eq.~\eqref{eq:guided_pst3}, we get that:
    \begin{equation}\label{eq:guided_pst4}
        \welfare(\profile^*)\geq(\rounds\cdot\reward+\revision{\stake})\cdot\rateMax+\tfrac{1}{\totalParties}\cdot(2\cdot\rounds\cdot\reward+\revision{\stake})\cdot\rateDiff-\negl(\secparam)\;.
    \end{equation}

    We argue that for all but a $\negl(\secparam)$ fraction of execution traces
    \wrt $\profile^*$, $\party_1$ performs an infraction. Indeed, assume that
    for the sake of contradiction, there is a non-negligible function $\phi$
    such that for a $\phi(\secparam)$ fraction of traces, $\party_1$ does not
    perform an infraction. Then, for a fraction
    $\phi(\secparam)-\negl(\secparam)$ fraction of traces, $\party_1$'s total
    rewards are no more than $\frac{4}{3}\cdot E[\msgNumber_{\party_1,
    \profile_\proto}]\cdot\reward\cdot\rateMax+0$ and at least $\frac{2}{3}\cdot E[\msgNumber_{\party_1,
    \profile_\proto}]\cdot\reward\cdot\rateMax+0$. The latter follows from the Chernoff
    concentration bounds of the random variable $\msgNumber_{\party_1,
    \profile_\proto}$, which imply that:
    \[ \Pr\big[\tfrac{2}{3}\cdot E[\msgNumber_{\party_1, \profile_\proto}]\leq\msgNumber_{\party_1, \profile_\proto}\leq\tfrac{4}{3}\cdot E[\msgNumber_{\party_1, \profile_\proto}]\big]=1-\negl(\secparam)\;. \]
    We define the strategy profile $\profile^{**}$ which is the same as
    $\profile^*$ for all parties except that $\party_1$ now always follows
    $\strategy_\infractionPredicate$. We observe that (i) for
    $1-\phi(\secparam)$ fraction of traces, the total rewards of $\party_1$ are
    the same in $\profile^*$ and $\profile^{**}$, since all parties' strategies
    are identical; (ii) for $\phi(\secparam)-\negl(\secparam)$ of traces where
    $\party_1$ unilaterally deviates by following
    $\strategy_\infractionPredicate$, the gain in total rewards of $\party_1$
    following  $\profile^{**}$ is at least:
    \begin{equation*}
        \begin{split}
            &\big(\revision{\stake_{\party_1}\cdot\rateMin}+\tfrac{2}{3}\cdot E[\msgNumber_{\party_1, \profile_\proto}]\cdot\reward\cdot\rateMin+\bribe_{\party_1}\big)-\\
            &-\big(\revision{\stake_{\party_1}\cdot\rateMax}+\tfrac{4}{3}\cdot E[\msgNumber_{\party_1, \profile_\proto}]\cdot\reward\cdot\rateMax+0\big)=\\
            =&\big(\tfrac{2}{3}\cdot \tfrac{1}{\totalParties}\cdot\rounds\cdot\reward\cdot\rateMin+\tfrac{2}{\totalParties}\cdot\rounds\cdot\reward\cdot(\rateMax-\rateMin)\big)-\\
            &-\big(\tfrac{4}{3}\cdot\tfrac{1}{\totalParties}\cdot\rounds\cdot\reward\cdot\rateMax+0\big)+\revision{\tfrac{1}{\totalParties}\cdot(\rateMin-\rateMax)}=\\
            =&\tfrac{1}{\totalParties}\cdot\rounds\cdot\reward\cdot\big(\tfrac{2}{3}\cdot\rateMax-\tfrac{4}{3}\cdot \rateMin\big)+\revision{\tfrac{1}{\totalParties}\cdot(\rateMin-\rateMax)}\geq\\
            \geq&\tfrac{1}{\totalParties}\cdot\rounds\cdot\reward\cdot\big(\tfrac{5}{3}\cdot\rateMin-\tfrac{4}{3}\cdot \rateMin\big)+\revision{\tfrac{1}{\totalParties}\cdot(\rateMin-\rateMax)}\\
            =&\frac{\rounds\cdot\reward\cdot\rateMin}{3\cdot\totalParties}+\revision{\tfrac{1}{\totalParties}\cdot(\rateMin-\rateMax)\geq\frac{\rounds\cdot\reward\cdot\rateMin}{6\cdot\totalParties}}\;,
        \end{split}
    \end{equation*}

    where we applied that $\frac{5}{2}\cdot\rateMin\leq\rateMax\revision{\leq\frac{\rounds\cdot\reward}{6}\cdot\rateMin}$. By the above:
    \[ \utility_{\party_1}(\profile^{**})\geq \utility_{\party_1}(\profile^*)+\phi(\secparam)\cdot\frac{\rounds\cdot\reward\cdot\rateMin}{6\cdot\totalParties}-\negl(\secparam)\;, \]
    which leads to contradiction, since $\profile^*$ is a
    $\negl(\secparam)$-Nash equilibrium. Therefore, for all but a
    $\negl(\secparam)$ fraction of execution traces \wrt $\profile^*$,
    $\party_1$ performs an infraction.

    Next, we argue that for all but a $\negl(\secparam)$ fraction of execution
    traces \wrt $\profile^*$, $\party_2$ does not perform an infraction.
    Indeed, assume that for the sake of contradiction, there is a
    non-negligible function $\psi$ such that for a $\psi(\secparam)$ fraction
    of traces, $\party_2$ performs an infraction. Since $\party_1$ almost
    always performs an infraction, for a $\psi(\kappa)-\negl(\secparam)$
    fraction of traces, both $\party_1$ and $\party_2$ perform an infraction.
    When this happens, since $\power_{\party_1}+\power_{\party_2}=\alpha$, the
    exchange rate drops to $\rateMin$ and the welfare is bounded by
    $(\rounds\cdot\reward+\revision{\stake})\cdot\rateMin+\bribe_{\party_1}+\bribe_{\party_2}$.
    Besides, for $1-\psi(\secparam)$ fraction of the traces (where $\party_2$
    does not perform an infraction), the welfare is bounded by
    $(\rounds\cdot\reward+\revision{\stake})\cdot\rateMax+\bribe_{\party_1}$. Therefore, we have
    that:
    \begin{equation*}
        \begin{split}
            &\welfare(\profile^*)\leq\\
             \leq&(\psi(\secparam)-\negl(\secparam))\cdot\Big((\rounds\cdot\reward+\revision{\stake})\cdot\rateMin+\bribe_{\party_1}+\bribe_{\party_2}\Big)+\\
            &+(1-\psi(\secparam))\cdot\Big((\rounds\cdot\reward+\revision{\stake})\cdot\rateMax+\bribe_{\party_1}\Big)+\negl(\secparam)\leq\\
            \leq&(\rounds\cdot\reward+\revision{\stake})\cdot\rateMax+\bribe_{\party_1}+\psi(\secparam)\cdot\Big(\bribe_{\party_2}-(\rounds\cdot\reward+\revision{\stake})\cdot\rateDiff\Big)+\\
            &+\negl(\secparam)=\\
            \leq&(\rounds\cdot\reward+\revision{\stake})\cdot\rateMax+\bribe_{\party_1}+\\
            &+\psi(\secparam)\cdot\Big((\threshold-\tfrac{2}{\totalParties})\cdot\rounds\cdot\reward\cdot\rateDiff+\revision{(\threshold-\tfrac{1}{\totalParties})\cdot\stake\cdot\rateDiff}-(\rounds\cdot\reward+\revision{\stake})\cdot\rateDiff\Big)+\\
            &+\negl(\secparam)\leq\\
            \leq&(\rounds\cdot\reward+\revision{\stake})\cdot\rateMax+\bribe_{\party_1}-\\
            &-\psi(\secparam)\cdot\Big(1-\threshold+\tfrac{2}{\totalParties}\Big)\cdot\rounds\cdot\reward\cdot\rateDiff-\revision{\psi(\secparam)\cdot\Big(1-\threshold+\tfrac{1}{\totalParties}\Big)\cdot\stake\cdot\rateDiff}+\\
            &+\negl(\secparam)<\\
            %
            %
            <&(\rounds\cdot\reward+\revision{\stake})\cdot\rateMax+\bribe_{\party_1}-\negl(\secparam)\;,
        \end{split}
    \end{equation*}
    \revision{since $\psi(\secparam)\cdot\Big(1-\threshold+\tfrac{2}{\totalParties}\Big)\cdot\rounds\cdot\reward\cdot\rateDiff$ and $\psi(\secparam)\cdot\Big(1-\threshold+\tfrac{1}{\totalParties}\Big)\cdot\stake\cdot\rateDiff$ are non-negligible functions}. However, the latter inequality contradicts to Eq.~\eqref{eq:guided_pst4}! Therefore, for all but a
    $\negl(\secparam)$ fraction of execution traces \wrt $\profile^*$,
    $\party_2$ does not perform an infraction.

    As a result, for all but a $\negl(\secparam)$ fraction of traces \wrt
    $\profile^*$, $\party_1$ performs an infraction whereas $\party_2$ does not
    perform an infraction. Hence, for all but a $\negl(\secparam)$ fraction of
    traces, the welfare is bounded by
    $(\rounds\cdot\reward+\revision{\stake})\cdot\rateMax+\bribe_{\party_1}$. The latter implies
    that 
    \begin{equation}\label{eq:guided_pst5}
        \welfare(\profile^*)\leq(\rounds\cdot\reward+\revision{\stake})\cdot\rateMax+\tfrac{1}{\totalParties}\cdot(2\cdot\rounds\cdot\reward+\revision{\stake})\cdot\rateDiff+\negl(\secparam)\;.
    \end{equation}
     
    Now consider the strategy profile $\hat{\profile}$ where $\party_2$ follows
    $\strategy_\infractionPredicate$ and all the other parties are honest.
    Since $\power_{\party_2}<\threshold$, the exchange rate is always
    $\rateMax$ and for all but a $\negl(\secparam)$ fraction of traces,
    $\party_2$ gets her external rewards. As a result:
    \begin{equation}\label{eq:guided_pst6}
        \begin{split}
            \welfare(\hat{\profile})&\geq(\rounds\cdot\reward+\revision{\stake})\cdot\rateMax+\bribe_{\party_2}-\negl(\secparam)=\\
            &=(\rounds\cdot\reward+\revision{\stake})\cdot\rateMax+\big(\threshold-\tfrac{2}{\totalParties}\big)\cdot\rounds\cdot\reward\cdot\rateDiff+\\
            &+\revision{\big(\threshold-\tfrac{1}{\totalParties}\big)\cdot\stake\cdot\rateDiff}-\negl(\secparam)\;.
        \end{split}
    \end{equation}
     
    By Eq.~\eqref{eq:guided_pst5} and~\eqref{eq:guided_pst6}, we can lower
    bound the price of stability as:
    \begin{equation*}
        \begin{split}
          &  \priceStab=\frac{\max_{\profile \in \strategySet^\totalParties} \welfare(\profile)}{\max_{\profile \in \equilibriaSet} \welfare(\profile)}\geq\\
            &\geq\frac{(\rounds\cdot\reward+\revision{\stake})\cdot\rateMax+\big(\threshold-\tfrac{2}{\totalParties}\big)\cdot\rounds\cdot\reward\cdot\rateDiff+\revision{\big(\threshold-\tfrac{1}{\totalParties}\big)\cdot\stake\cdot\rateDiff}-\negl(\secparam)}{(\rounds\cdot\reward+\revision{\stake})\cdot\rateMax+\tfrac{1}{\totalParties}\cdot(2\cdot\rounds\cdot\reward+\revision{\stake})\cdot\rateDiff+\negl(\secparam)}\geq\\
            &\geq\frac{(\rounds\cdot\reward+\revision{\stake})\cdot\rateMax+\big(\threshold-\tfrac{2}{\totalParties}\big)\cdot\rounds\cdot\reward\cdot\rateDiff+\revision{\big(\threshold-\tfrac{1}{\totalParties}\big)\cdot\stake\cdot\rateDiff}}{(\rounds\cdot\reward+\revision{\stake})\cdot\rateMax+\tfrac{1}{\totalParties}\cdot(2\cdot\rounds\cdot\reward+\revision{\stake})\cdot\rateDiff}-\negl(\secparam)\geq\\
                        &\geq\frac{(\rounds\cdot\reward+\revision{\stake})\cdot\rateMax+\big(\threshold-\tfrac{2}{\totalParties}\big)\cdot\rounds\cdot\reward\cdot\rateDiff+\revision{\big(\threshold-\tfrac{1}{\totalParties}\big)\cdot\stake\cdot\rateDiff}}{(\rounds\cdot\reward+\revision{\stake})\cdot\rateMax+\tfrac{2}{\totalParties}\cdot(\rounds\cdot\reward+\revision{\stake})\cdot\rateDiff}-\negl(\secparam)\geq\\
                        &\geq\frac{(\rounds\cdot\reward+\revision{\stake})\cdot\rateMax+\big(\threshold-\tfrac{2}{\totalParties}\big)\cdot\rounds\cdot\reward\cdot\rateDiff+\revision{\big(\threshold-\tfrac{2}{\totalParties}\big)\cdot\stake\cdot\rateDiff}}{(\rounds\cdot\reward+\revision{\stake})\cdot\rateMax+\tfrac{2}{\totalParties}\cdot(\rounds\cdot\reward+\revision{\stake})\cdot\rateDiff}-\negl(\secparam)=\\
                                                &=\frac{(\rounds\cdot\reward+\revision{\stake})\cdot\rateMax+\big(\threshold-\tfrac{2}{\totalParties}\big)\cdot(\rounds\cdot\reward+\revision{\stake})\cdot\rateDiff}{(\rounds\cdot\reward+\revision{\stake})\cdot\rateMax+\tfrac{2}{\totalParties}\cdot(\rounds\cdot\reward+\revision{\stake})\cdot\rateDiff}-\negl(\secparam)=\\
                                         &=\frac{\rateMax+\big(\threshold-\tfrac{2}{\totalParties}\big)\cdot\rateDiff}{\rateMax+\tfrac{2}{\totalParties}\cdot\rateDiff}-\negl(\secparam)=\\
            &=\frac{1+(\threshold-\frac{2}{\totalParties})\cdot\big(1-\frac{\rateMin}{\rateMax}\big)}{1+\frac{2}{\totalParties}\cdot\big(1-\frac{\rateMin}{\rateMax}\big)}-\negl(\secparam)\geq\\
            &\geq \frac{1+(\threshold-\frac{2}{\totalParties})\cdot\big(1-\frac{\rateMin}{\rateMax}\big)}{1+\frac{2}{\totalParties}}-\negl(\secparam)
        \end{split}
    \end{equation*}

    Finally, for any constant $\gamma\in(0,1)$ and
    $\totalParties\geq8^{\frac{1}{1-\gamma}}$, we get that
    \begin{equation*}
        \begin{split}
            &\frac{1+(\threshold-\frac{2}{\totalParties})\cdot\big(1-\frac{\rateMin}{\rateMax}\big)}{1+\frac{2}{\totalParties}}\geq1+\Big(\threshold-\frac{1}{\totalParties^\gamma}\Big)\cdot\Big(1-\frac{\rateMin}{\rateMax}\Big)\Leftrightarrow\\
            \Leftrightarrow&1+\Big(\threshold-\frac{2}{\totalParties}\Big)\cdot\Big(1-\frac{\rateMin}{\rateMax}\Big)\geq\\
            &\geq\Big(1+\frac{2}{\totalParties}\Big)\cdot\Big(1+\big(\threshold-\frac{1}{\totalParties^\gamma}\big)\cdot\big(1-\frac{\rateMin}{\rateMax}\big)\Big)\Leftrightarrow\\
            \Leftrightarrow&1+\threshold\cdot\Big(1-\frac{\rateMin}{\rateMax}\Big)-\frac{2}{\totalParties}\cdot\Big(1-\frac{\rateMin}{\rateMax}\Big)\geq\\
            &\geq1+\threshold\cdot\Big(1-\frac{\rateMin}{\rateMax}\Big)-\frac{1}{\totalParties^\gamma}\cdot\Big(1-\frac{\rateMin}{\rateMax}\Big)+\\
            &\quad+\frac{2}{\totalParties}+\frac{2}{\totalParties}\cdot\big(\threshold-\frac{1}{\totalParties^\gamma}\big)\cdot\big(1-\frac{\rateMin}{\rateMax}\big)\Leftarrow\\
            \Leftarrow&-\frac{2}{\totalParties}\cdot\Big(1-\frac{\rateMin}{\rateMax}\Big)\geq\\
            &\geq-\frac{1}{\totalParties^\gamma}\cdot\Big(1-\frac{\rateMin}{\rateMax}\Big)+\frac{2}{\totalParties}+\frac{2}{\totalParties}\cdot1\cdot\Big(1-\frac{\rateMin}{\rateMax}\Big)\Leftrightarrow\\
            \Leftrightarrow&\Big(\frac{1}{\totalParties^\gamma}-\frac{4}{\totalParties}\Big)\cdot\Big(1-\frac{\rateMin}{\rateMax}\Big)\geq\frac{2}{\totalParties}\Leftrightarrow\\
            \Leftrightarrow&\totalParties^{1-\gamma}-4\geq\frac{2}{1-\frac{\rateMin}{\rateMax}}\Leftarrow\\
            \Leftarrow&\totalParties^{1-\gamma}-4\geq4\Leftrightarrow\totalParties\geq8^{\frac{1}{1-\gamma}}\;.
        \end{split}
    \end{equation*}
     
    Hence, for any constant $\gamma\in(0,1)$ and
    $\totalParties\geq8^{\frac{1}{1-\gamma}}$, it holds that:
    \[ \priceStab\geq1+\Big(\threshold-\frac{1}{\totalParties^\gamma}\Big)\cdot\Big(1-\frac{\rateMin}{\rateMax}\Big)-\negl(\secparam)\;. \]

\end{proof}


\paragraph{\textsc{Theorem}~\ref{thm:guided_pan}}
\emph{    Assume the following:
    \begin{itemize}
        \item $\proto$ is a protocol run by parties in $\partySet$ in $\rounds$
            rounds, with block-proportional rewards (cf.
            Subsection~\ref{subsec:preliminaries_proportion}) with $\reward$
            rewards per block;
        \item \revision{each party's stake is $\stake_\party = \power_\party \cdot \stake$;}
        \item $\threshold$ is a security threshold \wrt $\infractionPredicate$;
        \item for every strategy profile $\profile$:
            \begin{itemize}
                \item the exchange rate $\exchangeRate_\profile$ \wrt
                    $\infractionPredicate, \threshold$ is defined as in
                    Eq.~\eqref{eq:exchange}; 
                \item the external rewards $\utilityBoost_{\party,\profile}$
                    due to guided bribing \wrt $\infractionPredicate$ are
                    defined as in Eq.~\eqref{eq:external_guided};
            \end{itemize}
    \end{itemize}
    Then, the following hold:
    \begin{enumerate}
        \item
            If for every party $\party \in \partySet$, it holds that
            $\power_\party \leq 1 - \threshold$, and
            $\sum_{\party \in \partySet} \bribe_\party$ is a non-negligible value
            upper bounded by
            $\threshold \cdot (\rounds \cdot \reward + \revision{\stake}) \cdot \rateDiff$, then the Price
            of Anarchy is bounded by:
            \[ \frac{1}{\threshold + (1 - \threshold) \cdot \frac{\rateMin}{\rateMax}} - \negl(\secparam) \leq \priceAnar < \infty\;. \]
        \item
            If for every party $\party \in \partySet$, it holds that
            $\power_\party < \threshold$, and
            $\sum_{\party \in \partySet} \bribe_\party = \negl(\secparam)$, then
            \revision{$\priceAnar \geq \Big( 1 + \frac{\rounds \cdot \reward}{\stake} \Big) \cdot \frac{\rateMax}{\rateMin}$}.
    \end{enumerate}
    }
\begin{proof}
    \emph{(\ref{item:guided_pan1}).}
    A straightforward lower bound on the maximum welfare is:
    \begin{equation}\label{eq:guided_pan1}
        \begin{split}
            &\max_{\profile \in \strategySet^\totalParties} \welfare(\profile) \geq \welfare(\profile_\proto) = (\rounds \cdot \reward + \revision{\stake}) \cdot \rateMax\;.
        \end{split}
    \end{equation}
     
    By Theorem~\ref{thm:guided} and given that for every party
    $\party\in\partySet$, it holds that $\power_\party\leq1-\threshold$ ,
    $\profile_\infractionPredicate$ is a $\negl(\secparam)$-Nash equilibrium.
    When all parties follow $\strategy_\infractionPredicate$, the attack is
    almost always successful, which implies that the welfare of
    $\profile_\infractionPredicate$ is:
    \begin{equation}\label{eq:guided_pan2}
        \begin{split}
            \welfare(\profile_\infractionPredicate) \leq (\rounds \cdot \reward + \revision{\stake}) \cdot \rateMin+\sum_{\party\in\partySet}\bribe_\party+\negl(\secparam)\;.
        \end{split}
    \end{equation}



    By Eq.~\eqref{eq:guided_pan1} and~\eqref{eq:guided_pan2}, and  given that
    $\negl(\secparam)<\sum_{\party\in\partySet}\bribe_\party\leq\threshold\cdot(\rounds\cdot\reward + \revision{\stake})\cdot\rateDiff$,
    we conclude that:
    \begin{equation*}
        \begin{split}
            \priceAnar&=\frac{\max_{\profile \in \strategySet^\totalParties} \welfare(\profile)}{\min_{\profile \in \equilibriaSet} \welfare(\profile)}\geq\\
            &\geq\frac{\max_{\profile \in \strategySet^\totalParties} \welfare(\profile)}{\welfare(\profile_\infractionPredicate)}\geq\\
            &\geq\frac{(\rounds \cdot \reward + \revision{\stake}) \cdot \rateMax}{(\rounds \cdot \reward + \revision{\stake}) \cdot \rateMin+\sum_{\party\in\partySet}\bribe_\party}-\negl(\secparam)\geq\\
            &\geq\frac{(\rounds \cdot \reward + \revision{\stake}) \cdot \rateMax}{(\rounds \cdot \reward + \revision{\stake}) \cdot \rateMin + \threshold\cdot(\rounds \cdot \reward + \revision{\stake}) \cdot \rateDiff}-\negl(\secparam)=\\
            &\geq\frac{1}{\frac{\rateMin}{\rateMax}+\threshold\cdot\big(1-\frac{\rateMin}{\rateMax}\big)}-\negl(\secparam)=\\
            &\geq\frac{1}{\threshold+(1-\threshold)\cdot\frac{\rateMin}{\rateMax}}-\negl(\secparam)\;.
        \end{split}
    \end{equation*}

    To show that $\priceAnar<\infty$, it suffices to prove that $\min_{\profile
    \in \equilibriaSet} \welfare(\profile)>0$. Let $\profile_0$ be a strategy
    profile \st $\welfare(\profile)=0$. We argue that
    $\profile_0\notin\equilibriaSet$. Indeed, $\welfare(\profile)=0$ implies
    that for every party $\party\in\partySet$:
    $\utility_{\party}(\profile_0)=0$. However,
    $\sum_{\party\in\partySet}\bribe_\party$ is non-negligible which implies
    that for at least for one party $\party^*$, $\bribe_{\party^*}$ is
    non-negligible. Thus, if $\party^*$ unilaterally deviates from $\profile_0$
    by following $\profile_\infractionPredicate$, she will almost always get
    her external rewards, so her gain of utility will be at least
    $\bribe_{\party^*}-\negl(\secparam)$ which is a non-negligible value. The
    latter implies that $\profile_0$ is not a $\negl(\secparam)$-Nash
    equilibrium, \ie $\profile_0\notin\equilibriaSet$.

    \paragraph{(\ref{item:guided_pan2}).} 
    If $\sum_{\party\in\partySet}\bribe_\party=\negl(\secparam)$, then
    $\bribe_\party=\negl(\secparam)$, for every $\party\in\partySet$. We argue
    that the all-abstain strategy profile $\profile_\text{abs}$ is a
    $\negl(\secparam)$-Nash equilibrium \wrt Reward under
    $\router_\text{sync}$. In particular, assume a unilateral deviation
    $\profile^*$ caused by some party $\party^*$ from $\profile_\text{abs}$.
    Since $\power_{\party^*}<\threshold$ and all other parties abstain from the
    execution in $\profile^*$, there will never be a block that will be signed
    by a committee with sufficiently large participation power. The latter
    implies that the ledger rewards of $\party^*$ in $\profile^*$ are $0$.
    Given that  $\bribe_{\party^*}=\negl(\secparam)$, we conclude that:
    \[ \utility_{\party^*}(\profile)= \revision{\stake_{\party^*} \cdot \rateMin}+0+\negl(\secparam)=\utility_{\party^*}(\profile_\text{abs})+\negl(\secparam)\;, \]
    which implies that $\profile_\text{abs}$ is a $\negl(\secparam)$-Nash
    equilibrium \wrt Reward under $\router_\text{sync}$.

    Consequently:
    \begin{equation*}
        \begin{split}
            &0\leq\min_{\profile \in \equilibriaSet} \welfare(\profile)\leq\welfare(\profile_\text{abs})=\revision{\stake \cdot \rateMin}
        \end{split}
    \end{equation*}
    Given that, as described above, $\max_{\profile \in \strategySet^\totalParties} \welfare(\profile) \geq (\revision{\stake} + \rounds \cdot \reward) \cdot \rateMax$,
    the bound for Price of Anarchy follows directly:
    \revision{$$\priceAnar \geq \Big( 1 + \frac{\rounds \cdot \reward}{\stake} \Big) \cdot \frac{\rateMax}{\rateMin}$$}
\end{proof}

\subsection{Effective Bribing}

\subsubsection{Equilibria}\label{app:eff_eq}

\paragraph{\textsc{Theorem}~\ref{thm:effective}}
  \emph{  Assume the following:
    \begin{itemize}
        \item $\proto$ is a protocol run by parties in $\partySet$ in $\rounds$
            rounds, with block-proportional rewards (cf.
            Subsection~\ref{subsec:preliminaries_proportion}) with $\reward$
            rewards per block;
        \item \revision{each party's stake is $\stake_\party = \power_\party \cdot \stake$;}
        \item $\threshold$ is a security threshold \wrt $\infractionPredicate$;
        \item $\profile_\proto$ denotes the all-honest strategy profile;
        \item $\profile_\text{abs}$ denotes the all-abstain strategy profile;
        \item for every strategy profile $\profile$:
            \begin{itemize}
                \item the exchange rate $\exchangeRate_\profile$ \wrt
                    $\infractionPredicate, \threshold$ is defined as in
                    Eq.~\eqref{eq:exchange}; 
                \item the external rewards $\utilityBoost_{\party, \profile}$
                    due to effective bribing \wrt $\infractionPredicate$ are
                    defined as in Eq.~\eqref{eq:external_effective};
            \end{itemize}
    \end{itemize}
    Then, the following hold:
    \begin{enumerate}
        \item
            If for every party $\party \in \partySet$, it holds that
            $\power_\party \leq 1 - \threshold$, then there is an $\epsilon$
            negligible in $\secparam$ \st the strategy profile
            $\profile_\infractionPredicate$ is an $\epsilon$-Nash equilibrium
            \wrt Reward under $\router_\text{sync}$.
        \item
            Assume that there exists a party $\party \in \partySet$ \st
            $\power_\party \geq \threshold$. Let $\theta > 0$ be a real value \st
            the probability 
            $\Pr \big[ \msgNumber_{\party, \profile_\proto} > (1 + \theta) \cdot E[\msgNumber_{\party, \profile_\proto}] \big]$ 
            is negligible in $\secparam$, where
            $\msgNumber_{\party, \profile_\proto}$ is the number of blocks
            created by $\party$ in $\profile_\proto$. If 
            $\bribe_\party \geq \big( (1 + \theta) \cdot E[\msgNumber_{\party, \profile_\proto}] \cdot \reward + \revision{\stake_\party} \big) \cdot \rateMax$, 
            then there is an $\epsilon$ negligible in $\secparam$ such that the
            strategy profile $\profile_\infractionPredicate$ is an
            $\epsilon$-Nash equilibrium \wrt Reward under
            $\router_\text{sync}$.
        \item
            If for every party $\party \in \partySet$, it holds that
            $\power_\party < \threshold$, then $\proto$ is a Nash equilibrium
            \wrt Reward under $\router_\text{sync}$.
        \item
            If for every party $\party \in \partySet$, it holds that
            $\power_\party < \threshold$, then $\profile_\text{abs}$ is a Nash
            equilibrium \wrt Reward under $\router_\text{sync}$.
    \end{enumerate}
    }
\begin{proof}
    \emph{(\ref{thm:effective_nash1}).} 
    The proof is similar to Proposition \ref{thm:guided_nash1} of Theorem~\ref{thm:guided}.
    Namely, when all other parties attempt the infraction, then exchange rate
    will drop to $\rateMin$ (with $1 - \negl(\secparam)$ probability) whatever
    the behavior of party $\party$ is. In addition, in
    $\profile_\infractionPredicate$, $\party$ almost always receives its
    external rewards. Therefore, any unilateral deviation from
    $\profile_\infractionPredicate$ will not increase $\party$'s utility by
    more than a $\negl(\secparam)$ gain.\\

    \emph{(\ref{thm:effective_nash2}).}
    The proof is similar to Proposition \ref{thm:guided_nash2} of Theorem~\ref{thm:guided}.
    Namely, given the proof of \ref{thm:effective_nash1} above, it suffices to
    examine the unilateral deviations of $\party$. Since
    $\power_\party \geq \threshold$, the value of the exchange rate in some trace
    $\trace$ depends  on whether $\party$ performs an infraction. Thus, if the
    bribe is sufficiently large, \ie 
    $\bribe_\party \geq \big( (1 + \theta) \cdot E[\msgNumber_{\party, \profile_\proto}] \cdot \reward + \revision{\stake_\party} \big) \cdot \rateMax$, 
    then $\party$ is incentivized to always attempt the infraction.\\

    \emph{(\ref{thm:effective_nash3}).}
    Let $\party^*$ be a party that unilaterally deviates from $\profile_\proto$
    according to strategy profile $\profile$. Since $\power_\party < \threshold$
    and no other party attempts an infraction in $\sigma$, the condition for
    the drop of exchange rate is never met. Namely, for every trace $\trace$,
    it holds that $\exchangeRate_\profile(\trace) = \rateMax$, so by
    Eq.~\eqref{eq:external_effective}, we have that
    $\utilityBoost_{\party, \profile}(\trace) = 0$. The latter and the fact that
    $\profile_\proto$ maximizes the ledger rewards of $\party^*$ implies that 

    \begin{equation*}
        \begin{split}
            \utility_{\party^*}(\profile)
            &=\revision{\stake_{\party^*} \cdot E[\exchangeRate_{\profile}]} + E[\reward_{\party^*, \profile}\cdot\exchangeRate_{\profile}]+E[\utilityBoost_{\party^*, \profile}]=\\
            &=(\revision{\stake_{\party^*}} + E[\reward_{\party^*, \profile}]) \cdot \rateMax+0\leq\\
            &\leq (\revision{\stake_{\party^*}} + E[\reward_{\party^*, \profile_\proto}]) \cdot \rateMax+0=\\
            &=\revision{\stake_{\party^*} \cdot E[\exchangeRate_{\profile_\proto}]} + E[\reward_{\party^*, \profile_\proto}\cdot\exchangeRate_{\profile_\proto}]+E[\utilityBoost_{\party^*, \profile_\proto}]=\\
            &=\utility_{\party^*}(\profile_\proto)\;.
        \end{split}
    \end{equation*}
    Therefore, $\profile_\proto$ is a Nash equilibrium \wrt Reward under
    $\router_\text{sync}$.\\

    \emph{(\ref{thm:effective_nash4}).}
    Let $\party^*$ be a party that unilaterally deviates from $\profile_\proto$
    according to strategy profile $\profile$. Since $\power_\party < \threshold$
    and all other parties abstain from the execution in $\sigma$, the following
    happen: (i) a block is never signed by a committee with at least
    $\threshold$ aggregate participation power, so the party gets no ledger
    rewards; (ii) the attack is never carried out, so the party gets no
    external rewards. Therefore:
    \[ \utility_{\party^*}(\profile)=\revision{\stake_{\party^*}\cdot\rateMin}=\utility_{\party^*}(\profile_\text{abs})\;, \]
    which implies that $\profile_\text{abs}$ is a Nash equilibrium \wrt Reward
    under $\router_\text{sync}$.
\end{proof}

\subsubsection{Price of \{ Stability, Anarchy \}}\label{app:eff_price}

\paragraph{\textsc{Theorem}~\ref{thm:effective_pst}}
  \emph{  Assume the following:
    \begin{itemize}
        \item $\proto$ is a protocol run by parties in $\partySet$ in $\rounds$
            rounds, with block-proportional rewards (cf.
            Subsection~\ref{subsec:preliminaries_proportion}) with $\reward$
            rewards per block;
        \item $\threshold$ is a security threshold \wrt $\infractionPredicate$;
        \item for every strategy profile $\profile$:
            \begin{itemize}
                \item the exchange rate $\exchangeRate_\profile$ \wrt
                    $\infractionPredicate, \threshold$ is defined as in
                    Eq.~\eqref{eq:exchange}; 
                \item the external rewards $\utilityBoost_{\party, \profile}$
                    due to effective bribing \wrt $\infractionPredicate$ are
                    defined as in Eq.~\eqref{eq:external_effective};
            \end{itemize}
        \item for every party $\party \in \partySet$, it holds that $\power_\party < \threshold$ \revision{and $\stake_\party = \power_\party \cdot \stake$};
        \item $\sum_{\party \in \partySet} \bribe_\party \leq (\rounds \cdot \reward + \revision{\stake}) \cdot \rateDiff$.\footnote{Note that this is a tighter bound than the one already applied by the bribing budget.}
    \end{itemize}
    Then, it holds that:
    \begin{enumerate}
        \item $\priceStab=1$
        \item \revision{$\priceAnar \geq (1 + \frac{\rounds \cdot \reward}{\stake}) \cdot \frac{\rateMax}{\rateMin}$}
    \end{enumerate}
    \wrt Reward under $\router_\text{sync}$.
    }
\begin{proof}
    \emph{(1).} 
    First, we upper bound the welfare of an arbitrary strategy profile
    $\profile$. For every execution trace \wrt $\profile$, \st the attack
    is successful, the rate drops to $\rateMin$ and the parties get their
    external rewards. Therefore, the welfare is bounded by:
    \[ (\rounds \cdot \reward + \revision{\stake}) \cdot \rateMin + \sum_{\party \in \partySet} \bribe_\party \leq (\rounds \cdot \reward + \revision{\stake}) \cdot \rateMin + (\rounds \cdot \reward + \revision{\stake}) \cdot \rateDiff = (\rounds \cdot \reward + \revision{\stake}) \cdot \rateMax\;. \]
     
    In the case where the attack is not successful, the rate remains
    $\rateMax$, yet no party gets external rewards. Thus, the welfare is
    bounded by $(\rounds \cdot \reward + \revision{\stake}) \cdot \rateMax$. 

    As a result, in every execution trace, the welfare is bounded by
    $(\rounds \cdot \reward + \revision{\stake}) \cdot \rateMax$. Since $\profile$ is arbitrary, we
    have that:
    \[\max_{\profile \in \strategySet^\totalParties} \welfare(\profile) \leq (\rounds \cdot \reward + \revision{\stake}) \cdot \rateMax\;. \]
     
    Next, we compute the welfare for the all-honesty strategy profile
    $\profile_\proto$. We directly get that:
    \[ \welfare(\profile_\proto) = \sum_{\party \in \partySet} (E[\msgNumber_{\party, \profile_\proto}] \cdot \reward + \revision{\stake_\party}) \cdot \rateMax + 0 = (\rounds \cdot \reward + \revision{\stake}) \cdot \rateMax\;. \]
     
    Since, for every party $\party\in\partySet$, it holds that
    $\power_\party \leq \threshold$ and, by Theorem~\ref{thm:effective},
    $\profile_\proto$ is a $\negl(\secparam)$-Nash equilibrium, we conclude
    that:
    \begin{equation*}
        \begin{split}
            & \rounds \cdot \reward \cdot \rateMax = \welfare(\profile_\proto) \leq \max_{\profile \in \equilibriaSet} \welfare(\profile) \leq \\
            \leq & \max_{\profile \in \strategySet^\totalParties} \welfare(\profile) \leq (\rounds \cdot \reward + \revision{\stake}) \cdot \rateMax\;,
        \end{split}
    \end{equation*}
    which implies that the Price of Stability is:
    \begin{equation*}
        \begin{split}
            \priceStab = &\frac{\max_{\profile \in \strategySet^\totalParties} \welfare(\profile)}{\max_{\profile \in \equilibriaSet} \welfare(\profile)} = 1\;.
        \end{split}
    \end{equation*}

    \noindent \emph{(2).}
    By Theorem~\ref{thm:effective}, $\profile_\text{abs}$ is a
    $\negl(\secparam)$-Nash equilibrium. Therefore:
    \begin{equation*}
        \begin{split}
            & 0 \leq \min_{\profile \in \equilibriaSet} \welfare(\profile) \leq \welfare(\profile_\text{abs}) = \revision{\stake \cdot \rateMin} \;,
        \end{split}
    \end{equation*}
    which implies that the Price of Anarchy is:
    \revision{$\priceAnar \geq (1 + \frac{\rounds \cdot \reward}{\stake}) \cdot \frac{\rateMax}{\rateMin}$}.
\end{proof}

\paragraph{\textsc{Theorem}~\ref{thm:effective_T-pan}}
 \emph{   Assume the following:
    \begin{itemize}
        \item $\proto$ is a protocol run by parties in $\partySet$ in $\rounds$ rounds, with block-proportional rewards (cf.
            Subsection~\ref{subsec:preliminaries_proportion}) with $\reward$
            rewards per block; 
        \item $\threshold$ is a security threshold \wrt $\infractionPredicate$;
        \item for every strategy profile $\profile$:
            \begin{itemize}
                \item the exchange rate $\exchangeRate_\profile$ \wrt
                    $\infractionPredicate, \threshold$ is defined as in
                    Eq.~\eqref{eq:exchange}; 
                \item the external rewards $\utilityBoost_{\party, \profile}$
                    due to effective bribing \wrt $\infractionPredicate$ are
                    defined as in Eq.~\eqref{eq:external_effective};
            \end{itemize}
        \item $\msgNumber_{\party, \profile_\proto}$ is the number of blocks
            created by $\party$ in $\profile_\proto$;
        \item for every party $\party \in \partySet$, it holds that
            $\power_\party \leq 1 - \threshold$, \revision{$\stake_\party = \power_\party \cdot \stake$}, and
            $\sum_{\party \in \partySet} \bribe_\party \leq \threshold \cdot (\rounds \cdot \reward + \revision{\stake}) \cdot \rateDiff$.
    \end{itemize}
    Then, the following hold:
    \begin{enumerate}
        \item
            The Price of Anarchy \wrt $\{\proto, \strategy_\infractionPredicate\}$
            is lower bounded by:
        \[ \{\proto, \strategy_\infractionPredicate\} \text{-} \priceAnar \geq \frac{1}{\threshold + (1 - \threshold) \cdot \frac{\rateMin}{\rateMax}} - \negl(\secparam)\;. \]
        \item
            The Price of Anarchy \wrt $\{\proto, \strategy_\infractionPredicate\}$ 
            is upper bounded by:
            \[ \{\proto, \strategy_\infractionPredicate\} \text{-} \priceAnar \leq \frac{(\revision{\stake} + \rounds \cdot \reward) \cdot \rateMax}{(\revision{\stake} + \rounds \cdot \reward) \cdot \rateMin + \Phi} + \negl(\secparam)\;, \]
        where $\Phi = \min_{\mathbb{A} \subseteq \partySet: \sum_{\party \in \mathbb{A}} \power_\party \geq \threshold} \sum_{\party \in \mathbb{A}} \bribe_\party$.
    \end{enumerate}
    }
\begin{proof}
    \emph{(\ref{item:effective_T-pan1}).}
    By Theorem~\ref{thm:effective}, since for every party
    $\party \in \partySet$, it holds that $\power_\party \leq 1 - \threshold$,
    we have that
    $\profile_\infractionPredicate \in \equilibriaSet \cap \big( \{\proto, \strategy_\infractionPredicate \} \big)^\totalParties$.
    The latter implies that
    $\equilibriaSet \cap \big( \{\proto, \strategy_\infractionPredicate \} \big)^\totalParties \neq \emptyset$.
    In addition, when the parties follow $\profile_\infractionPredicate$, the
    attack is almost always successful, hence:
     
    \begin{equation}\label{eq:effective_T-pan1}
        \begin{split}
            &\min_{\profile \in \equilibriaSet \cap(\{\proto, \strategy_\infractionPredicate\})^\totalParties} \welfare(\profile) \leq \welfare(\profile_\infractionPredicate)\\
            &\leq \revision{\stake \cdot \rateMin} + \rounds \cdot \reward \cdot \rateMin + \sum_{\party \in \partySet} \bribe_\party + \negl(\secparam)\;.
        \end{split}
    \end{equation}
     
    By Eq.~\eqref{eq:effective_T-pan1}, we get the following lower bound on
    $\{\proto, \strategy_\infractionPredicate\} \text{-} \priceAnar$:
     
    \begin{equation*}
        \begin{split}
            &\{\proto, \strategy_\infractionPredicate\} \text{-} \priceAnar =\\
            =&\frac{\max_{\profile \in \strategySet^\totalParties} \welfare(\profile)}{\min_{\profile \in \equilibriaSet\cap\big(\{\proto, \strategy_\infractionPredicate\}\big)^\totalParties} \welfare(\profile)} =\\
            =&\frac{(\revision{\stake} + \rounds \cdot \reward) \cdot \rateMax}{\min_{\profile \in \equilibriaSet\cap\big(\{\proto, \strategy_\infractionPredicate\}\big)^\totalParties} \welfare(\profile)} \geq\\
            \geq&\frac{(\revision{\stake} + \rounds \cdot \reward) \cdot \rateMax}{(\revision{\stake} + \rounds \cdot \reward) \cdot \rateMin + \threshold \cdot (\revision{\stake} + \rounds \cdot \reward) \cdot \rateDiff} - \negl(\secparam) =\\
            =&\frac{1}{\frac{\rateMin}{\rateMax} + \threshold \cdot \big(1 - \frac{\rateMin}{\rateMax}\big)} - \negl(\secparam) =\\
            =&\frac{1}{\threshold + (1 - \threshold) \cdot \frac{\rateMin}{\rateMax}} - \negl(\secparam)\;.
        \end{split}
    \end{equation*}

    \emph{(\ref{item:effective_T-pan2}).}
    Now, let
    $\profile^* \in \equilibriaSet \cap\big(\{\proto, \strategy_\infractionPredicate\}\big)^\totalParties$
    \st:
    \[ \welfare(\profile^*) = \min_{\profile \in \equilibriaSet \cap(\{\proto, \strategy_\infractionPredicate\})^\totalParties} \welfare(\profile).\]
     
    Let $\mathbb{A}^*$ be the subset of parties that follow
    $\strategy_\infractionPredicate$ in $\profile^*$
    (\ie $\profile^* = \profile_\infractionPredicate^{\mathbb{A}^*}$). 
    We argue that
    $\sum_{\party \in \mathbb{A}^*}\power_\party \geq \threshold$. Indeed, if
    $\sum_{\party \in \mathbb{A}^*}\power_\party < \threshold$, then the exchange
    rate never drops and the external rewards are always $0$, so:
     
    \begin{equation*}
        \begin{split}
            \welfare(\profile^*)& = (\revision{\stake} + \rounds \cdot \reward) \cdot \rateMax >\\
            &> (\revision{\stake} + \rounds \cdot \reward) \cdot \rateMin + \threshold \cdot (\revision{\stake} + \rounds \cdot \reward) \cdot \rateDiff + \negl(\secparam) \geq\\
            &\geq (\revision{\stake} + \rounds \cdot \reward) \cdot \rateMin + \sum_{\party \in \partySet} \bribe_\party + \negl(\secparam)
        \end{split}
    \end{equation*}
    which contradicts to Eq.~\eqref{eq:effective_T-pan1}.

    Besides, the fact that
    $\sum_{\party \in \mathbb{A}^*}\power_\party \geq \threshold$ implies that
    the attack is almost always successful. Thus:
    \begin{equation}\label{eq:effective_T-pan2}
        \welfare(\profile^*) \geq (\revision{\stake} + \rounds \cdot \reward) \cdot \rateMin + \sum_{\party \in \mathbb{A}^*} \bribe_\party - \negl(\secparam)\;.
    \end{equation}
     
    By Eq.~\eqref{eq:effective_T-pan2}, we get that:
     
    \begin{equation}\label{eq:effective_T-pan4}
        \begin{split}
            &\{\proto, \strategy_\infractionPredicate\} \text{-} \priceAnar = \\
            =& \frac{\max_{\profile \in \strategySet^\totalParties} \welfare(\profile)}{\min_{\profile \in \equilibriaSet \cap\big(\{\proto, \strategy_\infractionPredicate\}\big)^\totalParties} \welfare(\profile)} =\\
            =&\frac{(\revision{\stake} + \rounds \cdot \reward) \cdot \rateMax}{\welfare(\profile^*)} \leq\\
            \leq&\frac{(\revision{\stake} + \rounds \cdot \reward) \cdot \rateMax}{(\revision{\stake} + \rounds \cdot \reward) \cdot \rateMin + \sum_{\party \in \mathbb{A}^*} \bribe_\party} + \negl(\secparam) \leq\\
            \leq&\frac{(\revision{\stake} + \rounds \cdot \reward) \cdot \rateMax}{(\revision{\stake} + \rounds \cdot \reward) \cdot \rateMin + \Phi} + \negl(\secparam)\;,
        \end{split}
    \end{equation}
    where $\Phi = \min_{\mathbb{A} \subseteq \partySet: \sum_{\party \in \mathbb{A}} \power_\party \geq \threshold} \sum_{\party \in \mathbb{A}} \bribe_\party$.

\end{proof}

\subsection{Accountable Rewards Under Guided Bribing}

\subsubsection{Protocol Equilibrium and Negative Equilibrium}\label{app:acc_eq1}

\paragraph{\textsc{Theorem}~\ref{thm:accountable_protocol_nash}}
 \emph{   Assume the following:
    \begin{itemize}
        \item $\proto$ is a protocol with block-proportional rewards (cf.
            Section~\ref{subsec:preliminaries_proportion}), with $\reward$
            rewards per block;
        \item $\profile_\proto$ denotes the all-honest strategy profile;
        \item $\threshold$ is the security threshold \wrt an infraction
            predicate $\infractionPredicate$;
        \item for every strategy profile $\profile$: 
            \begin{itemize}
                \item the exchange rate $\exchangeRate_\profile$ \wrt
                    $\infractionPredicate, \threshold$ is defined as in
                    Eq.~\eqref{eq:exchange};
                \item the external rewards $\utilityBoost_{\party, \profile}$,
                    due to guided bribing \wrt $\infractionPredicate$, are
                    defined as in Eq.~\eqref{eq:external_guided};
               \item the compliance payout $ \deposit_{\party, \profile}$ \wrt $\infractionPredicate$ is defined as in Eq.~\eqref{eq:deposit};
            \end{itemize}
        \item For every party $\party \in \partySet$, it holds that $\power_\party < \threshold$ \revision{and $\stake_\party = \power_\party \cdot \stake$}.  
    \end{itemize}
    Then, it holds that there is an $\epsilon$ negligible in $\secparam$ \st
    $\proto$ is an $\epsilon$-Nash equilibrium \wrt Accountable Reward under
    $\router_\text{sync}$ if and only if
    for every $\party \in \partySet$: $\beta_\party - \power_\party \cdot \deposit \cdot \rateMax \leq \negl(\secparam)$.
    }
\begin{proof}
(If). Assume that for every $\party\in\partySet$, it holds that $\bribe_\party-\power_\party\cdot\deposit\cdot\rateMax\leq\negl(\secparam)$. The utility of $\party$ in the all-honest setting is
\begin{equation}\label{eq:accountable_protocol_nash1}
\begin{split}
&\utility_{\party}(\profile_\proto)=\\
    =&\revision{\stake_\party \cdot E[\exchangeRate_{\profile_\proto}]} + E[\reward_{\party, \profile_\proto}\cdot\exchangeRate_{\profile_\proto}]+E[\utilityBoost_{\party, \profile_\proto}]+E[\deposit_{\party, \profile_\proto}\cdot\exchangeRate_{\profile_\proto}]=\\
    =&\revision{\stake_\party \cdot \rateMax} + E[\reward_{\party, \profile_\proto}]\cdot\rateMax+0+E[\deposit_{\party, \profile_\proto}] \cdot \rateMax =\\
    =&\revision{\stake_\party \cdot \rateMax} + E[\reward_{\party, \profile_\proto}] \cdot \rateMax + \power_{\party} \cdot \deposit \cdot \rateMax \;.
\end{split}
\end{equation}
Now, let $\party^*$ be a party that unilaterally deviates from $\profile_\proto$ and let $\profile^*$ be the resulting strategy profile. Let $D$ be the event that $\party^*$ performs an infraction. Since $\power_{\party^*}<\threshold$, the exchange rate is always $\rateMax$ in $\profile^*$. When $D$ happens, $\party^*$ gets her external rewards but forfeits her compliance output. Therefore, by Eq.~\eqref{eq:accountable_protocol_nash1}, we have that 

\begin{equation*}
\begin{split}
&\utility_{\party^*}(\profile^*)=\\
    =&\revision{\stake_\party \cdot E[\exchangeRate_{\profile^*}]} + E[\reward_{\party^*, \profile^*}\cdot\exchangeRate_{\profile^*}]+E[\utilityBoost_{\party^*, \profile^*}]+E[\deposit_{\party^*, \profile^*}]=\\
    =&\sum_z z\cdot\Pr[(\revision{\stake_{\party^*}} + \reward_{\party^*, \profile^*})\cdot\exchangeRate_{\profile^*}+\utilityBoost_{\party^*, \profile^*}+\deposit_{\party^*, \profile^*}=z]=\\
    =&\sum_z z\cdot\Pr[(\revision{\stake_{\party^*}} + \reward_{\party^*, \profile^*})\cdot\exchangeRate_{\profile^*}+\utilityBoost_{\party^*, \profile^*}+\deposit_{\party^*, \profile^*}=z\wedge D]+\\
    &+\sum_z z\cdot\Pr[(\revision{\stake_{\party^*}} + \reward_{\party^*, \profile^*})\cdot\exchangeRate_{\profile^*}+\utilityBoost_{\party^*, \profile^*}+\\
  &\quad\quad+\deposit_{\party^*, \profile^*}=z\wedge \neg D]=\\
    =&\Pr[D]\cdot\sum_z z\cdot\Pr[(\revision{\stake_{\party^*}} + \reward_{\party^*, \profile^*})\cdot\exchangeRate_{\profile^*}+\utilityBoost_{\party^*, \profile^*}+\\
  &\quad\quad+\deposit_{\party^*, \profile^*}=z| D]+\\
    &+\Pr[\neg D]\cdot\sum_z z\cdot\Pr[(\revision{\stake_{\party^*}} + \reward_{\party^*, \profile^*})\cdot\exchangeRate_{\profile^*}+\utilityBoost_{\party^*, \profile^*}+\\
 &\quad\quad+\deposit_{\party^*, \profile^*}=z| \neg D]=\\
    =&\Pr[D]\cdot\sum_z z\cdot\Pr[(\revision{\stake_{\party^*}} + \reward_{\party^*, \profile^*})\cdot\rateMax+\bribe_{\party^*}+0=z| D]+\\
    &+\Pr[\neg D]\cdot\sum_z z\cdot\Pr[(\revision{\stake_{\party^*}} + \reward_{\party^*, \profile^*})\cdot\rateMax+0+\power_{\party^*}\cdot\deposit\cdot\rateMax=z| \neg D]=\\
    =&\Pr[D]\cdot\sum_z z\cdot\Pr[(\revision{\stake_{\party^*}} + \reward_{\party^*, \profile^*})\cdot\rateMax+\bribe_{\party^*}+0=z]+\\
    &+\Pr[\neg D]\cdot\sum_z z\cdot\Pr[(\revision{\stake_{\party^*}} + \reward_{\party^*, \profile^*})\cdot\rateMax+0+\power_{\party^*}\cdot\deposit\cdot\rateMax=z]=\\
    =&\Pr[D]\cdot \big((\revision{\stake_{\party^*}} + E[\reward_{\party^*, \profile^*}])\cdot\rateMax+\bribe_{\party^*}\big)+\\
 &+\Pr[\neg D]\cdot (\revision{\stake_{\party^*}} + E[\reward_{\party^*, \profile^*}]+\power_{\party^*}\cdot\deposit)\cdot\rateMax=\\
    =&\revision{\stake_{\party^*} \cdot \rateMax} + E[\reward_{\party^*, \profile^*}]\cdot\rateMax +\\
    &+\Pr[D]\cdot \bribe_{\party^*} + \Pr[\neg D]\cdot\power_{\party^*}\cdot\deposit\cdot\rateMax=\\
    =&\revision{\stake_{\party^*} \cdot \rateMax} + E[\reward_{\party^*, \profile^*}]\cdot\rateMax + \Pr[D]\cdot \bribe_{\party^*} +\\
 &+(1-\Pr[D])\cdot\power_{\party^*}\cdot\deposit\cdot\rateMax\leq\\
 =&\revision{\stake_{\party^*} \cdot \rateMax} + E[\reward_{\party^*, \profile^*}]\cdot\rateMax + \power_{\party^*}\cdot\deposit\cdot\rateMax +\\
 &+\Pr[D]\cdot(\bribe_{\party^*}-\power_{\party^*}\cdot\deposit\cdot\rateMax)\leq\\
 %
 %
 \leq&\utility_{\party^*}(\profile_\proto)+\negl(\secparam)\;.
\end{split}
\end{equation*}
Namely, there is an $\epsilon$ negligible in $\secparam$ s.t. $\Pi$ is an $\epsilon$-Nash equilibrium \wrt Accountable Reward under $\router_\text{sync}$.\\

(Only if). Assume that there is a party $\party^*$ such that $\bribe_{\party^*}-\power_{\party^*}\cdot\deposit\cdot\rateMax$ is a non-negligible value. Consider the case where $\party^*$ unilaterally deviates from $\profile_\proto$ by following $\strategy_\infractionPredicate$ and let $\profile^*$ be the resulting strategy profile. Since $\party^*$ will be almost always successful in performing the infraction, it holds that with $1-\negl(\secparam)$ probability, $\party^*$ will get its external rewards but no compliance payout when following $\profile^*$. In addition, $\power_\party^*<\threshold$, so the exchange rate will never drop in $\profile^*$. The above imply that the utility gain of $\party^*$ is $\bribe_{\party^*}-\power_{\party^*}\cdot\deposit\cdot\rateMax-\negl(\secparam)$, which is a non-negligible value. We conclude that $\Pi$ is not a $\negl(\secparam)$-Nash equilibrium \wrt Accountable Reward under $\router_\text{sync}$.

\end{proof}

\paragraph{\textsc{Theorem}~\ref{thm:accountable_negative_nash}}
 \emph{Assume the following:
    \begin{itemize}
        \item $\proto$: a protocol with block-proportional rewards (cf.
            Section~\ref{subsec:preliminaries_proportion}), with $\reward$
            rewards per block;
        \item $\threshold$: the security threshold \wrt an infraction predicate $\infractionPredicate$;
        \item $\profile_\infractionPredicate$: the strategy profile where all parties perform $\infractionPredicate$;
        \item for every strategy profile $\profile$: 
            \begin{itemize}
                \item the exchange rate $\exchangeRate_\profile$ \wrt
                    $\infractionPredicate, \threshold$ follows
                    Eq.~\eqref{eq:exchange};
                \item the external rewards $\utilityBoost_{\party, \profile}$,
                    due to guided bribing \wrt $\infractionPredicate$, 
                    follow Eq.~\eqref{eq:external_guided};
               \item the compliance payout $ \deposit_{\party, \profile}$ \wrt $\infractionPredicate$ follows Eq.~\eqref{eq:deposit};
            \end{itemize}
          \item For every party $\party \in \partySet$, it holds that $\power_\party < 1 - \threshold$ \revision{and $\stake_\party = \power_\party \cdot \stake$}.  
    \end{itemize}
    Then, it holds that there is an $\epsilon$ negligible in $\secparam$ \st
    $\profile_\infractionPredicate$ is an $\epsilon$-Nash equilibrium \wrt Accountable Reward under
    $\router_\text{sync}$ if
    for every $\party \in \partySet$: $\beta_\party - \power_\party \cdot \deposit \cdot \rateMin \geq \negl(\secparam)$.
    }
\begin{proof}
    The proof follows similarly to Theorem~\ref{thm:guided}. The only
    additional element is using the inequality 
    $\power_\party \cdot \deposit \cdot \rateMin \leq \bribe_\party$
    in order to show that deviating from the ``all bribed'' setting does not
    increase a party's utility. 
    For completeness, we described the proof in detail below.

    For some subset of parties $\infractionPartySet$, which follow
    $\strategy_\infractionPredicate$, let $F_\infractionPartySet$ be the event that the
    parties in $\infractionPartySet$ fail to perform the infraction. 
    By the discussion on Definition~\ref{def:attack-failure-event}, if
    $\sum_{\party \in \infractionPartySet} \power_\party \geq \threshold$, 
    then $\Pr[F_\infractionPartySet] = \negl(\secparam)$. 

    Let $Q_\party$ be the event that $\party$ does not perform an infraction.
    By Section~\ref{subsec:bribe_IPs}, we have that $\Pr[Q_\party] = \negl(\secparam)$. 
    Clearly, when $\neg Q_\party$ happens, the external rewards of $\party$ are $\bribe_\party$ and the compliance payout is $0$.

    Assume the strategy profile
    $\profile_\infractionPredicate = \langle \strategy_\infractionPredicate, \ldots, \strategy_\infractionPredicate \rangle$.
    It is straightforward that $\Pr[F_\partySet] = \negl(\secparam)$. 
    Therefore,
    $\neg F_\partySet$ happens with overwhelming probability, \ie all parties
    perform the infraction. Therefore, the exchange rate is $\rateMin$ and the parties
    receive their external rewards. By Eq.~\eqref{eq:exchange}
    and~\eqref{eq:external_guided}, and the fact that under block-proportional
    rewards, a party's maximum system's reward is polynomially bounded by
    $\rounds \cdot \reward$, where $\rounds$ is the number of rounds in the
    execution, the utility of some party $\party$ is:

    \begin{equation}\label{eq:utility_ds_all-accountable}
        \begin{split}
           & \utility_\party(\profile_\infractionPredicate)=\\
            =&\revision{\stake_\party \cdot E[\exchangeRate_{\profile_\infractionPredicate}]} + E[\reward_{\party, \profile_\infractionPredicate}\cdot\exchangeRate_{\profile_\infractionPredicate}]+E[\utilityBoost_{\party, \profile_\infractionPredicate}]+E[\deposit_{\party, \profile_\infractionPredicate}\cdot\exchangeRate_{\profile_\infractionPredicate}]=\\
            =&\revision{\stake_\party \cdot \sum_z z\cdot\Pr[\exchangeRate_{\profile_\infractionPredicate}=z]} + \sum_z z\cdot\Pr[\reward_{\party, \profile_\infractionPredicate}\cdot \exchangeRate_{\profile_\infractionPredicate}=z]+\\
            &+\sum_z z\cdot\Pr[\utilityBoost_{\party, \profile_\infractionPredicate}=z]+\sum_z z\cdot\Pr[\deposit_{\party, \profile_\infractionPredicate}\cdot \exchangeRate_{\profile_\infractionPredicate}=z]\geq\\
            \geq&\revision{\stake_\party \cdot \sum_z z\cdot\Pr[\exchangeRate_{\profile_\infractionPredicate}=z\wedge (\neg F_\partySet)]}+\\
            &+\sum_z z\cdot\Pr[\reward_{\party, \profile_\infractionPredicate}\cdot \exchangeRate_{\profile_\infractionPredicate}=z\wedge (\neg F_\partySet)]+\\
            &+\sum_z z\cdot\Pr[\utilityBoost_{\party, \profile_\infractionPredicate}=z\wedge (\neg Q_\party)]+\\
            &+\sum_z z\cdot\Pr[\deposit_{\party, \profile_\infractionPredicate}\cdot \exchangeRate_{\profile_\infractionPredicate}=z\wedge (\neg Q_\party)]=\\
            =&\revision{\stake_\party \cdot \sum_z z\cdot\Pr[\neg F_\partySet]\cdot\Pr[\exchangeRate_{\profile_\infractionPredicate}=z|\neg F_\partySet]}+\\
            &+\sum_z z\cdot\Pr[\neg F_\partySet]\cdot\Pr[\reward_{\party, \profile_\infractionPredicate}\cdot \exchangeRate_{\profile_\infractionPredicate}=z|\neg F_\partySet]+\\
            &+\sum_z z\cdot\Pr[\neg Q_\party]\cdot\Pr[\utilityBoost_{\party, \profile_\infractionPredicate}=z|\neg Q_\party]+\\
            &+\sum_z z\cdot\Pr[\neg Q_\party]\cdot\Pr[\deposit_{\party, \profile_\infractionPredicate}\cdot \exchangeRate_{\profile_\infractionPredicate}=z| \neg Q_\party]=\\
            =&\revision{\stake_\party \cdot \rateMin \cdot\Pr[\neg F_\partySet]}+\\
            &+\sum_z z\cdot\Pr[\neg F_\partySet]\cdot\Pr[\reward_{\party, \profile_\infractionPredicate}\cdot \rateMin=z|\neg F_\partySet]+\\
            &+\sum_z z\cdot\Pr[\neg Q_\party]\cdot\Pr[\utilityBoost_{\party, \profile_\infractionPredicate}=z|\neg Q_\party]+0=\\
            =&\revision{\stake_\party \cdot \rateMin \cdot\Pr[\neg F_\partySet]}+\\
            &+\sum_z z\cdot\Pr[\reward_{\party, \profile_\infractionPredicate}\cdot \rateMin=z\wedge\neg F_\partySet]+\\
            &+\sum_z z\cdot\Pr[\neg Q_\party]\cdot\Pr[\utilityBoost_{\party, \profile_\infractionPredicate}=z|\neg Q_\party]\geq\\
            \geq&\revision{\stake_\party \cdot \rateMin \cdot (1 - \negl(\secparam))}+\\
            &+\sum_z z\cdot\big(\Pr[\reward_{\party, \profile_\infractionPredicate}\cdot \rateMin=z]-\negl(\secparam)\big)+\\
            &+\sum_z z\cdot\Pr[\neg Q_\party]\cdot\Pr[\utilityBoost_{\party, \profile_\infractionPredicate}=z|\neg Q_\party]\geq\\
            \geq&\revision{\stake_\party \cdot \rateMin \cdot (1 - \negl(\secparam))}+\\
            &+\sum_z z\cdot\Pr[\reward_{\party, \profile_\infractionPredicate}\cdot \rateMin=z]-\sum_z z\cdot\negl(\secparam)+\\
            &+(1-\negl(\secparam))\cdot\sum_z z\cdot\Pr[\utilityBoost_{\party, \profile_\infractionPredicate}=z|\neg Q_\party]\geq\\
            \geq&\revision{\stake_\party \cdot \rateMin} + E[\reward_{\party, \profile_\infractionPredicate}]\cdot\rateMin-\big(\revision{\stake_\party\cdot\rateMin} + (\rounds \cdot \reward)^2\big)\cdot\negl(\secparam)+\\
            &+(1-\negl(\secparam))\cdot\bribe_\party=\\
            =&\revision{\stake_\party \cdot \rateMin} + E[\reward_{\party, \profile_\infractionPredicate}]\cdot\rateMin+\bribe_\party-\negl(\secparam).
        \end{split}
    \end{equation}

    Assume now that some party $\party^*$ unilaterally deviates from
    $\profile_\infractionPredicate$ to some strategy profile $\profile$. 

    By assumption, $\power_{\party^*} \leq 1 - \threshold$. Observe that, under
    $\profile$, all other parties, which collectively hold at least
    $\threshold$ of participating power, perform an infraction whenever they
    can. So when $\neg F_{\partySet \setminus \{ \party^* \}}$ happens, the
    exchange rate is $\rateMin$ irrespective of the strategy of $\party^*$.
    Since $\sum_{\party \in \partySet \setminus \{ \party^* \}} \power_\party \geq \threshold$,
    we have that $\Pr \big[ F_{\party \in \partySet \setminus \{ \party^* \}} \big] = \negl(\secparam)$.
    In addition, a party's expected ledger rewards are maximized when it
    follows $\strategy_\infractionPredicate$. 
    Finally, by not performing the infraction, party $\party^*$ receives the compliance payout 
    following Eq.~\ref{eq:compliance_rewards}.

    Given the fact that, under block-proportional
    rewards, a party's maximum system's reward is polynomially bounded by
    $\rounds \cdot \reward$, we get that:
     
    \begin{equation}\label{eq:utility_ds_all_deviate-accountable}
        \begin{split}
           & \utility_{\party^*}(\profile)=\\
            =&\revision{\stake_{\party^*} \cdot E[\exchangeRate_\profile]} + E[\reward_{\party^*, \profile}\cdot\exchangeRate_\profile]+E[\utilityBoost_{\party^*, \profile}] + E[\deposit_{\party^*, \profile}\cdot\exchangeRate_{\profile}] =\\
            =&\revision{\stake_{\party^*} \cdot \sum_z z\cdot\Pr[\exchangeRate_{\profile}=z]} + \sum_z z\cdot\Pr[\reward_{\party^*, \profile}\cdot \exchangeRate_{\profile}=z]+\\
            &+\sum_z z\cdot\Pr[\utilityBoost_{\party^*, \profile}=z]+\sum_z z\cdot\Pr[\deposit_{\party^*, \profile}\cdot \exchangeRate_{\profile}=z]\leq\\
            =&\revision{\stake_{\party^*} \cdot \sum_z z\cdot\Pr[\exchangeRate_\profile=z]}+\sum_z z\cdot\Pr[\reward_{\party^*, \profile}\cdot\exchangeRate_\profile=z]+\\
            &+\sum_z z\cdot\big(\Pr\big[\utilityBoost_{\party^*, \profile}=z\wedge\big(\neg F_{\party\in\partySet\setminus\{\party^*\}}\big)\big]+\negl(\secparam)\big)+\\
            &+\sum_z z\cdot\big(\Pr\big[\deposit_{\party^*, \profile}\cdot \exchangeRate_{\profile}=z\wedge\big(\neg F_{\party\in\partySet\setminus\{\party^*\}}\big)]+\negl(\secparam)\big)\leq\\
            \leq&\revision{\stake_{\party^*} \cdot \sum_z z\cdot\Pr[\exchangeRate_\profile=z]} + \sum_z z\cdot\Pr[\reward_{\party^*, \profile}\cdot\exchangeRate_\profile=z]+\\
            &+\sum_z z\cdot\Pr\big[\utilityBoost_{\party^*, \profile}=z\big|\neg F_{\party\in\partySet\setminus\{\party^*\}}\big]+\\
            &+\sum_z z\cdot\Pr\big[\deposit_{\party^*, \profile}\cdot \exchangeRate_{\profile}=z\big|\neg F_{\party\in\partySet\setminus\{\party^*\}}]+\negl(\secparam)\leq\\
            \leq&\revision{\stake_{\party^*} \cdot \sum_z z\cdot\Pr[\exchangeRate_\profile=z]} + \sum_z z\cdot\Pr[\reward_{\party^*, \profile}\cdot\exchangeRate_\profile=z]+\\
            &\mathsf{max}\{\bribe_{\party^*},\power_{\party^*} \cdot \deposit \cdot \rateMin\}+\negl(\secparam)\leq\\
            =&\revision{\stake_{\party^*} \cdot \sum_z z\cdot\Pr\big[\exchangeRate_\profile=z\wedge\big(\neg F_{\party\in\partySet\setminus\{\party^*\}}\big)\big]}+\\
            &+\revision{\stake_{\party^*} \cdot \sum_z z\cdot\Pr\big[\exchangeRate_\profile=z\wedge F_{\party\in\partySet\setminus\{\party^*\}}\big]}+\\
            &+\sum_z z\cdot\Pr\big[\reward_{\party^*, \profile}\cdot\exchangeRate_\profile=z\wedge\big(\neg F_{\party\in\partySet\setminus\{\party^*\}}\big)\big]+\\
            &+\sum_z z\cdot\Pr\big[\reward_{\party^*, \profile}\cdot\exchangeRate_\profile=z\wedge F_{\party\in\partySet\setminus\{\party^*\}}\big]+\\
            &+\bribe_{\party^*}+\negl(\secparam)\leq\\
            \leq&\revision{\stake_{\party^*} \cdot \sum_z \Big(z\cdot\Pr\big[\neg F_{\party\in\partySet\setminus\{\party^*\}}\big]\cdot\Pr\big[\exchangeRate_\profile=z\big|\neg F_{\party\in\partySet\setminus\{\party^*\}}\big]\Big)}+\\
              &+\revision{\stake_{\party^*} \cdot \sum_z z\cdot\Pr\big[ F_{\party\in\partySet\setminus\{\party^*\}}\big]}+\\
            &+\sum_z \Big(z\cdot\Pr\big[\neg F_{\party\in\partySet\setminus\{\party^*\}}\big]\cdot\\
            &\quad\cdot\Pr\big[\reward_{\party^*, \profile}\cdot \exchangeRate_\profile=z\big|\neg F_{\party\in\partySet\setminus\{\party^*\}}\big]\Big)+\\
            &+\sum_z z\cdot\Pr\big[ F_{\party\in\partySet\setminus\{\party^*\}}\big]+\bribe_{\party^*}+\negl(\secparam)=\\
                    \end{split}
    \end{equation}
               \begin{equation*}
        \begin{split}
            =&\revision{\stake_{\party^*} \cdot \rateMin \cdot \big[\neg F_{\party\in\partySet\setminus\{\party^*\}}\big]}+\\
              &+\revision{\stake_{\party^*} \cdot \sum_z z\cdot\negl(\secparam)}+\\
            &+\sum_z \Big(z\cdot\Pr\big[\neg F_{\party\in\partySet\setminus\{\party^*\}}\big]\cdot\\
            &\quad\cdot\Pr\big[\reward_{\party^*, \profile}\cdot \rateMin=z\big|\neg F_{\party\in\partySet\setminus\{\party^*\}}\big]\Big)+\\
            &+\sum_z z\cdot\Pr\big[ F_{\party\in\partySet\setminus\{\party^*\}}\big]+\bribe_{\party^*}+\negl(\secparam)\leq\\
            \leq&\revision{\stake_{\party^*} \cdot \rateMin \cdot (1 - \negl(\secparam))}+\\
            &+\revision{\stake_{\party^*} \cdot \rateMax^2\cdot\negl(\secparam)}+\\
            &+\sum_z z\cdot\Pr\big[\reward_{\party^*, \profile}\cdot \rateMin=z\wedge\neg F_{\party\in\partySet\setminus\{\party^*\}}\big]+\\
            &+\sum_z z\cdot\negl(\secparam)+\bribe_{\party^*}+\negl(\secparam)\leq\\
            \leq&\revision{\stake_{\party^*} \cdot \rateMin \cdot (1 - \negl(\secparam))}+\\
            &+\sum_z z\cdot\Pr\big[\reward_{\party^*, \profile}\cdot \rateMin=z\big]+\\
            &+\sum_z z\cdot\negl(\secparam)+\bribe_{\party^*}+\negl(\secparam)\leq\\
            \leq&\revision{\stake_{\party^*} \cdot \rateMin} + E[\reward_{\party^*, \profile}]\cdot\rateMin +\big((\revision{\rounds\cdot\reward)^2 - \stake_{\party^*}\cdot\rateMin} \big)\cdot\negl(\secparam)+\\
            &+\bribe_{\party^*}+\negl(\secparam)\leq\\
            \leq&\revision{\stake_{\party^*} \cdot \rateMin} + E[\reward_{\party^*, \profile}]\cdot\rateMin+\bribe_{\party^*}+\negl(\secparam)\leq\\
            \leq& \revision{\stake_{\party^*} \cdot \rateMin} + E[\reward_{\party^*, \profile_\infractionPredicate}]\cdot\rateMin+\bribe_{\party^*}+\negl(\secparam).
        \end{split}
    \end{equation*}
    Therefore, by Equations~\eqref{eq:utility_ds_all-accountable}
    and~\eqref{eq:utility_ds_all_deviate-accountable}, we have that:
    \[ \utility_{\party^*}(\profile) \leq \utility_{\party^*}(\profile_\infractionPredicate) + \negl(\secparam) .\]

    So, $\profile_\infractionPredicate$ is an $\epsilon$-Nash equilibrium \wrt
    Reward under $\router_\text{sync}$, for some $\epsilon = \negl(\secparam)$.
\end{proof}

\subsubsection{Promising Parties}\label{app:acc_prom}

\paragraph{\textsc{Lemma}~\ref{lem:acc_maximal_existence}}
  \emph{  Assume the following:
    \begin{itemize}
        \item  $\proto$ is a protocol run by parties in $\partySet$.
        \item $\threshold$ is a security threshold \wrt $\infractionPredicate$;
        \item for every strategy profile $\profile$:
            \begin{itemize}
                \item the external rewards $\utilityBoost_{\party, \profile}$
                    due to guided bribing \wrt $\infractionPredicate$ are
                    defined as in Eq.~\eqref{eq:external_guided};
                \item the compliance payout $ \deposit_{\party, \profile}$ \wrt $\infractionPredicate$ 
                    is defined as in Eq.~\eqref{eq:deposit};
            \end{itemize}
        \item $\sum_{\party \in \partySet} \bribe_\party \leq \threshold \cdot \deposit\cdot\rateMax$.
    \end{itemize}
    Then, it holds that $\sum_{\party \in \partySet_\text{prom}^{\{ \bribe_\party \}, \deposit}} \power_\party < \threshold$.}
\begin{proof}
    For the sake of contradiction, assume that
    $\sum_{\party \in \partySet_\text{prom}^{\{\bribe_\party\},\deposit}} \power_\party \geq \threshold$.
    We have that    %
    \begin{equation*}
        \begin{split}
            &\sum_{\party\in\partySet}\bribe_\party\geq\sum_{\party\in\partySet_\text{prom}^{\{\bribe_\party\},\deposit}}\bribe_\party>\\
            >&\sum_{\party \in \partySet_\text{prom}^{\{\bribe_\party\},\deposit}} \power_\party\cdot\deposit\cdot\rateMax\geq\threshold\cdot\deposit\cdot\rateMax
        \end{split}
    \end{equation*}
    which contradicts to the condition $\sum_{\party \in \partySet} \bribe_\party \leq \threshold\cdot\deposit\cdot\rateMax$.\smallskip

   So, we have that
    $\sum_{\party \in \partySet_\text{prom}^{\{\bribe_\party\},\deposit}} \power_\party < \threshold$.
\end{proof}


\paragraph{\textsc{Theorem}~\ref{thm:acc_maximal_sufficient}}
   \emph{ Assume the following:
    \begin{itemize}
        \item $\proto$ is a protocol run by parties in $\partySet$;
        \item \revision{each party's stake is $\stake_\party = \power_\party \cdot \stake$;}
        \item $\threshold$ is the security threshold \wrt an infraction
            predicate $\infractionPredicate$;
        \item for every strategy profile $\profile$: 
            \begin{itemize}
                \item the exchange rate $\exchangeRate_\profile$ \wrt
                    $\infractionPredicate, \threshold$ is defined as in
                    Eq.~\eqref{eq:exchange};
                \item the external rewards $\utilityBoost_{\party, \profile}$,
                    due to guided bribing \wrt $\infractionPredicate$, are
                    defined as in Eq.~\eqref{eq:external_guided}.
                \item the compliance payout $ \deposit_{\party, \profile}$ \wrt $\infractionPredicate$ 
                    is defined as in Eq.~\eqref{eq:deposit};
            \end{itemize}
    \end{itemize}
    If $\sum_{\party \in \partySet_\text{prom}^{\{\bribe_\party\} ,\deposit}} \power_\party < \threshold$, 
    then there exists an $\epsilon$ negligible in $\secparam$ \st
    $\profile_\infractionPredicate^{\partySet_\text{prom}^{\{\bribe_\party\}, \deposit}}$ is an $\epsilon$-Nash
    equilibrium \wrt utility Accountable Reward under $\router_\text{sync}$.
    }
\begin{proof}
    Let $\party^*$ be a party that unilaterally deviates from
    $\profile_\infractionPredicate^{\partySet_\text{prom}^{\{\bribe_\party\},\deposit}}$ to strategy profile
    $\profile$. Since
    $\sum_{\party \in\partySet_\text{prom}^{\{\bribe_\party\},\deposit}}\power_\party < \threshold$, the
    exchange rate is always $\rateMax$ in
    $\profile_\infractionPredicate^{\partySet_\text{prom}^{\{\bribe_\party\},\deposit}}$. 

    We distinguish the following cases.

    \underline{Case 1. $\party^*\in\partySet\setminus\partySet_\text{prom}^{\{\bribe_\party\},\deposit}$:} 
    By the definition of $\partySet_\text{prom}^{\{\bribe_\party\},\deposit}$, we have that
     $\bribe_{\party^*}\leq \power_{\party^*}\cdot\deposit\cdot \rateMax$.  In addition, $\party^*$ is honest in $\partySet_\text{prom}^{\{\bribe_\party\},\deposit}$, so we have that      
    \[\utility_{\party^*}(\profile_\infractionPredicate^{\partySet_\text{prom}^{\{\bribe_\party\},\deposit}})=\revision{\stake_{\party^*} \cdot \rateMax} + E[\reward_{\party^*, \profile_\proto}]\cdot \rateMax+\power_{\party^*}\cdot\deposit\cdot\rateMax.\]

    Let $D$ be the event that $\party^*$ performs an infraction \wrt
    $\infractionPredicate$ by following $\profile$. Since $\profile$ is a
    unilateral deviation from
    $\profile_\infractionPredicate^{\partySet_\text{prom}^{\{\bribe_\party\},\deposit}}$, all the parties in
    $\partySet_\text{prom}^{\{\bribe_\party\},\deposit}$ will perform an infraction when they are able to. For
    every $\party \in \partySet_\text{prom}^{\{\bribe_\party\},\deposit}$, let $F_\party$ be the event that
    $\party$ (following $\profile$) fails to perform an infraction. Since
    $\party$ behaves according to
    $\strategy_\infractionPredicate$, we have that
    $\Pr[F_\party] = \negl(\secparam)$. 

    Thus, we have that:

    \begin{equation*}
        \begin{split}
            &\utility_{\party^*}(\profile)=\\
            =&\revision{\stake_{\party^*} \cdot E[\exchangeRate_{\profile}]} + E[\reward_{\party^*, \profile}\cdot\exchangeRate_{\profile}]+E[\utilityBoost_{\party^*, \profile}]+E[\deposit_{\party^*, \profile}\cdot\exchangeRate_{\profile}]=\\
            =&\sum_z z\cdot\Pr[\revision{\stake_{\party^*} \cdot \exchangeRate_{\profile}} + \reward_{\party^*, \profile}\cdot\exchangeRate_{\profile}+\utilityBoost_{\party^*, \profile}+\deposit_{\party^*, \profile}\cdot\exchangeRate_{\profile}=z]=\\
            =&\sum_z z\cdot\Pr\Big[\revision{\stake_{\party^*} \cdot \exchangeRate_{\profile}} + \reward_{\party^*, \profile}\cdot\exchangeRate_{\profile}+\utilityBoost_{\party^*, \profile}+\deposit_{\party^*, \profile}\cdot\exchangeRate_{\profile}=z\wedge D\\
            &\quad\wedge(\bigwedge_{\party\in\partySet_\text{prom}^{\{\bribe_\party\},\deposit}}\neg F_\party)\Big]+\\
            &+\sum_z z\cdot\Pr\Big[\revision{\stake_{\party^*} \cdot \exchangeRate_{\profile}} + \reward_{\party^*, \profile}\cdot\exchangeRate_{\profile}+\utilityBoost_{\party^*, \profile}+\deposit_{\party^*, \profile}\cdot\exchangeRate_{\profile}=z\wedge D\\
            &\quad\wedge(\bigvee_{\party\in\partySet_\text{prom}^{\{\bribe_\party\},\deposit}} F_\party))\Big]+\\
            &+\sum_z z\cdot\Pr\big[\revision{\stake_{\party^*} \cdot \exchangeRate_{\profile}} + \reward_{\party^*, \profile}\cdot\exchangeRate_{\profile}+\utilityBoost_{\party^*, \profile}+\\
            &\quad+\deposit_{\party^*, \profile}\cdot\exchangeRate_{\profile}=z\wedge \neg D\big]\leq\\
            \leq&\sum_z z\cdot\Pr\Big[D\wedge(\bigwedge_{\party\in\infractionPartySet}\neg F_\party)\Big]\cdot \Pr\Big[\revision{\stake_{\party^*} \cdot \exchangeRate_{\profile}} + \reward_{\party^*, \profile}\cdot\exchangeRate_{\profile}+\utilityBoost_{\party^*, \profile}+\\
            &\quad+\deposit_{\party^*, \profile}\cdot\exchangeRate_{\profile}=z\Big| D\wedge(\bigwedge_{\party\in\partySet_\text{prom}^{\{\bribe_\party\},\deposit}}\neg F_\party)\Big]+\\
            &+\sum_z z\cdot\Pr\Big[\bigvee_{\party\in\partySet_\text{prom}^{\{\bribe_\party\},\deposit}} F_\party\Big]+\\
            &+\sum_z z\cdot\Pr\big[\neg D\big]\cdot\Pr\big[\revision{\stake_{\party^*} \cdot \exchangeRate_{\profile}} + \reward_{\party^*, \profile}\cdot\exchangeRate_{\profile}+\utilityBoost_{\party^*, \profile}+\\
            &\quad+\deposit_{\party^*, \profile}\cdot\exchangeRate_{\profile}=z\big| \neg D\big]\leq\\
            \leq&\sum_z z\cdot\Big(\Pr[D]-\Pr\Big[D\wedge(\bigvee_{\party\in\partySet_\text{prom}^{\{\bribe_\party\},\deposit}} F_\party)\Big]\Big)\cdot\\
            &\quad\cdot\Big(\Pr\Big[\revision{\stake_{\party^*} \cdot \exchangeRate_{\profile}} + \reward_{\party^*, \profile}\cdot\exchangeRate_{\profile}+\bribe_{\party^*}+0=z\Big| D\wedge\\
            &\quad\quad\wedge(\bigwedge_{\party\in\partySet_\text{prom}^{\{\bribe_\party\},\deposit}}\neg F_\party)\Big]+\negl(\secparam)\Big)\\
            &+\sum_z z\cdot\negl(\secparam)+\\
            &+\sum_z z\cdot\Pr[\neg D]\cdot\\
            &\quad\cdot\Pr\big[\revision{\stake_{\party^*} \cdot \rateMax} + \reward_{\party^*, \profile}\cdot \rateMax+0+\power_{\party^*}\cdot\deposit\cdot \rateMax=z\big|\neg D\big]\leq\\
                    \end{split}
    \end{equation*}
               \begin{equation*}
        \begin{split}
            \leq&\sum_z z\cdot\big(\Pr[D]-\negl(\secparam)\big)\cdot\big(\Pr\big[\revision{\stake_{\party^*} \cdot \exchangeRate_{\profile}} + \reward_{\party^*, \profile}\cdot\exchangeRate_{\profile}+\bribe_{\party^*}=z\big]+\\
            &\quad+\negl(\secparam)\big)+(\rounds\cdot\reward)^2\cdot\negl(\secparam)+\\
            &+\sum_z z\cdot\Pr[\neg D]\cdot\Pr\big[\revision{\stake_{\party^*}\cdot \rateMax}+\reward_{\party^*, \profile}\cdot \rateMax+\\
            &\quad+\power_{\party^*}\cdot\deposit\cdot\rateMax=z\big]\leq\\
            \leq&\negl(\secparam)+\\
            &+\sum_z z\cdot\Pr[D]\cdot\Pr\big[\revision{\stake_{\party^*}\cdot\exchangeRate_{\profile}}+\reward_{\party^*, \profile}\cdot\exchangeRate_{\profile}+\bribe_{\party^*}=z\big]+\\
            &+(\rounds\cdot\reward)^2\cdot\negl(\secparam)+\\
            &+\sum_z z\cdot\Pr[\neg D]\cdot\Pr\big[\revision{\stake_{\party^*}\cdot \rateMax}+\reward_{\party^*, \profile}\cdot \rateMax+\\
            &\quad+\power_{\party^*}\cdot\deposit\cdot\rateMax=z\big]\leq\\
            \leq&\negl(\secparam)+\Pr[D]\cdot\big(\revision{\stake_{\party^*}\cdot E[\exchangeRate_{\profile}]} + E[\reward_{\party^*, \profile}\cdot\exchangeRate_{\profile}]+\bribe_{\party^*}\big)+\\
            &+(\rounds\cdot\reward)^2\cdot\negl(\secparam)+\\
            &+(1-\Pr[ D])\cdot\big(\revision{\stake_{\party^*}\cdot \rateMax} + E[\reward_{\party^*, \profile}]\cdot \rateMax+\power_{\party^*}\cdot\deposit\cdot\rateMax\big)\leq\\
            \leq&\negl(\secparam)+\Pr[D]\cdot\big(\revision{\stake_{\party^*}\cdot\rateMax}+E[\reward_{\party^*, \profile_\proto}]\cdot\rateMax+\bribe_{\party^*}\big)+\\
            &+(\rounds\cdot\reward)^2\cdot\negl(\secparam)+\\
            &+(1-\Pr[ D])\cdot\big(\revision{\stake_{\party^*}\cdot \rateMax}+E[\reward_{\party^*, \profile_\proto}]\cdot \rateMax+\power_{\party^*}\cdot\deposit\cdot\rateMax\big)\leq\\
            \leq& \revision{\stake_{\party^*} \cdot \rateMax} + E[\reward_{\party^*, \profile_\proto}]\cdot \rateMax+\power_{\party^*}\cdot\deposit\cdot\rateMax\\
            &+\Pr[D]\cdot\big(\bribe_{\party^*}-(\revision{\stake_{\party^*}} + E[\reward_{\party^*, \profile_\proto}])\cdot\\
            &\quad\cdot(\rateMax-\rateMax)-\power_{\party^*}\cdot\deposit\cdot\rateMax\big)+\\
            &+\negl(\secparam)=\\
            =&\revision{\stake_{\party^*} \cdot \rateMax} + E[\reward_{\party^*, \profile_\proto}]\cdot \rateMax+\power_{\party^*}\cdot\deposit\cdot\rateMax+\\
            &+\Pr[D]\cdot\big(\bribe_{\party^*}-\power_{\party^*}\cdot\deposit\cdot\rateMax\big)+ \negl(\secparam)\leq\\
            \leq& \revision{\stake_{\party^*} \cdot \rateMax} + E[\reward_{\party^*, \profile_\infractionPredicate^{\partySet_\text{prom}^{\{\bribe_\party\},\deposit}}}]\cdot \rateMax+\power_{\party^*}\cdot\deposit\cdot\rateMax+\\
            &+\Pr[D]\cdot 0+\negl(\secparam)\leq\\
            =&\utility_{\party^*}(\profile_\infractionPredicate^{\partySet_\text{prom}^{\{\bribe_\party\},\deposit}})+\negl(\secparam).
        \end{split}
    \end{equation*}

    \underline{Case 2. $\party^*\in\partySet_\text{prom}^{\{\bribe_\party\},\deposit}$:} 
    In this case, $\bribe_{\party^*}>\power_{\party^*}\cdot\deposit\cdot \rateMax$ while the probability that $\party^*$ performs an infraction when
    following $\profile$ is no more than when following
    $\profile_\infractionPredicate^{\partySet_\text{prom}^{\{\bribe_\party\},\deposit}}$ (by the definition of
    $\strategy_\infractionPredicate$).   Besides, the ledger rewards of
    $\party^*$ are maximized in
    $\profile_\infractionPredicate^{\partySet_\text{prom}^{\{\bribe_\party\},\deposit}}$ and the exchange rate
    is always $\rateMax$.   Let $D$  be the event that $\party^*$ performs an infraction \wrt
    $\infractionPredicate$ by following $\profile$, and $F_{\party^*}$ be the event that $\party^*$ fails to perform an infraction when following $\profile_\infractionPredicate^{\partySet_\text{prom}^{\{\bribe_\party\},\deposit}}$. Since $\Pr[F_{\party^*}]=\negl(\secparam)$, we conclude that 
    \begin{equation*}
    \begin{split}
    &\utility_{\party^*}(\profile)=\\
        =&\revision{\stake_{\party^*} \cdot E[\exchangeRate_{\profile}]} + E[\reward_{\party^*, \profile}\cdot\exchangeRate_{\profile}]+\Pr[D]\cdot\bribe_{\party^*}+\\
        &+\Pr[\neg D]\cdot\power_{\party^*}\cdot\deposit\cdot \rateMax<\\
    <& \revision{\stake_{\party^*} \cdot E[\exchangeRate_{\profile}]} + E[\reward_{\party^*, \profile}\cdot\exchangeRate_{\profile}]+\bribe_{\party^*}\leq\\
    \leq& \revision{\stake_{\party^*} \cdot \rateMax} +  E[\reward_{\party^*, \profile_\infractionPredicate^{\partySet_\text{prom}^{\{\bribe_\party\},\deposit}}}]\cdot \rateMax+\Pr[\neg F_{\party^*}]\cdot\bribe_{\party^*}+\\
    &+\Pr[F_{\party^*}]\cdot\bribe_{\party^*}\leq\\
    \leq& \utility_{\party^*}(\profile_\infractionPredicate^{\partySet_\text{prom}^{\{\bribe_\party\},\deposit}})+\negl(\secparam).
    \end{split}
    \end{equation*}
    

    Therefore, in any case, there is an $\epsilon$ negligible in $\secparam$ \st
    $\utility_{\party^*}(\profile) \leq \utility_{\party^*}(\profile_\infractionPredicate^{\partySet_\text{prom}^{\{\bribe_\party\},\deposit}}) + \epsilon$,
    \ie $\profile_\infractionPredicate^{\partySet_\text{prom}^{\{\bribe_\party\},\deposit}}$ is an
    $\epsilon$-Nash equilibrium \wrt Accountable Reward under $\router_\text{sync}$.

\end{proof}

\subsubsection{Price of Stability}\label{app:acc_pst}

\paragraph{\textsc{Theorem}~\ref{thm:accountable_protocol_pst}}
\emph{    Assume the following:
    \begin{itemize}
        \item $\proto$ is a protocol with block-proportional rewards (cf.
            Section~\ref{subsec:preliminaries_proportion}), with $\reward$
            rewards per block;
        \item \revision{each party's stake is $\stake_\party = \power_\party \cdot \stake$;}
        \item $\profile_\proto$ denotes the all-honest strategy profile;
        \item $\threshold$ is the security threshold \wrt an infraction
            predicate $\infractionPredicate$;
        \item for every strategy profile $\profile$: 
            \begin{itemize}
                \item the exchange rate $\exchangeRate_\profile$ \wrt
                    $\infractionPredicate, \threshold$ is defined as in
                    Eq.~\eqref{eq:exchange};
                \item the external rewards $\utilityBoost_{\party, \profile}$,
                    due to guided bribing \wrt $\infractionPredicate$, are
                    defined as in Eq.~\eqref{eq:external_guided};
               \item the compliance payout $ \deposit_{\party,\profile}$ \wrt $\infractionPredicate$ 
                   is defined as in Eq.~\eqref{eq:deposit};
            \end{itemize}
          \item For every party $\party \in \partySet$, it holds that
              (i) $\power_\party < \threshold$ and 
              (ii) $\beta_\party - \power_\party \cdot \deposit\cdot\rateMax \leq \negl(\secparam)$.  
    \end{itemize}
Then, it holds that $\priceStab\leq 1+\negl(\secparam)$.
}
\begin{proof}
    Let $\profile$ be an arbitrary strategy profile. We provide an upper bound on $\welfare(\profile)$. We know that there is an $\epsilon$ negligible in $\secparam$ such that for every $\party\in\partySet$, it holds that  $\beta_\party-\power_\party\cdot\deposit\cdot\rateMax\leq\epsilon$. Besides, for every execution trace $\trace$, the random variable $\reward_{\party, \profile}\cdot\exchangeRate_{\profile}+\utilityBoost_{\party, \profile}+\deposit_{\party, \profile}$ can be either (i) $\reward_{\party, \profile}(\trace)\cdot\exchangeRate_{\profile}(\trace)+\beta_\party$, if $\party$ performs an infraction, or (ii) $\reward_{\party, \profile}(\trace)\cdot\exchangeRate_{\profile}(\trace)+\deposit_{\party, \profile}(\trace)\cdot\exchangeRate_{\profile}(\trace)$, otherwise. Thus, in any case, the utiilty of $\party$ is upper bounded by $\revision{\stake_\party \cdot \rateMax} + \reward_{\party, \profile}(\trace)\cdot\rateMax+\power_\party\cdot\deposit\cdot\rateMax+\epsilon$, i.e.,
\begin{equation*}
\begin{split}
&\revision{\stake_\party \cdot \exchangeRate_{\profile}}+\reward_{\party, \profile}\cdot\exchangeRate_{\profile}+\utilityBoost_{\party, \profile}+\deposit_{\party, \profile}\leq\\
\leq&\revision{\stake_\party \cdot \rateMax}+\reward_{\party, \profile}\cdot\rateMax+\power_\party\cdot\deposit\cdot\rateMax+\epsilon\;,
\end{split}
\end{equation*}
which implies that for every $\party\in\partySet$:
\begin{equation*}
\begin{split}
    \utility_\party(\profile)&\leq \revision{\stake_\party \cdot \rateMax} + E[\reward_{\party, \profile}]\cdot\rateMax+\power_\party\cdot\deposit\cdot\rateMax+\epsilon\leq\\
 &\leq \revision{\stake_\party \cdot \rateMax} + E[\reward_{\party, \profile_\proto}]\cdot\rateMax+\power_\party\cdot\deposit\cdot\rateMax+\epsilon=\\
 &=\utility_\party(\profile_\proto)+\epsilon\;.
\end{split}
\end{equation*}
Therefore, we get that
\begin{equation}\label{eq:accountable_protocol_pst1}
\begin{split}
&\max_{\profile \in \strategySet^\totalParties} \welfare(\profile)=\sum_{\party\in\partySet}\utility_\party(\profile)\leq\\
\leq&\sum_{\party\in\partySet}(\utility_\party(\profile_\proto)+\epsilon)=\sum_{\party\in\partySet}\utility_\party(\profile_\proto)+\totalParties\cdot\epsilon\leq\\
\leq&\welfare(\profile_\proto)+\negl(\secparam)\;.
\end{split}
\end{equation}

Moreover, for every party $\party\in\partySet$, it holds that (i) $\power_\party<\threshold$ and (ii) $\beta_\party-\power_\party\cdot\deposit\cdot\rateMax\leq\negl(\secparam)$, so by Theorem~\ref{thm:accountable_protocol_nash}, we get that $\profile_\proto$ is a $\negl(\secparam)$-Nash equilibrium \wrt Accountable Reward under $\router_\text{sync}$. This fact and Eq.~\eqref{eq:accountable_protocol_pst1} imply that
\begin{equation*}
\begin{split}
 \priceStab=&\frac{\max_{\profile \in \strategySet^\totalParties} \welfare(\profile)}{\max_{\profile \in \equilibriaSet} \welfare(\profile)}\leq\\
 &\leq\frac{\welfare(\profile_\proto)+\negl(\secparam)}{\welfare(\profile_\proto)}\leq 1+\negl(\secparam).
\end{split}
\end{equation*}

\end{proof}

\paragraph{\textsc{Theorem}~\ref{thm:accountable_maximal_pst}}
  \emph{  Assume the following:
    \begin{itemize}
        \item $\proto$ is a protocol run by parties in $\partySet$ in $\rounds$
            rounds, with block-proportional rewards (cf.
            Subsection~\ref{subsec:preliminaries_proportion}) with $\reward$
            rewards per block;
        \item \revision{each party's stake is $\stake_\party = \power_\party \cdot \stake$;}
        \item $\threshold$ is a security threshold \wrt $\infractionPredicate$;
        \item for every strategy profile $\profile$:
            \begin{itemize}
                \item the exchange rate $\exchangeRate_\profile$ \wrt
                    $\infractionPredicate, \threshold$ is defined as in
                    Eq.~\eqref{eq:exchange}; 
                \item the external rewards $\utilityBoost_{\party, \profile}$
                    due to guided bribing \wrt $\infractionPredicate$ are
                    defined as in Eq.~\eqref{eq:external_guided};
                     \item the compliance payout $ \deposit_{\party,\profile}$ \wrt $\infractionPredicate$ is defined as in Eq.~\eqref{eq:deposit};
            \end{itemize}
        \item the condition  $\sum_{\party \in \partySet} \bribe_\party \leq \threshold\cdot\deposit\cdot\rateMax$ in
            Lemma~\ref{lem:acc_maximal_existence} holds, so  $\sum_{\party \in \partySet_\text{prom}^{\{\bribe_\party\},\deposit}} \power_\party < \threshold$. 
    \end{itemize}
Then, it holds that $\priceStab \leq 1 +\negl(\secparam)$.
}
\ignore{
    Then, the following hold:
    \begin{enumerate}
        \item
            $\priceStab \leq 1 +\frac{\threshold\cdot\deposit}{\rounds\cdot\reward\cdot\rateMax}$.

        \item
            Assume that $\rateMax \geq 2 \cdot \rateMin$ and that for some
            $\hat{\rounds} \geq \frac{\threshold - \frac{3}{\totalParties}}{\threshold - \frac{2}{\totalParties}} \cdot \rounds$,
            the set of conditions in Lemma~\ref{lem:acc_maximal_existence}
            holds. Then, there exist a participation power allocation
            $\{\power_\party\}_{\party \in \partySet}$ and a bribe allocation
            $\{\bribe_\party\}_{\party \in \partySet}$ \st for any constant
            $\gamma \in (0,1)$ and $\totalParties \geq 8^{\frac{1}{1 - \gamma}}$, 
            it holds that:
        \[ \priceStab \geq 1 + \Big( \threshold - \frac{1}{\totalParties^\gamma} \Big) \cdot \Big( 1 - \frac{\rateMin}{\rateMax} \Big) - \negl(\secparam)\;. \]
    \end{enumerate}
   } 
\begin{proof}
    Let $\profile$ be a strategy profile. For every party $\party\in\partySet$, the term $E[\reward_{\party, \profile}\cdot\exchangeRate_{\profile}]$ of the utility of $\party$ is upper bounded by the value  $E[\reward_{\party, \profile_\proto}]\cdot\rateMax$. \revision{Similarly, the term $\stake_\party \cdot E[\exchangeRate_{\profile}]$ is upper bounded by $\stake_\party \cdot \rateMax$.} In addition, if $\party\in\partySet_\text{prom}^{\{\bribe_\party\},\deposit}$, i.e., $\bribe_\party>\power_\party\cdot\deposit\cdot\rateMax$, then the term $E[\utilityBoost_{\party, \profile}]+E[\deposit_{\party, \profile}\cdot\exchangeRate_{\profile}]$ is upper bounded by $\bribe_\party$ (as if $\party$ always performs an infraction). On the other hand, if $\party\notin\partySet_\text{prom}^{\{\bribe_\party\},\deposit}$, then the term $E[\utilityBoost_{\party, \profile}]+E[\deposit_{\party, \profile}\cdot\exchangeRate_{\profile}]$ is upper bounded by $\power_\party\cdot\deposit\cdot\rateMax$ (as if $\party$ never performs an infraction). By the  above, the welfare of $\profile$ is upper bounded by 
\begin{equation}\label{eq:accountable_maximal_pst1}
\begin{split}
&\welfare(\profile)=\sum_{\party \in \partySet}\utility_\party(\profile)=\\
=&\sum_{\party \in \partySet_\text{prom}^{\{\bribe_\party\},\deposit}}\utility_\party(\profile)+\sum_{\party \notin \partySet_\text{prom}^{\{\bribe_\party\},\deposit}}\utility_\party(\profile)\leq\\
    \leq&\sum_{\party \in \partySet_\text{prom}^{\{\bribe_\party\},\deposit}}\big((\revision{\stake_\party} + E[\reward_{\party, \profile_\proto}])\cdot\rateMax+\bribe_\party\big)+\\
    &+\sum_{\party \notin \partySet_\text{prom}^{\{\bribe_\party\},\deposit}}\big((\revision{\stake_\party} + E[\reward_{\party, \profile_\proto}])\cdot\rateMax+\power_\party\cdot\deposit\cdot\rateMax\big)=\\
    =&\sum_{\party \in \partySet} (\revision{\stake_\party} + E[\reward_{\party, \profile_\proto}])\cdot\rateMax+\\
    &+\sum_{\party \in \partySet_\text{prom}^{\{\bribe_\party\},\deposit}}\bribe_\party+\sum_{\party \notin \partySet_\text{prom}^{\{\bribe_\party\},\deposit}}\power_\party\cdot\deposit\cdot\rateMax=\\
    =&(\revision{\stake} + \rounds\cdot\reward)\cdot\rateMax+\sum_{\party \in \partySet_\text{prom}^{\{\bribe_\party\},\deposit}}\bribe_\party+\sum_{\party \notin \partySet_\text{prom}^{\{\bribe_\party\},\deposit}}\power_\party\cdot\deposit\cdot\rateMax\;.
\end{split}
\end{equation}
    In addition, $\sum_{\party \in \partySet_\text{prom}^{\{\bribe_\party\},\deposit}} \power_\party < \threshold$ holds. So, by Theorem~\ref{thm:acc_maximal_sufficient},
    the strategy profile $\profile_\infractionPredicate^{\partySet_\text{prom}^{\{\bribe_\party\},\deposit}}$ is a
    $\negl(\secparam)$-Nash equilibrium. Moreover, the exchange rate
    when the parties follow $\profile_\infractionPredicate^{\partySet_\text{prom}^{\{\bribe_\party\},\deposit}}$ is
    always $\rateMax$. Since (i) compared to $\profile_\proto$, the number of blocks on the ledger created by each party is not affected when the parties follow  $\profile_\infractionPredicate^{\partySet_\text{prom}^{\{\bribe_\party\},\deposit}}$, and (ii) the subset of parties that perform the infraction (except from $\negl(\secparam)$ failure probability) is exactly $\profile_\infractionPredicate^{\partySet_\text{prom}^{\{\bribe_\party\},\deposit}}$, we get that
     \begin{equation}\label{eq:accountable_maximal_pst2}
     \begin{split}
     &\welfare(\profile_\infractionPredicate^{\partySet_\text{prom}^{\{\bribe_\party\},\deposit}})=\\
     =&(\revision{\stake} + \rounds\cdot\reward)\cdot\rateMax+\sum_{\party \in \partySet_\text{prom}^{\{\bribe_\party\},\deposit}}\bribe_\party+\\
     &+\sum_{\party \notin \partySet_\text{prom}^{\{\bribe_\party\},\deposit}}\power_\party\cdot\deposit\cdot\rateMax-\negl(\secparam)\;.
             \end{split}
    \end{equation}   
     
    By Eq.~\eqref{eq:accountable_maximal_pst1} and~\eqref{eq:accountable_maximal_pst2}, we conclude that:
    \begin{equation*}
        \begin{split}
            &\priceStab=\frac{\max_{\profile \in \strategySet^\totalParties} \welfare(\profile)}{\max_{\profile \in \equilibriaSet} \welfare(\profile)}\leq\\
            \leq&\tfrac{(\revision{\stake} + \rounds\cdot\reward)\cdot\rateMax+\sum_{\party \in \partySet_\text{prom}^{\{\bribe_\party\},\deposit}}\bribe_\party+\sum_{\party \notin \partySet_\text{prom}^{\{\bribe_\party\},\deposit}}\power_\party\cdot\deposit\cdot\rateMax}{(\revision{\stake} + \rounds\cdot\reward)\cdot\rateMax+\sum_{\party \in \partySet_\text{prom}^{\{\bribe_\party\},\deposit}}\bribe_\party+\sum_{\party \notin \partySet_\text{prom}^{\{\bribe_\party\},\deposit}}\power_\party\cdot\deposit\cdot\rateMax-\negl(\secparam)}=\\
            =&1+\negl(\secparam)\;.
        \end{split}
    \end{equation*}

\end{proof}


\paragraph{\textsc{Theorem}~\ref{thm:accountable_pan}}
  \emph{Assume the following:
    \begin{itemize}
        \item $\proto$: a protocol run by parties in $\partySet$ in $\rounds$ rounds with block-proportional rewards (cf.
            Section~\ref{subsec:preliminaries_proportion}), with $\reward$
            rewards per block;
        \item \revision{each party's stake is $\stake_\party = \power_\party \cdot \stake$;}
        \item $\threshold$: the security threshold \wrt an infraction predicate $\infractionPredicate$;
        \item $\profile_\infractionPredicate$: the strategy profile where all parties perform $\infractionPredicate$;
        \item for every strategy profile $\profile$: 
            \begin{itemize}
                \item the exchange rate $\exchangeRate_\profile$ \wrt
                    $\infractionPredicate, \threshold$ follows
                    Eq.~\eqref{eq:exchange};
                \item the external rewards $\utilityBoost_{\party, \profile}$,
                    due to guided bribing \wrt $\infractionPredicate$, 
                    follow Eq.~\eqref{eq:external_guided};
               \item the compliance payout $ \deposit_{\party, \profile}$ \wrt $\infractionPredicate$ follows Eq.~\eqref{eq:deposit};
            \end{itemize}
          \item For every party $\party \in \partySet$, it holds that:
              (i) $\power_\party < 1 - \threshold$;
              (ii) $\beta_\party - \power_\party \cdot \deposit \cdot \rateMin \geq \negl(\secparam)$;
              (iii)  $\beta_\party - \power_\party \cdot \deposit\cdot\rateMax \leq \negl(\secparam)$.
    \end{itemize}
    Then, it holds that $\priceAnar \geq \frac{(\revision{\stake} + \rounds\cdot\reward + \deposit)\cdot\rateMax}{(\revision{\stake} + \rounds\cdot\reward)\cdot\rateMin+\budget}- \negl(\secparam)$.
   } 
\begin{proof}
    The proof follows directly from Theorem~\ref{thm:accountable_negative_nash}.

    Since  for every party $\party \in \partySet$, it holds that $\beta_\party - \power_\party \cdot \deposit\cdot\rateMax \leq \negl(\secparam)$, as shown in the proof of Theorem~\ref{thm:accountable_protocol_pst},
    welfare is upper bounded by:
    \begin{equation*}
    \begin{split}
    \welfare(\profile_\proto) + \negl(\secparam) =\\
    \sum_{\party\in\partySet} (\revision{\stake_\party \cdot \rateMax} + E[\reward_{\party, \profile_\proto}]\cdot\rateMax +\power_\party \cdot \deposit \cdot \rateMax) + \negl(\secparam) = \\
    \rateMax \cdot (\sum_{\party\in\partySet} (\revision{\stake_\party} + E[\reward_{\party, \profile_\proto}]+\power_\party \cdot \deposit ) + \negl(\secparam)=\\
    (\revision{\stake} + \rounds\cdot\reward + \deposit)\cdot\rateMax+ \negl(\secparam).
    \end{split}
    \end{equation*}

    Second, the welfare for the (negative) equilibrium of
    Theorem~\ref{thm:accountable_negative_nash} is:
    \begin{equation*}
    \begin{split}
    \welfare(\profile_\infractionPredicate) = \sum_{\party\in\partySet}\utility_\party(\profile_\infractionPredicate) =\\
    \sum_{\party\in\partySet}\big (\revision{\stake_\party \cdot \rateMin} + E[\reward_{\party, \profile_\infractionPredicate}]\cdot\rateMin+\bribe_\party - \negl(\secparam)\big) = \\
    \sum_{\party\in\partySet} ( \revision{\stake_\party \cdot \rateMin} + E[\reward_{\party, \profile_\infractionPredicate}]\cdot\rateMin+\bribe_\party ) - \negl(\secparam) = \\
    (\revision{\stake} + \rounds\cdot\reward) \cdot\rateMin+\budget- \negl(\secparam).
    %
    %
    \end{split}
    \end{equation*}

    Therefore, the Price of Anarchy is bounded by:
        \begin{equation*}
    \begin{split}
  \priceAnar \geq 
        \frac{(\revision{\stake} + \rounds\cdot\reward + \deposit)\cdot\rateMax+ \negl(\secparam)}{ (\revision{\stake} + \rounds\cdot\reward) \cdot \rateMin + \budget - \negl(\secparam)} \geq 
        \frac{(\revision{\stake} + \rounds\cdot\reward + \deposit)\cdot\rateMax}{ (\revision{\stake} + \rounds\cdot\reward) \cdot\rateMin+\budget}- \negl(\secparam)\;.
    \end{split}
    \end{equation*}
    %
\end{proof}

\fi

\end{document}